\documentclass[iop]{aastex62}
\usepackage{longtable,graphicx}

\shorttitle{Stellar structure in M31's halo}
\shortauthors{McConnachie et al.}

\begin{document}

\title{The large-scale structure of the halo of the Andromeda galaxy II. Hierarchical structure in the Pan-Andromeda Archaeological Survey}

\author{Alan W. McConnachie}
\affiliation{NRC Herzberg Astronomy and Astrophysics, Dominion Astrophysical Observatory, 5071 West Saanich Road, Victoria, B.C., V9E 2E7, Canada}

\author{Rodrigo Ibata}
\affiliation{Universit{\'e} de Strasbourg, CNRS, Observatoire astronomique de Strasbourg, UMR 7550, F-67000 Strasbourg, France}

\author{Nicolas Martin} 
\affiliation{Universit{\'e} de Strasbourg, CNRS, Observatoire astronomique de Strasbourg, UMR 7550, F-67000 Strasbourg, France}

\author{Annette M. N. Ferguson}
\affiliation{Institute for Astronomy, University of Edinburgh, Royal Observatory, Blackford Hill, Edinburgh EH9 3HJ, UK}

\author{Michelle Collins}
\affiliation{Department of Physics, University of Surrey, Guildford, GU2 7XH, Surrey, UK}

\author{Stephen Gwyn}
\affiliation{NRC Herzberg Astronomy and Astrophysics, Dominion Astrophysical Observatory, 5071 West Saanich Road, Victoria, B.C., V9E 2E7, Canada} 

\author{Mike Irwin}
\affiliation{Institute of Astronomy, University of Cambridge, Cambridge CB3 0HA, UK}

\author{Geraint F. Lewis}
\affiliation{Sydney Institute for Astronomy, School of Physics, A28, The University of Sydney, Sydney, NSW 2006, Australia}

\author{A. Dougal Mackey}
\affiliation{Research School of Astronomy and Astrophysics, Australian National University, Canberra, ACT 2611, Australia}

\author{Tim Davidge}
\affiliation{NRC Herzberg Astronomy and Astrophysics, Dominion Astrophysical Observatory, 5071 West Saanich Road, Victoria, B.C., V9E 2E7, Canada} 

\author{Veronica Arias}
\affiliation{Departamento de F{\'i}sica, Universidad de los Andes, Cra. 1 No. 18A-10, Edificio Ip, Bogot{\'a}, Colombia}
\affiliation{Sydney Institute for Astronomy, School of Physics, A28, The University of Sydney, Sydney, NSW 2006, Australia}

\author{Anthony Conn}
\affiliation{Sydney Institute for Astronomy, School of Physics, A28, The University of Sydney, Sydney, NSW 2006, Australia}

\author{Patrick C{\^o}t{\'e}}
\affiliation{NRC Herzberg Astronomy and Astrophysics, Dominion Astrophysical Observatory, 5071 West Saanich Road, Victoria, B.C., V9E 2E7, Canada}

\author{Denija Crnojevic}
\affiliation{Texas Tech University, Physics and Astronomy Department, Box 41051, Lubbock, TX 79409}

\author{Avon Huxor}
\affiliation{HH Wills Physics Laboratory, University of Bristol, Tyndall Avenue, Bristol, BS8 1TL, UK} 

\author{Jorge Penarrubia}
\affiliation{Institute for Astronomy, University of Edinburgh, Royal Observatory, Blackford Hill, Edinburgh EH9 3HJ, UK}

\author{Chelsea Spengler}
\affiliation{Physics \& Astronomy Department, University of Victoria, 3800 Finnerty Rd, Victoria BC, Canada V8P 5C2}

\author{Nial Tanvir}
\affiliation{Department of Physics and Astronomy, University of Leicester, University Road, Leicester, LE1 7RH, UK}

\author{David Valls-Gabaud}
\affiliation{LERMA, CNRS UMR 8112, Observatoire de Paris, PSL, 61 Avenue de l'Observatoire, 75014 Paris, France} 

\author{Arif Babul}
\affiliation{Physics \& Astronomy Department, University of Victoria, 3800 Finnerty Rd, Victoria BC, Canada V8P 5C2}

\author{Pauline Barmby}
\affiliation{Faculty of Science \& Department of Physics and Astronomy, Western University, 1151 Richmond Street, London, Ontario, Canada, N6A 3K7}

\author{Nicholas F. Bate}
\affiliation{Institute of Astronomy, University of Cambridge, Cambridge CB3 0HA, UK}
\affiliation{Sydney Institute for Astronomy, School of Physics, A28, The University of Sydney, Sydney, NSW 2006, Australia}

\author{Edouard Bernard}
\affiliation{Universit{\'e} C{\^o}te d'Azur, Observatoire de la C{\^o}te d'Azur, CNRS, Laboratoire Lagrange, Bd de l'Observatoire, CS 34229, 06304 Nice cedex 4, France}

\author{Scott Chapman}
\affiliation{Department of Physics and Atmospheric Science, Dalhousie University, Halifax, NS B3H 3J5 Canada}

\author{Aaron Dotter}
\affiliation{Harvard-Smithsonian Center for Astrophysics, Cambridge, MA 02138, USA}
 
\author{William Harris}
\affiliation{Department of Physics \& Astronomy, McMaster University, Hamilton, Ontario, L8S 4K1, Canada}
 
\author{Brendan McMonigal}
\affiliation{Sydney Institute for Astronomy, School of Physics, A28, The University of Sydney, Sydney, NSW 2006, Australia}
 
\author{Julio Navarro}
\affiliation{Physics \& Astronomy Department, University of Victoria, 3800 Finnerty Rd, Victoria BC, Canada V8P 5C2}

\author{Thomas H. Puzia}
\affiliation{Institute of Astrophysics, Pontificia Universidad Cat\'olica de Chile, Av.~Vicu\~na Mackenna 4860, 7820436 Macul, Santiago, Chile}

\author{R. Michael Rich}
\affiliation{Department of Physics and Astronomy, UCLA, PAB 430 Portola Plaza, Box 951547, Los Angeles, CA 90095-1547}

\author{Guillaume Thomas}
\affiliation{NRC Herzberg Astronomy and Astrophysics, Dominion Astrophysical Observatory, 5071 West Saanich Road, Victoria, B.C., V9E 2E7, Canada}

\author{Lawrence M. Widrow}
\affiliation{Department of Physics, Engineering Physics, and Astronomy, Queen's University, Kingston, ON K7L 3N6 Canada}

\correspondingauthor{Alan W. McConnachie}
\email{alan.mcconnachie@nrc-cnrc.gc.ca}

\begin{abstract}
The Pan-Andromeda Archaeological Survey is a survey of $>400$ square degrees centered on the Andromeda (M31) and Triangulum (M33) galaxies that has provided the most extensive panorama of a $L_\star$ galaxy group to large projected galactocentric radii. Here, we collate and summarise the current status of our knowledge of the substructures in the stellar halo of M31, and discuss connections between these features. We estimate that the 13 most distinctive substructures were produced by at least 5 different accretion events, all in the last 3 or 4 Gyrs. We suggest that a few of the substructures furthest from M31 may be shells from a single accretion event. We calculate the luminosities of some prominent substructures for which previous estimates were not available, and we estimate the stellar mass budget of the outer halo of M31. We revisit the problem of quantifying the properties of a highly structured dataset; specifically, we use the OPTICS clustering algorithm to quantify the hierarchical structure of M31's stellar halo, and identify three new faint structures. M31's halo, in projection, appears to be dominated by two ``mega-structures'', that can be considered as the two most significant branches of a merger tree produced by breaking M31's stellar halo into smaller and smaller structures based on the stellar spatial clustering. We conclude that OPTICS is a powerful algorithm that could be used in any astronomical application involving the hierarchical clustering of points. The publication of this article coincides with the public release of all PAndAS data products. 
\end{abstract}

\keywords{galaxies: general -- Local Group -- galaxies: structure -- catalogs}

\section{Introduction}

It is more than  90 years since Edwin Hubble calculated a distance to the Great Andromeda Nebula from analysis of Cepheid variable stars, and in so doing crowned it as the first confirmed ``island universe'' beyond the Milky Way (\citealt{hubble1926}). It is one of the most photographed objects in the sky and its proximity and general morphological similarity to the Milky Way has ensured that it remains the subject of intense study for a broad swath of astrophysical research.

The Andromeda Galaxy (M31) is the most luminous, and possibly the most
massive, of the galaxies in the Local Group (for a good review of Milky Way mass estimates, see the discussion in \citealt{eadie2016}; for M31, see discussion in \citealt{watkins2010} and \citealt{penarrubia2014} ). In its vicinity (some 15
degrees away on the sky) is the third most luminous Local Group
member, the Triangulum Galaxy (M33). At last count, there are at least 36 other
known galaxies that make up what is in effect a sub-group of the Local
Group, and that includes a compact elliptical galaxy, three dwarf
elliptical galaxies, at least 30 dwarf spheroidal galaxies, one (and possibly two)
low mass dwarf irregular/transition-type dwarfs, and a dwarf starburst galaxy.

Images of the Andromeda Galaxy are well known, all concentrating on the photogenic disk that is home to the overwhelming fraction of the stellar mass of the galaxy. What has been less photographed is the distant, very low surface brightness surroundings of the galaxy, in large part because there is almost nothing there. But what little is there turns out to be extremely interesting, particularly with respect to understanding aspects of Andromeda's formation.

The Pan-Andromeda Archaeological Survey (PAndAS) has, at the time of writing, formed the basis for a total of 38 peer reviewed papers on the subjects of the stellar populations, structure and evolution of M31 and its satellite systems. A large number of these papers have focused on either the discovery of, or quantification of the properties of, new stellar substructures, dwarf galaxies, globular clusters and features presumably relating to the accretion history of M31. The primary purpose of the current paper is to collect these results together, to fill in a few remaining holes, and to describe, as best as possible, our current understanding of the accretion history of the halo of M31. This includes offering a new perspective on how to quantify the spatial properties of something as vast, complex and structured as the stellar halo of an $L\star$ galaxy out to nearly half of its virial radius.

This paper is structured as follows. Section 2 provides a summary of the observational strategies adopted by the M31 survey programs that contribute to the PAndAS dataset. In section 3, we provide a complete description of the data reduction and processing procedure, including a description of the data products released to the international community though interfaces provided by the Canadian Astronomical Data Center (CADC). Section 4 conducts a census of all prominent stellar sub-structures in the halo of M31, including globular clusters and dwarf galaxies as well as stellar streams and other tidal debris, and discusses possible associations between these features. We estimate the stellar mass budget of the M31 halo in Section 5. This includes the derivation of integrated properties for a few stellar substructures for which literature estimates are currently lacking. In Section 6, we re-examine the concept of stellar substructure and attempt to provide a first quantitative description of the structure of the M31 stellar halo in terms of a hierarchy based on objective spatial clustering properties. Section 7 summarises our results. In the Appendix, we provide details on the clustering algorithm that we use to identify hierarchical structures in Section 6.

Throughout, we adopt a distance to M31 of $783 \pm 25$\,kpc and to M33 of $809 \pm 24$\,kpc (\citealt{mcconnachie2005a}). At these distances, 1 degree is equivalent to 13.7\,kpc and 14.1\,kpc, respectively.

\section{Observing Strategy and Data Acquisition}

\begin{figure}
  \begin{center}
    \includegraphics[angle=270, width=8cm]{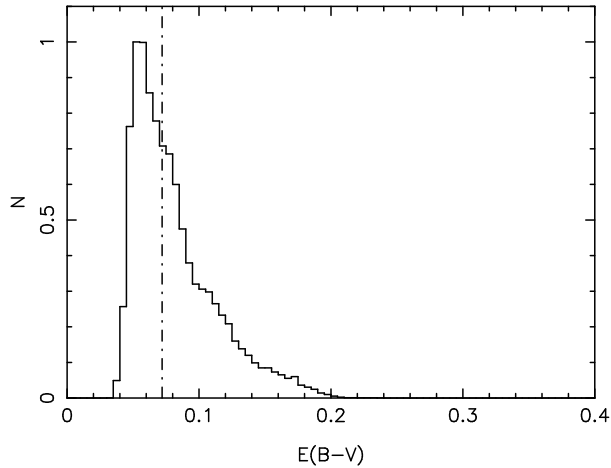}
    \caption{Distribution of the applied $E(B-V)$ values for every stellar source within the PAndAS dataset, excluding sources within the inner 2 degrees around M31 and the inner 1 degree around M33. Values are those derived by \cite{schlegel1998}, with the corrections defined by \cite{schlafly2011}.  Extinction is generally low, with a median value of 0.072\,mags, as indicated by the dot-dashed line.}\label{extinct}
  \end{center}
\end{figure}

The primary observing goal of PAndAS was to provide contiguous mapping of the resolved stellar content of the halo of M31 out to approximately half of its halo virial radius (in projection), and it ran from 2008 to 2011. Observations were planned to be sufficiently deep to provide coverage of at least the top $\sim3$ mags of the red giant branch (RGB). For M31, the virial radius is estimated to be $R_{vir} \sim 300$\,kpc (\citealt{klypin2002}).  In addition, we aimed to map M33 out to a broadly equivalent radius ($\sim 50$\,kpc), accounting for the difference in mass between these galaxies (e.g., \citealt{corbelli2003}) and so we tried to ensure that the transition region between M31 and M33 was well covered. In the north of the survey, Galactic extinction and stellar foreground contamination becomes significant due to the rapidly increasing influence of the Galactic disk. As such, we adopted an initial northern limit of $b\simeq-14$ degrees for the survey region. During the course of the survey, the decision was made to obtain some data further to the north to more fully map the region surrounding the NGC147/185 subgroup, and also to extend the survey southward to include the area surrounding the dwarf galaxy Andromeda XIV (see \citealt{majewski2007}). The histogram in Figure~\ref{extinct} shows the $E(B-V)$ values for all stellar sources in the PAndAS dataset, excluding sources within 2 degrees of M31 and within 1 degree of M33. Values are those derived by \cite{schlegel1998}, with the corrections defined by \cite{schlafly2011}. The median $E(B-V)$ in PAndAS measured in this way is 0.072\,mags. In what follows, the subscript ``$_0$'' indicates extinction-corrected magnitudes and colors.

\begin{figure}
  \begin{center}
    \includegraphics[angle=270, width=8cm]{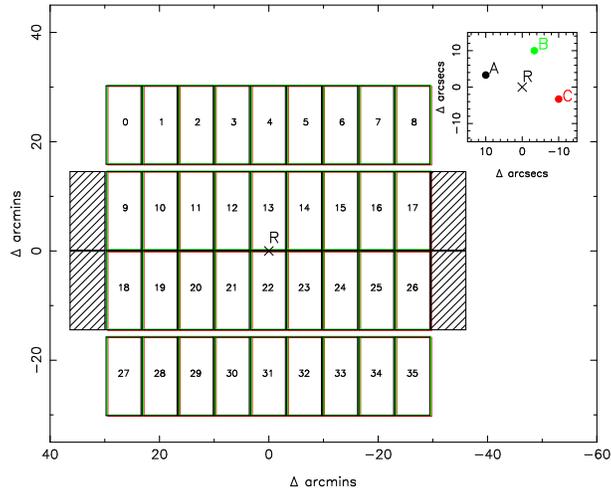}
    \caption{The CFHT/MegaCam footprint showing the layout of the 40 CCDs and their numbering system. The four CCDs that form the "ears" of MegaCam (hatched) were not in use during the era of PAndAS. Most PAndAS fields consist of a dithered set of three exposures, with the relative positions indicated by the black, green, and red footprints. The inset panel shows a zoom-in of the relative positions of the centers of the three dithered exposures (A, B and C) relative to the reference position, R.}\label{megacam}
  \end{center}
\end{figure}

PAndAS used the CFHT/MegaCam wide field camera (\citealt{boulade2003}). MegaCam consists of 40, 2048 $\times$ 4612 pixel, CCDs with a pixel scale of 0.185"/pixel, arranged in the geometry shown in Figure~\ref{megacam}. Prior to 2015, and for the entire duration of the observations discussed in this article, the four CCDs that form the ``ears'' of MegaCam (hatched pattern in Figure~\ref{megacam}) were not in use, resulting in an effective rectangular field of view of $0.96^\circ \times 0.94^\circ$. 

\begin{figure}
  \begin{center}
    \includegraphics[angle=270, width=8cm]{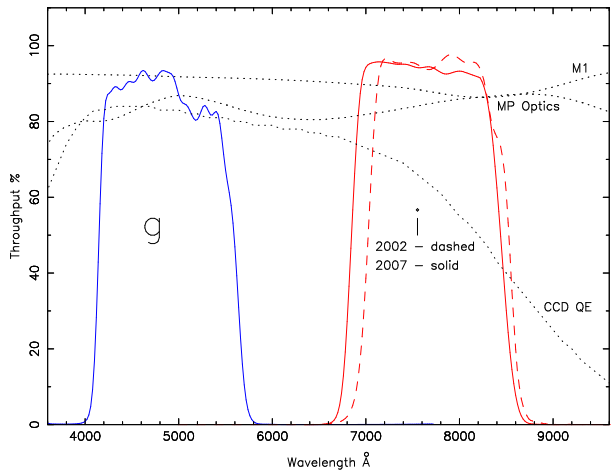}
    \caption{The filter transmission curves for the CFHT/MegaCam filters used during the course of PAndAS. Due to an accident at CFHT in 2007, the original $i-$band filter (dashed red line) was replaced with a new filter with very similar but not identical transmission (solid red line). Also shown as dotted lines are the reflectivity of the primary (M1), the transmission of the MegaPrime optics (MP optics) and the average quantum efficiency of the CCDs (CCD QE), each of which is a significant contributor to the overall effective transmission in each band. Note that the MegaCam filters were all replaced in 2014, and that the filters shown in this plot are no longer in standard use at the telescope.}\label{filters}
  \end{center}
\end{figure}

PAndAS employed $g$ and $i-$band filters in order to provide good color discrimination of RGB stars. All broad-band MegaCam filters were replaced in 2014 with physically larger filters (leading to the use of all 40 MegaCam CCDs); the filters that PAndAS used are therefore no longer in regular operation at CFHT. In addition, an accident at CFHT in 2007 resulted in the original MegaCam $i-$band filter being damaged, and so some of our earlier observations used a slightly different $i-$filter than the bulk of the observations that contribute to the final dataset. The filter transmission curves of the relevant $g-$band filter (CFHT filter ID 9401) and the two relevant $i-$band filters (pre-2007 - CFHT filter ID 9701; post-2007 - CFHT filter ID 9702) for PAndAS are shown in Figure~\ref{filters}, along with the throughputs and quantum efficiency of the significant optics and CCDs. 

\begin{figure*}
  \begin{center}
    \includegraphics[angle=0, width=18cm]{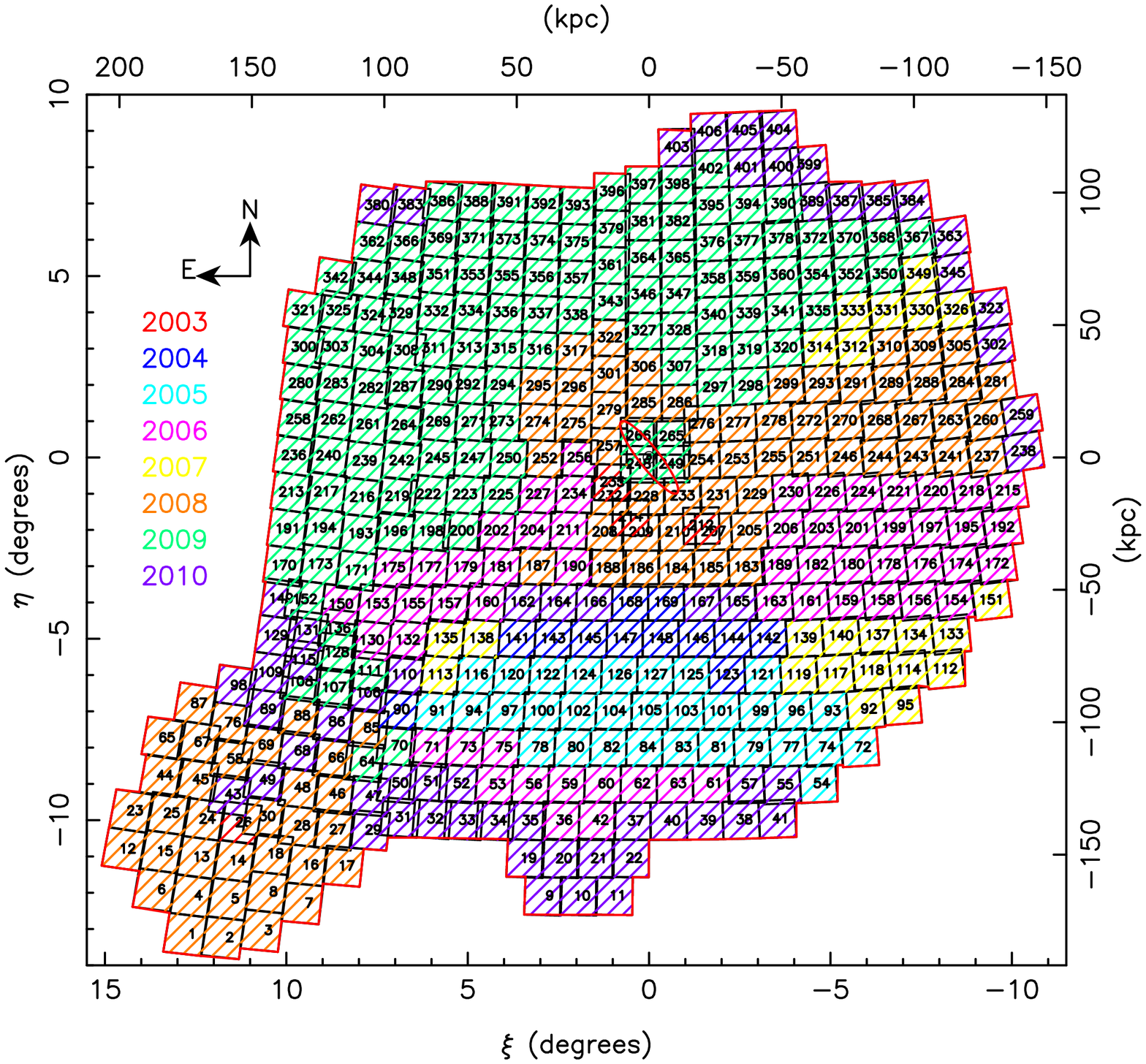}
    \caption{Tangent plane projection centered on M31 showing all individual CFHT/MegaCam pointings that contribute to the PAndAS footprint. These have been color-coded by the year in which they were observed. North is to the top and east is to the left. The inner red ellipse represents a disk of inclination $77^\circ$ and radius $1.25^\circ$ (17kpc), the approximate edge of the ``classical'' regular M31 stellar disk. Duplicate of Figure 1 from Paper I. }\label{fields}
  \end{center}
\end{figure*}

The PAndAS Large Program ran from Semester 2008B to Semester 2010B (B semesters only) as a queue program,  with observations ending formally at the end of Semester 2010B in January 2011. However, the Large Program built upon several precursor programs obtained through normal observing programs, and the final dataset also includes some archival data. The final PAndAS dataset is therefore contributed to by several observing programs from 2003 -- 2010. Figure~\ref{fields} is a copy of Figure~1 from \cite{ibata2014a} (hereafter Paper I). It shows a tangent plane projection of the region surrounding M31, centered on the galaxy, and with the locations of every PAndAS field that contributes to the final dataset. These are color-coded by the year in which they were obtained. The overlap between adjacent fields is typically of order 0.1 degrees. Each observing program contributing to the dataset is listed in Table~\ref{programs}, along with notes indicating the P.I.s and any changes from the baseline exposure times. 

\begin{table}
\begin{center}
\begin{tabular*}{0.48\textwidth}{cccl}
\hline
Program I.D. & Semester & $N_{exp}$ & Notes\\
\hline
10BP01 &2010B  &428 &McConnachie\\
10BP02 &2010B  &262& McConnachie\\
09BP01 &2009B  &303& McConnachie\\
09BP02 &2009B  &270 &McConnachie\\
08BP01 &2008B  &308& McConnachie\\
08BP02 &2008B  &275 &McConnachie\\
07BC02 &2007B  &151& McConnachie\\
06BC17 &2006B  &160& McConnachie\\
06BF37 &2006B  &125 &Ibata\\
06BF99 &2006B  &57 & Ibata\\
05BF48 &2005B  &220 &Ibata, 289s\\
05BF99 &2005B  &162 &Ibata, 289s\\
04BF20 &2004B  &160 &Ibata, 289s\\
04BF26 &2004B  &32 & Beaulieu, 480s (g), 600s (i)\\
04BH20& 2004B & 17 & Hodapp, 500s (g)\\
04BH98& 2004B & 15 & Hodapp, 500s (i)\\
03BF15 &2003B  &23  &Beaulieu, 530s (g), 660s (i)\\
03BC20 &2003B  &17 & Guhathakurta, 1160s\\
\hline
\end{tabular*}
\caption{List of observing programs that contribute to the final PAndAS dataset, including both the Large Program allocations and the precursor survey programs. Some archival programs that overlap with the PAndAS area in the appropriate filters are also incorporated into the final dataset.}\label{programs}
\end{center}
\end{table}

For the Large Program, individual exposures per field were 1350 seconds in $g$ and $i$, split as three subexposures of 450 seconds each. The black, green and red outlines in Figure~\ref{megacam} show the three-point dithering pattern that was adopted. The inset panel shows a zoom-in of the central regions of this dither pattern, where A, B and C show the positions of each subexposure relative to the central reference point R. This pattern is sufficient to cover the small gaps between each CCD, but does not cover the large gaps between the CCDs in rows 1 and 2, and between the CCDs in rows 3 and 4. Some exposures were repeated if the original observing conditions were poor, and all viable exposures contribute to the final dataset. 

\begin{figure}
  \begin{center}
    \includegraphics[angle=270, width=8cm]{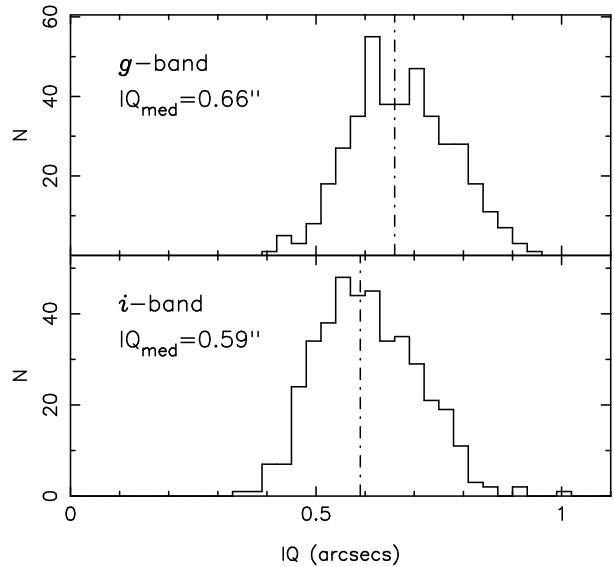}
    \caption{Distribution of the median measured image quality for all PAndAS exposures for the $g-$band (upper panel) and $i-$band (lower panel). The medians of the distributions are 0.66'' and 0.59'' for the $g-$ and $i-$band, respectively, and are indicated by the vertical dot-dashed lines.}\label{IQ}
  \end{center}
\end{figure}

The image quality of all PAndAS observations is excellent. Figure~\ref{IQ} shows the distribution of the measured median IQ for each sub-exposure, split for the $g$ and $i$ bands. The median IQ values are indicated by the dashed lines, and are $0.66"$ and $0.59"$ (85th percentile values of $0.79"$ and $0.71"$), for the $g$ and $i$ bands, respectively.

Table~\ref{obslog} reports, for each exposure contributing to the final PAndAS stacks and catalogs: the epoch of its observation (start date, start time, start MJD); the position in celestial coordinates; the CFHT/MegaCam exposure ID (odometer number); the field ID in the final PAndAS naming scheme shown in Figure~\ref{fields}; the original field ID; the relevant observing program (see Table~\ref{programs}); the filter used; the exposure time in seconds. In total, there are 2985 exposures contributing to the PAndAS stacks/catalogs, and a full version of this table is available in the online version of the journal.

\begin{table*}
\begin{center}
\begin{tabular*}{0.99\textwidth}{ccccccccccc}
\hline
Date-start & UT-start &MJD-start   &  R.A. & Dec. & Exp. ID & Field ID & Other & Prog. ID & Filter & Exp. (s)\\
\hline
2003-08-22 & 11:29:38.27 & 52873.4789152 & 1:33:51.00 & 30:39:36.8 & 714745 & m026 & M33C & 03BF15 & g.MP9401 & 530 \\ 
2003-08-22 & 11:53:39.54 & 52873.4955965 & 1:33:51.00 & 30:39:36.8 & 714747 & m026 & M33C & 03BF15 & i.MP9701 & 660 \\ 
2003-08-24 & 12:51:34.76 & 52875.5358189 & 1:33:50.99 & 30:39:37.0 & 715147 & m026 & M33C & 03BF15 & g.MP9401 & 530 \\ 
2003-08-24 & 13:15:17.02 & 52875.5522804 & 1:33:50.99 & 30:39:36.9 & 715149 & m026 & M33C & 03BF15 & i.MP9701 & 660 \\ 
2003-08-24 & 13:28:28.93 & 52875.5614459 & 1:33:51.00 & 30:39:37.0 & 715150 & m026 & M33C & 03BF15 & g.MP9401 & 530 \\ 
2003-08-24 & 13:55:05.14 & 52875.5799206 & 1:33:51.00 & 30:39:36.9 & 715152 & m026 & M33C & 03BF15 & i.MP9701 & 660 \\ 
2003-08-29 & 10:05:23.22 & 52880.4204076 & 1:33:51.00 & 30:39:36.9 & 715706 & m026 & M33C & 03BF15 & g.MP9401 & 171 \\ 
2003-08-30 & 10:38:17.82 & 52881.4432618 & 1:33:51.00 & 30:39:37.0 & 715797 & m026 & M33C & 03BF15 & g.MP9401 & 530 \\ 
2003-08-30 & 11:02:04.54 & 52881.4597748 & 1:33:51.00 & 30:39:37.0 & 715799 & m026 & M33C & 03BF15 & i.MP9701 & 660 \\ 
2003-08-31 & 9:52:31.48 & 52882.4114754 & 1:33:51.00 & 30:39:37.1 & 715935 & m026 & M33C & 03BF15 & g.MP9401 & 530 \\ 
2003-08-31 & 10:16:16.05 & 52882.4279635 & 1:33:51.00 & 30:39:37.0 & 715937 & m026 & M33C & 03BF15 & i.MP9701 & 660 \\ 
2003-08-31 & 10:29:18.25 & 52882.4370168 & 1:33:50.99 & 30:39:37.1 & 715938 & m026 & M33C & 03BF15 & g.MP9401 & 530 \\ 
2003-08-31 & 10:53:05.62 & 52882.4535373 & 1:33:51.00 & 30:39:36.8 & 715940 & m026 & M33C & 03BF15 & i.MP9701 & 660 \\ 
2003-09-03 & 13:50:37.71 & 52885.5768254 & 1:33:51.00 & 30:39:37.1 & 716134 & m026 & M33C & 03BF15 & g.MP9401 & 530 \\ 
2003-09-03 & 14:14:25.08 & 52885.5933459 & 1:33:51.00 & 30:39:36.9 & 716136 & m026 & M33C & 03BF15 & i.MP9701 & 660 \\ 
2003-09-23 & 8:06:13.32 & 52905.3376542 & 0:35:17.27 & 39:22:09.4 & 718878 & m212 & M31-3 & 03BC20 & g.MP9401 & 1160 \\ 
2003-09-23 & 8:26:23.70 & 52905.3516632 & 0:35:17.70 & 39:21:54.4 & 718879 & m212 & M31-3 & 03BC20 & g.MP9401 & 1160 \\ 
2003-09-23 & 8:46:34.08 & 52905.3656722 & 0:35:18.56 & 39:22:04.4 & 718880 & m212 & M31-3 & 03BC20 & g.MP9401 & 1160 \\ 
2003-09-23 & 9:49:09.72 & 52905.4091403 & 0:35:17.70 & 39:21:54.3 & 718884 & m212 & M31-3 & 03BC20 & i.MP9701 & 1160 \\ 
2003-09-23 & 10:09:20.00 & 52905.4231481 & 0:35:18.56 & 39:22:04.4 & 718885 & m212 & M31-3 & 03BC20 & i.MP9701 & 1160 \\ 
2003-09-24 & 8:24:29.27 & 52906.3503388 & 0:48:03.62 & 40:34:09.5 & 719008 & m235 & M31-1 & 03BC20 & g.MP9401 & 1160 \\ 
2003-09-24 & 8:46:32.44 & 52906.3656533 & 0:48:04.06 & 40:33:54.2 & 719009 & m235 & M31-1 & 03BC20 & g.MP9401 & 1160 \\ 
2003-09-24 & 9:08:41.62 & 52906.3810373 & 0:48:04.93 & 40:34:04.3 & 719010 & m235 & M31-1 & 03BC20 & g.MP9401 & 1160 \\ 
2003-09-24 & 9:30:00.60 & 52906.3958402 & 0:48:03.62 & 40:34:09.4 & 719011 & m235 & M31-1 & 03BC20 & i.MP9701 & 1160 \\ 
2003-09-24 & 9:51:48.72 & 52906.4109806 & 0:48:04.06 & 40:33:54.3 & 719012 & m235 & M31-1 & 03BC20 & i.MP9701 & 1160 \\ 
2003-09-24 & 10:13:24.60 & 52906.4259791 & 0:48:04.93 & 40:34:04.3 & 719013 & m235 & M31-1 & 03BC20 & i.MP9701 & 1160 \\ 
2003-09-24 & 10:34:48.53 & 52906.4408394 & 0:45:23.97 & 39:37:09.4 & 719014 & m214 & M31-2 & 03BC20 & g.MP9401 & 1160 \\ 
2003-09-24 & 10:54:59.21 & 52906.4548519 & 0:45:24.40 & 39:36:54.4 & 719015 & m214 & M31-2 & 03BC20 & g.MP9401 & 1160 \\ 
2003-09-24 & 11:15:10.44 & 52906.4688708 & 0:45:25.27 & 39:37:04.4 & 719016 & m214 & M31-2 & 03BC20 & g.MP9401 & 1160 \\ 
2003-09-28 & 12:22:23.21 & 52910.5155464 & 0:45:23.97 & 39:37:09.4 & 719867 & m214 & M31-2 & 03BC20 & i.MP9701 & 1160 \\ 
2003-09-28 & 13:02:53.52 & 52910.543675 & 0:45:24.40 & 39:36:54.5 & 719868 & m214 & M31-2 & 03BC20 & i.MP9701 & 1160 \\ 
2003-09-28 & 13:23:05.75 & 52910.5577054 & 0:45:25.27 & 39:37:04.4 & 719869 & m214 & M31-2 & 03BC20 & i.MP9701 & 1160 \\ 
2003-10-19 & 6:28:05.80 & 52931.2695116 & 1:33:51.00 & 30:39:36.9 & 723478 & m026 & M33C & 03BF15 & g.MP9401 & 530 \\ 
2003-10-19 & 6:51:53.12 & 52931.2860315 & 1:33:51.00 & 30:39:37.1 & 723480 & m026 & M33C & 03BF15 & i.MP9701 & 660 \\ 
2003-10-19 & 7:05:53.42 & 52931.2957572 & 1:33:51.00 & 30:39:37.0 & 723481 & m026 & M33C & 03BF15 & g.MP9401 & 530 \\ 
2003-10-19 & 7:29:39.84 & 52931.3122667 & 1:33:51.00 & 30:39:37.0 & 723483 & m026 & M33C & 03BF15 & i.MP9701 & 660 \\ 
2003-10-19 & 7:42:52.55 & 52931.3214415 & 1:33:51.00 & 30:39:37.0 & 723484 & m026 & M33C & 03BF15 & g.MP9401 & 530 \\ 
2003-10-19 & 9:41:26.24 & 52931.4037759 & 1:33:51.00 & 30:39:37.0 & 723487 & m026 & M33C & 03BF15 & i.MP9701 & 660 \\ 
2003-10-19 & 9:54:46.99 & 52931.4130439 & 1:33:51.00 & 30:39:37.0 & 723488 & m026 & M33C & 03BF15 & g.MP9401 & 530 \\ 
2003-10-19 & 10:18:34.26 & 52931.4295632 & 1:33:51.00 & 30:39:36.9 & 723490 & m026 & M33C & 03BF15 & i.MP9701 & 660 \\ 
2004-08-15 & 11:12:58.51 & 53232.4673439 & 1:33:50.80 & 30:39:37.0 & 757859 & m026 & M33 & 04BH98 & i.MP9701 & 500 \\ 
2004-08-15 & 11:22:12.88 & 53232.4737602 & 1:33:49.09 & 30:39:07.0 & 757860 & m026 & M33 & 04BH98 & i.MP9701 & 500 \\ 
2004-08-15 & 11:31:28.80 & 53232.4801945 & 1:33:51.42 & 30:38:06.9 & 757861 & m026 & M33 & 04BH98 & i.MP9701 & 500 \\ 
2004-08-15 & 11:40:43.12 & 53232.4866102 & 1:33:52.50 & 30:40:07.0 & 757862 & m026 & M33 & 04BH98 & i.MP9701 & 500 \\ 
2004-08-15 & 11:49:57.79 & 53232.49303 & 1:33:50.18 & 30:41:07.0 & 757863 & m026 & M33 & 04BH98 & i.MP9701 & 500 \\ 
2004-08-15 & 11:59:12.76 & 53232.4994532 & 1:33:49.64 & 30:40:37.0 & 757864 & m026 & M33 & 04BH98 & i.MP9701 & 500 \\ 
2004-08-15 & 12:08:27.27 & 53232.5058712 & 1:33:51.96 & 30:38:37.0 & 757865 & m026 & M33 & 04BH98 & i.MP9701 & 500 \\ 
2004-08-15 & 12:24:50.82 & 53232.5172548 & 1:33:50.80 & 30:39:37.0 & 757867 & m026 & M33 & 04BH20 & g.MP9401 & 500 \\ 
2004-08-15 & 12:34:05.99 & 53232.5236804 & 1:33:49.09 & 30:39:07.0 & 757868 & m026 & M33 & 04BH20 & g.MP9401 & 500 \\ 
\hline
\end{tabular*}
\caption{Observing log for all of the 2985 CFHT/MegaCam exposures contributing to the final PAndAS stacks and catalogs. The complete table is available in the on-line journal.}
\label{obslog}
\end{center}
\end{table*}

\section{Data Processing and Calibration}

\subsection{Reprocessing of the Data}

\begin{figure*}
  \begin{center}
    \includegraphics[angle=90, width=15cm]{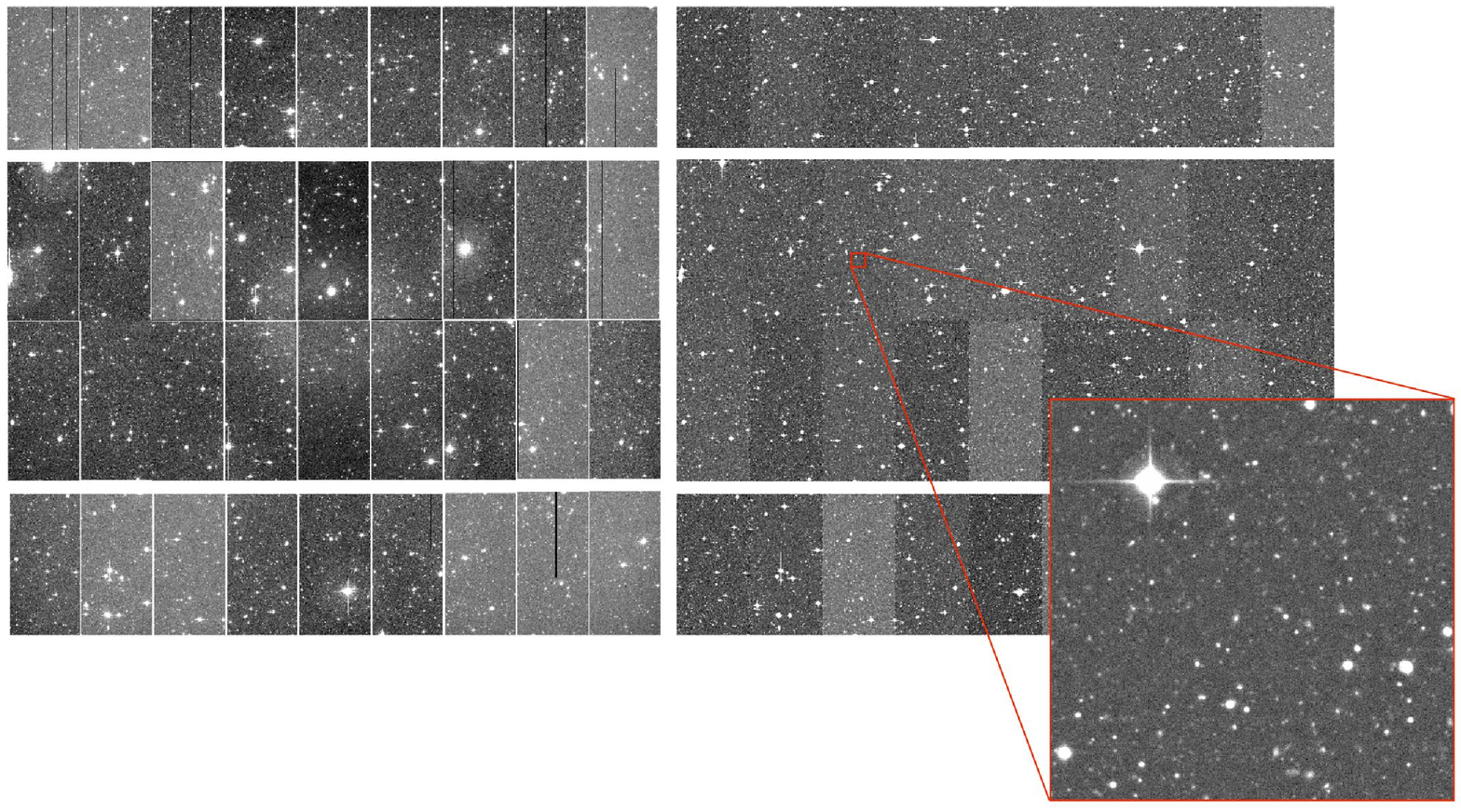}
    \caption{Left: WCS mosaic of a typical MegaCam exposure (1022197p, $g$-band) more than 2 degrees from the center of M31, processed with Elixir. This image shows all 36 CCDs, the gaps between the detectors, and scattered light effects due to bright stars and MegaCam optics. Right: WCS mosaic of the same field (m232) after stacking and additional processing, as described in the text. The dithering pattern removes the small gaps between the CCDs but leaves the large gaps. The zoom-in panel shows a $\sim 2$\,square arcmin segment of the field and highlights the large number of stellar sources that are resolved.}\label{m232_image}
  \end{center}
\end{figure*}

We have taken the opportunity afforded by the advent of the Pan-STARRS DR1 (\citealt{flewelling2016}) and Gaia DR1 (\citealt{brown2016}) surveys to re-reduce the PAndAS dataset with the aim to recalibrate the astrometry onto the excellent Gaia system, and to use the Pan-STARRS photometry to correct for large-scale zero-point uncertainties in our survey. Otherwise, the reduction steps were similar to those described in Paper~I. The images are first pre-processed with the Elixir pipeline (\citealt{magnier2004}), which removes the bias and flat-fields the mosaic. The left panel of Figure~\ref{m232_image} shows a single PAndAS $g$-band exposure (1022197p, which contributes to field m232) with Elixir processing. We employed the Cambridge Astronomical Survey Unit (CASU) pipeline (\citealt{irwin2004}) to identify sources on the images. In an initial pass, we set the detection threshold at $10\sigma$, and selected point-like sources to derive astrometric solutions with respect to Gaia on a chip-by-chip basis. Typically, the residuals were better than $0.02$~arcsec rms. The astrometrically-recalibrated images were then combined with the CASU software: this performs a seeing-weighted sum of the images, where deviant pixels are rejected based on their statistical difference with respect to the median. When stacking the images, we choose to not remove the background pedestal level for each CCD so that the stacked image retains this information. The right panel of Figure~\ref{m232_image} shows the stacked version of the same field as in the left panel, where the scattered light effects have been removed on individual frames through additional processing involving an iterative mapping and correction of residuals relative to the stack. 

\begin{figure*}
  \begin{center}
    \includegraphics[angle=270, width=18cm]{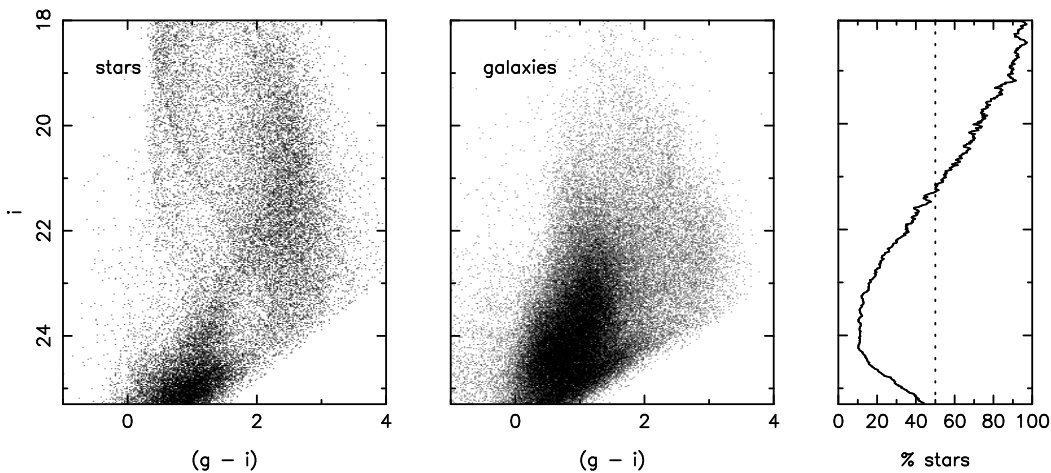}
    \caption{Left panel: CMD of all high-confidence PAndAS stellar sources in a single, unremarkable, MegaCam field (m247) located in the eastern halo of M31, approximately half-way to the survey boundaries. Middle panel: Corresponding CMD of all PAndAS extended sources (galaxies) for the same field. Right panel: Fraction of all sources identified as stellar as a function of $i-$band magnitude. Notice that the majority of detected sources fainter than $i \gtrsim 21$ are not stellar. The turnover at $i \sim 24$ is due to an increasingly large number of galaxies being misidentified as stars at faint magnitudes.}\label{sources}
  \end{center}
\end{figure*}

The source detection was then re-run with a $3\sigma$ threshold on a summed ${\rm g+i}$ frame, and we re-measured the photometry at the locations of the resulting source list on the (stacked) $g$ and $i$-band frames separately (i.e. we performed forced-photometry).  The zoom-in panel of Figure~\ref{m232_image} shows the large number of sources, even in this field that is not close to the center of M31. In each band, the sources were then classified into point sources, extended objects or noise using the CASU software. This classification is based on the flux contained within various apertures, where stellar sources form a tight locus. Figure~\ref{sources} shows a color-magnitude diagram (CMD) of objects confidently classified as stars in the left panel, and objects confidently classified as galaxies in the middle panel. This field (m247) is located approximately 5 degrees from the center of M31. It is worth stressing the critical importance of star -- galaxy classification, as illustrated in the right panel of Figure~\ref{sources}; even at $i \simeq 21$, approximately half of all objects in the field are background galaxies.

\begin{figure}
  \begin{center}
    \includegraphics[angle=270, width=10cm]{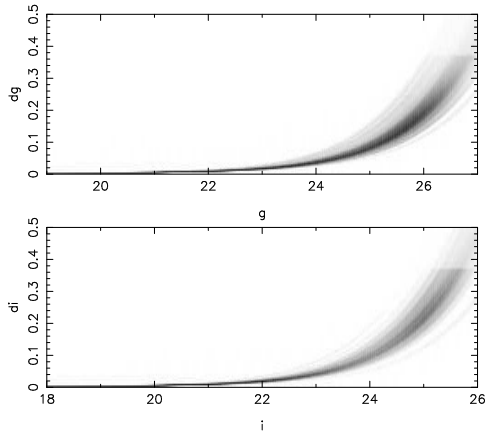}
    \caption{Photometric uncertainties as a function of magnitude for PAndAS stellar sources identified in the $g-$band (top panel) and $i-$band (bottom panel), excluding sources within the inner 2 degrees around M31 and the inner 1 degree around M33.}\label{errors}
  \end{center}
\end{figure}

Finally, the data were calibrated onto the Pan-STARRS photometric system. Figure~\ref{errors} shows the photometric uncertainties as a function of magnitude for PAndAS stellar sources identified in the $g-$band (top panel) and $i-$band (bottom panel), excluding sources within the inner 2 degrees around M31 and the inner 1 degree around M33. Generally, our photometric errors are better than 0.1\,mags at $g \simeq 25, i \simeq 24$.

\subsection{Data release}

\begin{figure*}
  \begin{center}
    \includegraphics[angle=270, width=15cm]{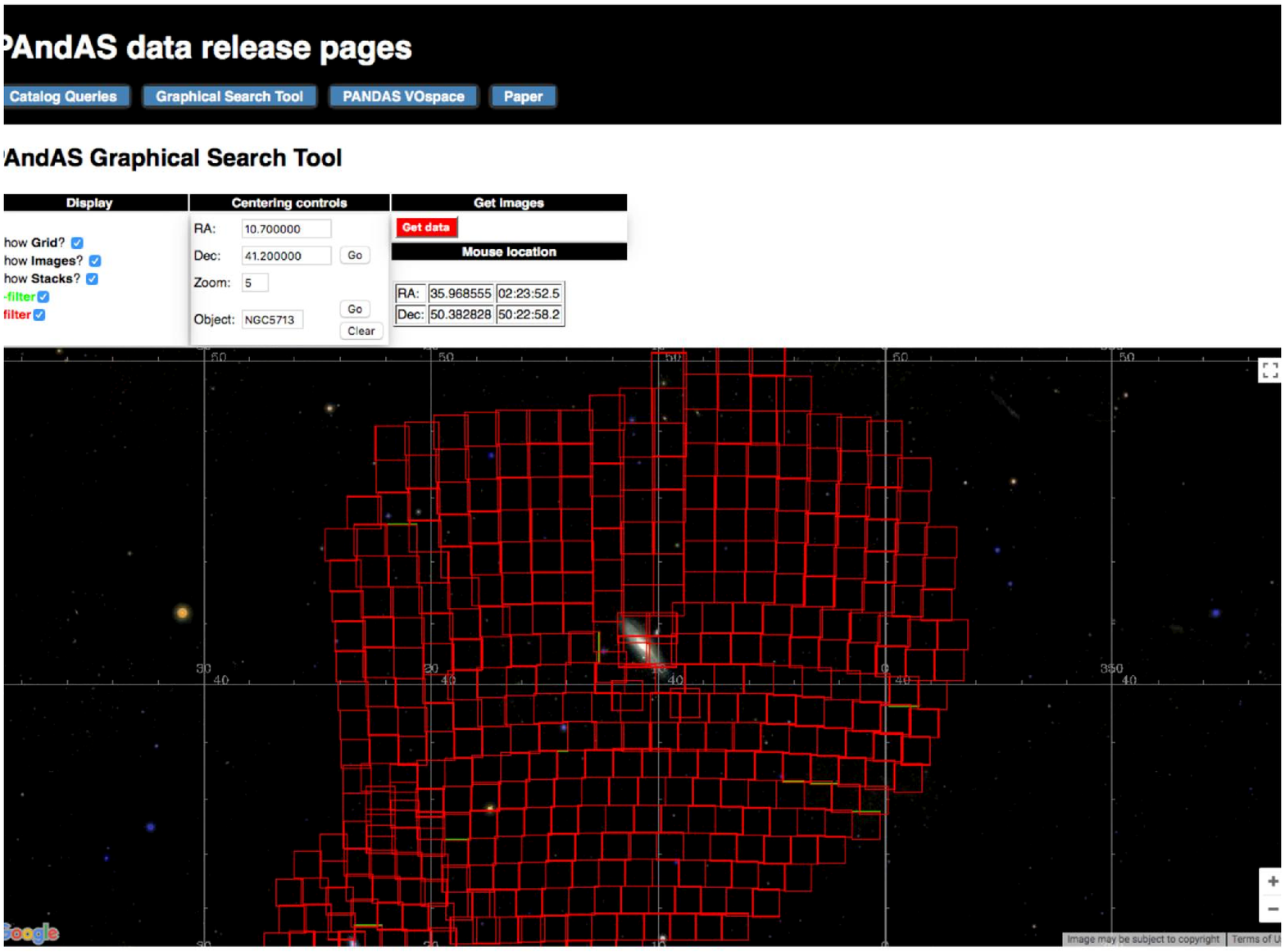}
    \caption{Screenshot of the main portal into the PAndAS data release pages at the Canadian Astronomical Data Center (CADC; http://www.cadc-ccda.hia-iha.nrc-cnrc.gc.ca/en/community/pandas/query.html) . A graphical interface allows easy interaction with the dataset. Direct queries of the combined stellar catalog are also possible, and the entire dataset is available for direct download from the PAndAS VOspace. These options are all linked off the main homepage.}\label{interface}
  \end{center}
\end{figure*}

The publication of this paper coincides with the public data release of all relevant PAndAS data products to the international community. These data are made available with the support of the CADC and can be accessed via the web at http://www.cadc-ccda.hia-iha.nrc-cnrc.gc.ca/en/community/pandas/query.html.

Specifically, the following data products are available:

\begin{enumerate}
\item Individual Exposures: all individual MegaCam exposures, processed via Elixir, in the $g-$ and $i-$bands (2985 exposures, corresponding to each of the individual rows of Table~2);
\item Stacked Images: stacked exposures in each of the $g-$ and $i-$ bands for all 406 fields in PAndAS (corresponding to each of the fields shown in Figure~4 and referenced by the ``Field ID'' in Table~2;
\item Individual Object Catalogs: combined $gi$ object catalogs for each of the 406 PAndAS fields, including astrometry, photometry and morphological classification (98,914,490 objects);
\item Combined Stellar Catalog: stellar object catalog for the entire survey, combining all 406 fields with duplicate objects removed, and including all objects classified as potentially stellar in at least one band (38,372,664 objects).
\end{enumerate}  

The main portal is via a graphical interface, shown in Figure~\ref{interface}. This interface allows the user to conduct coordinates-based searches of the dataset. It is also possible to search the combined stellar catalog directly via a catalog-query. Finally, the entire dataset can be accessed in the PAndAS VOspace for direct download. All of these download options are linked off the main graphical interface page.

\subsubsection{Individual Exposures}

All individual exposures correspond to the output of the Elixir preprocessing pipeline that is standard for CFHT/MegaCam, and with no additional processing. As such, all pixel values are in their native bands (i.e.,  no transformation has been applied at the pixel level to convert the pre-2007 $i-$filter, i.MP9701, to the post-2007 $i$-filter, i.MP9702) and none of the exposures have been corrected on a pixel level for the photometric corrections derived from PS1 as discussed in Section 3.1. 

\subsubsection{Stacked Images}

Stacked images corresponding to the Elixir-processed individual frames. The stacks have not been corrected on a pixel level for the photometric corrections derived from PS1 as discussed in Section 3.1, and all pixel values are in their native bands.

\subsubsection{Individual Object Catalogs}

The catalog photometry has been transformed into the post-2007 i-band (i.MP9702), and has been corrected using the PS1 data for the effect described in Section 3.1. All sources identified by the CASU software are listed. Columns are:

\begin{itemize} 
\item Right Ascension (hexadesimal), declination (hexadecimal);
\item $i_{CCD}$ (MegaCam CCD number from 1 -- 36);
\item $x_g, y_g$ (x,y position on CCD in pixels, measured in the $g-$band);
\item $g, dg$ ($g-$band photometry);
\item $i_g$ (morphology flat in the $g-$band);
\item $x_i, y_i$ (x,y position on CCD in pixels, measured in the $i-$band);
\item $i, di$ ($i-$band photometry);
\item $i_i$ (morphology flat in the $i-$band);
\item $ia$ (redundant);
\item Field ID.
\end{itemize}

For the morphology flags, negative values indicate a point source, positive vales indicate an extended source, and ``0'' indicates a noise source. ``-1'' indicates a source that is within 1-sigma of the locus of point sources (i.e., very probably a point source), and ``-2'' indicates a source that is within 2-sigma of the locus of point sources (i.e., probably a point source). Similarly for extended sources with positive values of the morphology flag.

\subsubsection{Combined Stellar Catalog}

The combined stellar catalog merges all the individual object catalogs and removes duplicates. Further, only sources that are classified as stellar (``-1'' or ``-2'') in at least one band are listed. Column headings refer to the columns as described for the individual fields.

\section{A census of M31 stellar halo substructure}

\begin{figure}
  \begin{center}
    \includegraphics[angle=270, width=10cm]{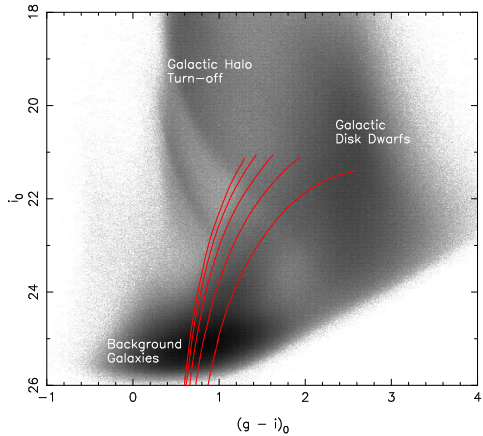}
    \caption{$i_0$ vs $(g-i)_0$ CMD of all PAndAS stellar sources exluding the inner 2 degrees around M31 and the inner 1 degree around M33, shown as a Hess diagram with logarithmic scaling. The loci of various foreground and background contamination sources are indicated with the white text. Isochrones corresponding to a 13Gyr stellar population at the distance of M31, with [Fe/H]$ = -2.5, -2.0, -1.5, -1.0, -0.5$\,dex, are overlaid (from left to right; \citealt{dotter2007,dotter2008}).}\label{cmds}
  \end{center}
\end{figure}

\begin{figure*}
  \begin{center}
    \includegraphics[angle=270, width=18cm]{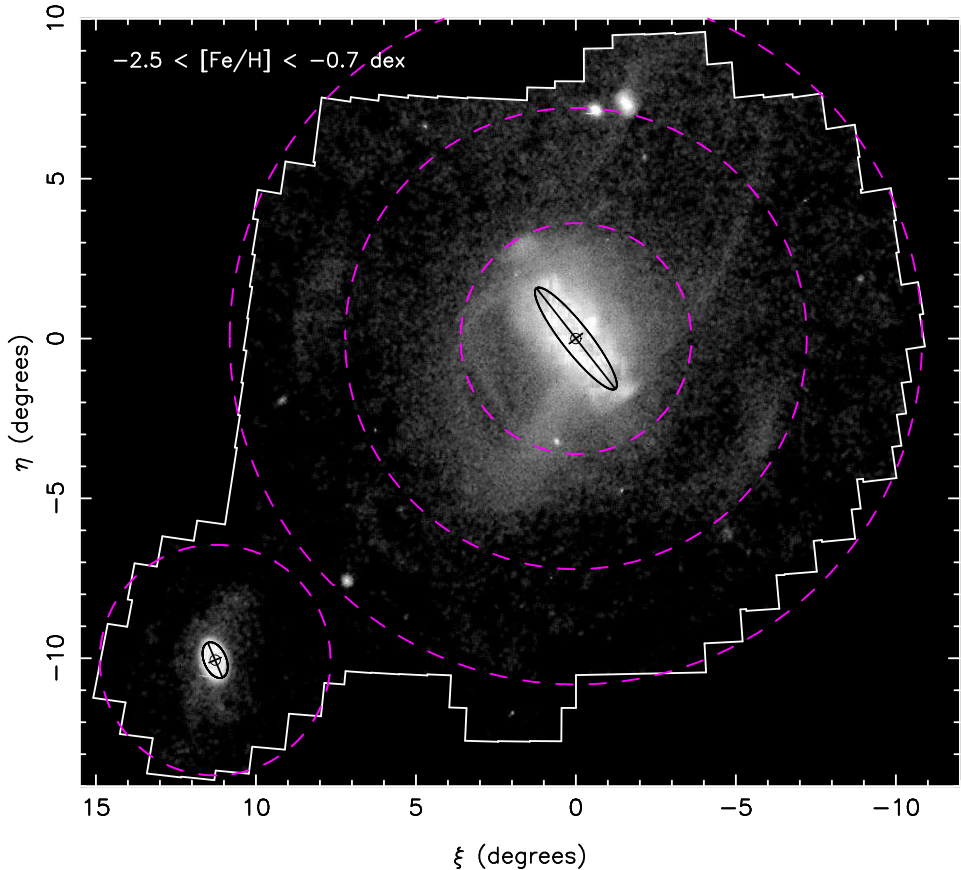}
    \caption{Tangent plane projection centered on M31 showing the spatial density distribution of candidate red giant branch stars using a logarithmic scaling. The dashed circles correspond to projected radii of 50\,kpc, 100\,kpc and 150\,kpc from M31, and 50\,kpc from M33. Pixels are $1 \times 1$\,arcmins and the image is smoothed using a top-hat filter whose size scales as a function of local density as described in the text. In the central few degrees of M31, the width of the filter is less than the size of the cell, and no smoothing is applied; in the most sparse outer regions of M31, the width of the filter is $\sim1$ degree. }\label{adapt}
  \end{center}
\end{figure*}

Figure~\ref{cmds} shows the $i_0$ versus $(g-i)_0$ CMD of all stellar sources in the PAndAS footprint, excluding the inner 2 degrees around M31 and the inner 1 degree around M33. Overlaid are Dartmouth isochrones (\citealt{dotter2007,dotter2008}) showing the theoretical RGB locus for old (13Gyr) stellar populations with metallicities of [Fe/H] $= -2.5, -2, -1.5, -1, -0.5$\,dex (left to right, respectively) and with $[\alpha/Fe] = 0.0$ at the distance of M31. The positions in the CMD corresponding to the locus of Galactic main sequence turn-off halo stars and dwarfs in the Galactic disk are indicated. Note that the sequences in the foreground halo stars correspond to foreground substructures at specific distances in the Galactic halo, and are discussed in \citet[see also \citealt{martin2007a}]{martin2014}. Also indicated at faint magnitudes is contamination due to misidentified compact background galaxies.

Figure~\ref{adapt} shows a logarithmic density map of the spatial distribution of RGB star candidates from PAndAS selected based on photometric metallicity estimates from Figure~\ref{cmds}. Here, we follow Paper~I and use a version of the dataset in which the foreground has been statistically removed. Specifically, \cite{martin2013c} develop a foreground contamination model for PAndAS that allows the generation of synthetic CMDs for the foreground at any position in the survey. We use this model to statistically predict the colors and magnitudes of contaminating foreground stars (i.e., foreground stars that lie in the same color-magnitude locus as RGB stars at the distance of M31) at every position in the survey. We then remove from the dataset the appropriate number of stars at each position that are closest in color-magnitude space to stars in this foreground population. The final ``foreground subtracted'' dataset is confined to $i_0 <23.5$, $(g-i)_0 <1.8$ and $-2.5 \le$ [Fe/H] $\le 0$\,dex. 

In Figure~\ref{adapt}, we have used 1\,arcmin pixels and have smoothed the image using a top-hat filter whose size scales as a function of local density. Specifically, the width of the filter applied to each cell is proportional to the square of the average distance to the $k^{th}$ nearest neighbour of all the stars in the cell. Here, $k=11$, although we note that the relative widths of the filters applied to high and low density regions is not strongly sensitive to the exact value of $k$. In the central few degrees of Figure~\ref{adapt}, the width of the filter calculated in this manner is less than the size of the cell, and no smoothing is applied. In the most sparse outer regions of the figure, the width of the filter is $\sim1$ degree. Figure~\ref{adapt} has the advantage that the majority of all the major stellar substructures in the M31 outer halo are visible in a single figure. 

\begin{figure*}
  \begin{center}
    \includegraphics[angle=270, width=17cm]{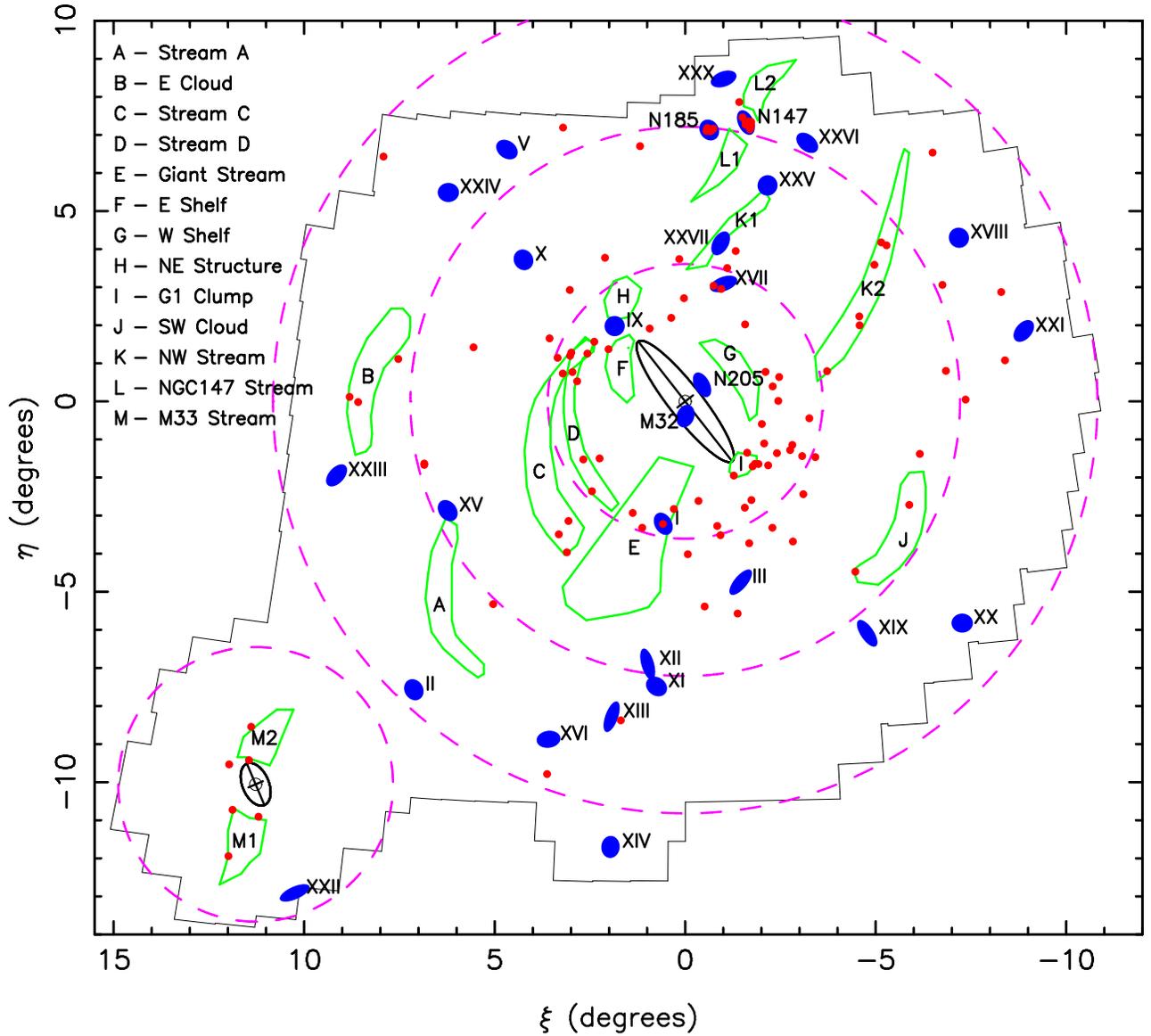}
    \caption{Tangent plane projection centered on M31 with the positions of all major stellar substructures, globular clusters and dwarf galaxies highlighted. The dashed circles correspond to projected radii of 50\,kpc, 100\,kpc and 150\,kpc from M31, and 50\,kpc from M33. The outline of the complete PAndAS footprint is shown in gray. Red dots correspond to all known globular clusters at projected radii of greater than 2 degrees from M31 as given by \cite{mackey2018}, and includes globular clusters belonging to NGC147, NGC185, Andromeda~I and M33, as described in the text. Blue ellipses correspond to all known Local Group dwarf galaxies in the PAndAS footprint (all but one of which is a probable satellite of M31). The ellipticity and position angle of each ellipse corresponds to the measured ellipticity and position angle of each galaxy; the areas of each ellipse are set to be equal to each other for clarity. Green polygons highlight the approximate positions of each of the major stellar substructures that have been identified in the halo of M31, and which trace the major features visible in Figure~\ref{adapt}.}\label{schematic}
  \end{center}
\end{figure*}

\begin{figure*}
  \begin{center}
    \includegraphics[angle=270, width=15cm]{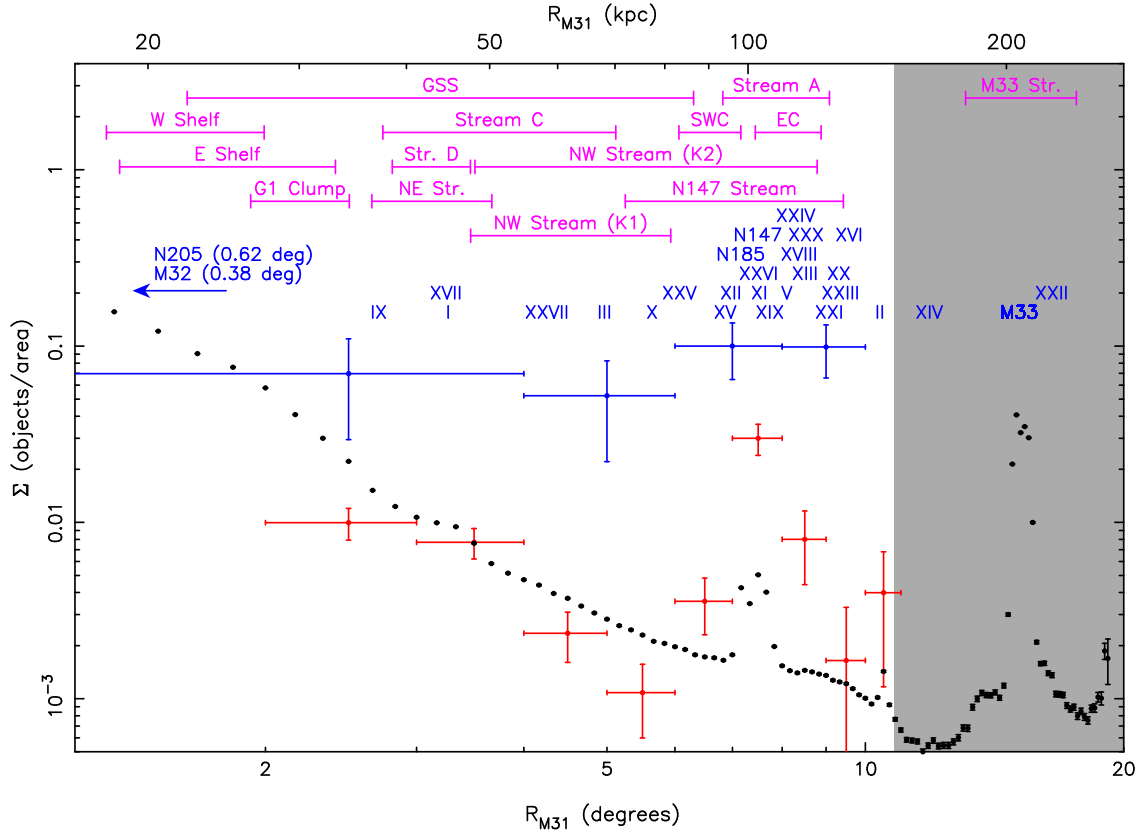}
    \caption{Projected radial profiles of all stellar associations in the PAndAS footprint, assuming circular annuli. The radial range covered by each stellar substructure highlighted in Figure~\ref{schematic} is shown by the magenta bars. The blue points indicate the radial profile of dwarf galaxies surrounding M31 (and the radial positions of each dwarf are indicated by their names). Horizontal bars indicate the width of the bin and vertical bars indicate the poisson uncertainty in each bin. Only dwarfs in the PAndAS footprint are shown. Red points correspond to the radial profile of all known globular clusters at projected radii of greater than 2 degrees from M31 as given by \cite{mackey2018}, and includes globular clusters belonging to NGC147, NGC185 and Andromeda~I. Horizontal bars indicate the width of the bin and vertical bars indicate the poisson uncertainty in each bin. The black points show the radial profile of red giant branch stars, selected with $i_0 \le 23.5$ and $-2.5 \le$ [Fe/H] $\le 0.0$. All RGB stars that meet the selection are shown, including those that are associated with dwarf galaxies or stellar substructures. The gray region corresponds to the area beyond a projected radius of 150kpc from M31, where PAndAS has extremely limited azimuthal coverage of M31. A vertical offset has been applied to each population for plotting purposes.}\label{profile}
  \end{center}
\end{figure*}

Figure~\ref{schematic} highlights the spatial positions of the most prominent substructures visible in Figure~\ref{adapt}, along with the population of dwarf galaxies (blue ellipses) and outer halo globular clusters (defined as globular clusters that are more than $2$\,degrees in projection from the center of M31; red dots). The outlines of the stellar substructures are approximate and are intended for reference only. Figure~\ref{profile} shows the same spatial information as Figure~\ref{schematic}, but now as a function of projected radius and assuming circular annuli. Here, the projected radial distribution of dwarf galaxies is shown as blue points (along with the names of each dwarf positioned at the appropriate radius), the projected radial distribution of outer halo globular clusters is shown as red dots, and the projected radial range covered by each stellar substructure is indicated as a magenta bar. Finally, the black points show the projected radial distribution of RGB stars ($i_0 \le 23.5$ and $-2.5 \le$ [Fe/H] $\le 0.0$.  All RGB stars that meet the selection are shown, including those that are associated with dwarf galaxies or stellar substructures. Here, we have used the same foreground-subtracted catalog as used in Figure~\ref{adapt}.

All of the labelled substructures shown in Figure~\ref{schematic} and \ref{profile} have been discussed or referenced previously in the literature. We only show those substructures that are at reasonably large (projected) radius from the center of M31 (more than a degree or so). In addition to these stellar substructures, there are numerous associations of M31 dwarf galaxies and globular clusters (such as the NGC147/185 subgroup) which can be considered as substructures within these populations. 

In this section, we first provide a summary of the status of knowledge regarding the individual properties of each luminous substructure that has been identified in the halo of M31 (whether it consists of stars, dwarf galaxies, and/or globular clusters). Much of this is a review of existing literature, and we also refer the reader to the related discussion in \cite{ferguson2016}. We go on to discuss possible causal connections between these different individual substructures, and determine a ``best guess'' for the lower limit on the number of distinct accretion events in the recent history of M31.

\subsection{Individual substructures}

\subsubsection{Substructure in the stellar distribution}

The following stellar substructures have previously been identified in the halo of M31 and discussed in the literature. They are all highlighted in Figures~\ref{schematic} and \ref{profile}:

\begin{itemize}

\item {\bf A -- Stream A:} This structure was first identified in an earlier version of the MegaCam dataset presented in \cite{ibata2007}. The region outlined in Figure~\ref{schematic} is $\sim 5$\,degrees long and is located more than 100\,kpc  from the center of M31. \cite{ibata2007} estimate an integrated luminosity of $M_V \simeq -11.1$ and a surface brightness of approximately $\Sigma_V \sim 32\,$mags\,arcsec$^{-2}$.  Spectroscopic observations of fields overlapping serendipitously with Stream A were obtained by \cite{gilbert2006} and \cite{koch2007c} and analyzed by \cite{chapman2008}. The latter authors identify three stars that could plausibly belong to M31 based on their radial velocities. These stars have  [Fe/H] $\sim -1.3$\,dex, in good agreement with a photometric estimate obtained for Stream A by \cite{ibata2007}. They also note that the measured velocities of the stars are closer to that of M33 than M31, and they suggest Stream A could in fact be associated primarily with M33's halo. Clearly, more spectroscopic observations of candidate members of Stream A are required to better understand its origin.

\item {\bf B -- East cloud:} The East Cloud is visible in the version of the PAndAS dataset first presented in \cite{richardson2011} and a dedicated analysis of this structure is presented in \cite{mcmonigal2016}. These authors estimate a luminosity of $M_V \sim -10.7$. They suggest that this luminosity accounts for at most 20\% of the stellar mass that is implied from using the mass -- metallicity relation, given their measured photometric metallicity of [Fe/H] $\sim -1.2$\,dex.

\item {\bf C -- Stream C:} Stream C was identified in \cite{ibata2007}, although its full extent to the north was not apparent until the version of the data presented in \cite{richardson2011}. In the south, it overlaps with the GSS and so its path is difficult to follow; in the north, Stream C overlaps with Stream D (see next bullet). It is a metal-rich structure, and the luminosity estimate by \cite{ibata2007} implies a massive (larger than Fornax) dwarf progenitor. 

Subsequent spectroscopy of Stream C from  Keck/DEIMOS is presented in \cite{chapman2008}. These authors refine the analysis by \cite{ibata2007} in light of these new data and show that the morphology of Stream C changes significantly with metallicity. They argue that Stream C is in fact two distinct structures that happen to overlap spatially; a dominant metal-rich stream and a less well-defined metal-poor stream. Specifically, a metal-rich feature is measured to have a metallicity of [Fe/H] $\sim -0.6$ and a velocity dispersion of $\sigma \simeq 5$\,km/s. Its systemic velocity of $v_r \simeq -349$\,km/s is close to the M31 velocity, consistent with a tangential orbit. A metal-poor feature ([Fe/H] $\sim -1.3$\,dex)  is also detected via a grouping of 5 stars with a velocity dispersion around 5\,km/s at a systemic velocity of $v_r \simeq -286$\,km/s. This interpretation is backed up by an independent spectroscopic analysis by \cite{gilbert2009}.  

\item {\bf Stream D:} Like Stream C, part of Stream D was identified in \cite{ibata2007}, although its northern extension was not revealed until \cite{mcconnachie2009b}, which shows that Streams C and D appear to merge in projection. Like Stream C, the southern extent of Stream D is unclear due to the presence of the Giant Stellar Stream. Intriguingly, the relatively luminous dwarf spheroidal Andromeda I aligns with Stream D at similar projected radius from M31 as the main body of the substructure. It has previously been argued by \cite{mcconnachie2006b} that Andromeda I is in the process of being tidally disrupted, since its outer isophotes show the distinctive "S" shape that is indicative of tidal stripping. However, a firm association between Stream D and Andromeda I has not been made. \cite{ibata2007} estimate a median photometric metallicity of [Fe/H] $\simeq -1.2$\,dex (cf [Fe/H] $\simeq -1.2$\,dex for Andromeda I; \citealt{kalirai2010}). 

\cite{chapman2008} analysed Keck/DEIMOS spectroscopy of stars in a field overlapping with Stream D. They identified 5 stars that may plausibly be associated with this feature, but overall they could not determine any robust properties for Stream D based on these data. 

\item {\bf D -- The Giant Stellar Stream / F-- the Eastern Shelf / G -- the Western Shelves:} The Giant Stellar Stream (GSS) was first identified by \cite{ibata2001a} from the Isaac Newton Telescope Wide Field Camera (INT WFC) survey of M31. It is easily the most massive stellar substructure in the halo of M31, and has been the subject of considerable follow-up observations. The first follow-up was photometry with CFHT\,12K (\citealt{mcconnachie2003}), followed by extensive spectroscopy using Keck/DEIMOS (\citealt{ibata2004, guhathakurta2006,gilbert2009}). Numerous dynamical models have also been produced (\citealt{ibata2004,fardal2006,fardal2007,fardal2008,fardal2013,sadoun2014,miki2016,kirihara2017a}). Observations and modeling suggest that the stream is on a radial orbit that passes close to the center of M31. According to \cite{fardal2006}, the progenitor is constrained to have a stellar mass of order $10^{9.5}\,M_\odot$, at the last pericentric passage, and this passage was $760 \pm 50$\,Myrs ago. Thus, it is a relatively young feature. The spatial morphology of the stream, in particular the relatively well-defined northern edge and the ``fanning'' of material to the south, suggests that the progenitor might have been a disk galaxy (\citealt{fardal2008,kirihara2017a}). It is worth noting that the age of the stream is reasonably close to the age of the star forming ring in M31 (\citealt{davidge2012a, dalcanton2012}).

Hubble Space Telescope (HST) Advanced Camera for Surveys (ACS) observations by \cite{richardson2008} and \cite{bernard2015b} of 14  fields located at projected radii of $11.5 < R_p < 45$\,kpc from the center of M31 reveal that the stellar populations of the GSS are very similar to the stellar populations of some other prominent substructures, in particular the Eastern and Western Shelves (see also \citealt{ferguson2005}). This is also suggested by the PAndAS RGB maps split by photometric metallicity, whereby all three of these features are most prominent in the most metal-rich RGB selections, and a connection between the features was originally suggested by \cite{ferguson2002}. The model by \cite{fardal2007} naturally explains this coincidence due to the very close, and hence highly destructive, last pericentric passage of the GSS progenitor. The progenitor is nearly totally destroyed by this passage, and two prominent radial ``shell'' features are created that we subsequently identify as the Eastern and Western Shelves. Kinematic observations of the Western Shelf support this interpretation (\citealt{fardal2012}). If a progenitor is still present, it is likely in the vicinity of the Eastern Shelf, close to the M31 disk. Note that the Eastern Shelf is sometimes referred to as the North-East Shelf (e.g., \citealt{ferguson2005}).

\item {\bf H -- North-East Structure:} \cite{zucker2004b} identify this diffuse structure in photometry from an SDSS scan of M31, and it is also visible in the  INT WFC footprint of M31 presented by \cite{lewis2004}. \cite{zucker2004b} called it ``Andromeda North-East'' and it is also occasionally referred to as the ``NE Clump'' (e.g., \citealt{richardson2008, lewis2013, ferguson2016}), although this should not be confused with a seperate structure that was found by \cite{davidge2012b} that this author also referred to as the NE Clump.  This latter feature is only about 3.5kpc from the center of M31 in projection, and is possibly the remnant of a dwarf galaxy, or a fossil star formation region in the disk of M31.

\cite{zucker2004b} measure an integrated SDSS $g$ band magnitude for the NE Structure of $M_g \sim -11.6$ and a central surface brightness of $\Sigma_g \sim 29$\,mags\,arcsec$^{-2}$. They suggest that it could possibly be an extremely diffuse dwarf galaxy, or a tidal remnant of a dwarf since destroyed. The structure has a similar mean RGB color to the G1 clump and the outer disk structures studied by \cite{richardson2008}. These authors argue that all these features are probably outer disk material that has been heated and disrupted into the halo region of M31 due to accretion events (such as that which gave rise to the GSS and related structures). This idea is discussed further in \cite{bernard2015b}.

\item {\bf I -- The G1 Clump:} \cite{ferguson2002} identified the G1 clump in early data from the INT WFC survey of M31, and it is located on the outer fringes of the main body of M31. It spatially overlaps with the massive globular cluster G1, but kinematics show that it is unrelated to this unusual globular cluster (\citealt{reitzel2004}). \cite{ferguson2002} measure an integrated magnitude in the INT V band of $M_V \simeq -12.6$. \cite{ferguson2005} and \cite{richardson2008} measure the stellar populations of this field using HST ACS and determine them to be ``disk-like''; a member of the set of substructures that these authors argue originate from the heating and/or disruption of the thin disk.

\item {\bf J -- The South West Cloud:} This structure was identified in the PAndAS dataset presented in \cite{mcconnachie2009b}, and a dedicated analysis based on these data is presented in \cite{bate2014}. The SW Cloud is estimated to lie at very similar distance to M31 (i.e., not notably in the foreground or background of the halo) and has a photometric metallicity of [Fe/H] $\simeq -1.3$\,dex. \cite{bate2014} measure an integrated luminosity of $M_V = -12.1$\,mags, which they estimate to be $\sim 75\%$ of the luminosity that is implied from the metallicity and the luminosity-metallicity relation. Kinematic analysis of stars in this substructure using Keck/DEIMOS by \cite{mackey2014} allows the robust identification of candidate members in a cold peak ($\Delta v \lesssim 20$\,km/s) at around $v_r \sim -400$\,km/s. Overall, the relatively high luminosity and probable association of globular clusters with this feature (see later) imply that its progenitor would have been relatively massive in comparison to the known dwarf galaxy population of M31 (considerably more massive than Andromeda II, which has a similar luminosity to the substructure but no globular clusters).

\item {\bf K -- The North West Stream:} The southern fragment of  the north-west stream  was first discovered in the PAndAS dataset presented in \cite{mcconnachie2009b}. However,  it was not until the more complete version of the dataset presented in \cite{richardson2011} became available that its full possible extent was realised. The fragment labelled ``K1'' in Figure~\ref{schematic} is nearly $3$\,degrees long in projection, reaching between $\sim 50 - 80$ kpc in projected radius from the center of M31, and contains the dwarf galaxy Andromeda~XXVII. The fragment labelled ``K2'' in Figure~\ref{schematic} is $\sim 6$\,degrees long in projection ($\sim 80$\,kpc), reaching between $\sim 50 - 120$\,kpc in projected radius from the center of M31. The version of the data presented in \cite{richardson2011} suggests that these two fragments are part of the same, enormous, stellar stream, and they suggest that Andromeda~XXVII is the progenitor. A kinematic analysis by \cite{collins2013} suggests that Andromeda~XXVII is almost certainly not in dynamical equilibrium.

The fact that it appeared as if K1 and K2 were part of the same stream,  albeit with at least one prominent ``gap'', led \cite{carlberg2011} to analyse the density profile of the full feature within the context of gap-production via interactions of the stellar stream with dark matter substructures (see, for example, \citealt{ibata2002b, johnston2002}). Recently, \cite{kirihara2017b} have attempted to produce N-body models of the NW stream, concentrating on the K2 segment, where velocity information is available via a population of globular clusters that appear to be  associated with the stream (\citealt{veljanoski2013b}).

The reprocessed data shown in Figure~\ref{adapt}, for which the photometry is both more uniform and more precise, does not lend itself so readily to the interpretation first put forward by \cite{richardson2011}. Rather, it is wholly unclear from Figure~\ref{adapt} if K2 loops around to reconnect with K1. Instead, there is a tentative suggestion that K2 may extent in the general direction of the NGC147/185 subgroup. Spectroscopic observations currently being analyzed by our group show that that K1 is in fact a distinct feature, and is not plausibly associated with K2 (Preston et al., {\it in preparation}).

\item {\bf L -- The NGC147 Stream:} The tidal stream emanating from NGC147 was not discovered until late in the PAndAS observing program because of its northern locale. The southern part of the stream is visible in \cite{richardson2011}, and this prompted a set of fields to the north of NGC147 to be added to the program. The full extent of the feature as traced by PAndAS data was first presented in \cite{lewis2013}. Its origin, as it relates to both the orbit of NGC147 around M31, and the relative orbit of NGC147 with respect to its closest neighbours was analyzed in \cite{arias2016}. However, no detailed observational analysis of the stream has yet been presented in the literature, although an analysis of NGC147 extending to large radius is presented in \cite{crnojevic2014}. The NGC147 stream, as seen in Figure~\ref{adapt} and shown schematically in Figure~\ref{schematic} extends approximately 2 degrees to the south and north, for a total projected length $\simeq 40$\,kpc (assuming a distance to NGC147 of $676 \pm 28$\,kpc; see Table~\ref{dwarfs}). In Section~\ref{n147}, we derive an integrated magnitude of $M_V = -12.2$\,mags, implying that the stream possesses $\sim4$\% of the stellar mass of the main body of NGC147. There is no HI detection associated with the stream (\citealt{lewis2013}).

\item {\bf M -- The M33 Stream:} \cite{mcconnachie2009b} presented the discovery of the large low surface brightness stellar substructure that surrounds the M33 disk, and which extends out to nearly 40\,kpc in projection. The feature was measured to have an integrated luminosity of $M_V \simeq -12.7$ by \cite{mcconnachie2010b}. 

\cite{mcconnachie2009b} argue that the morphology of the feature is very similar to what is expected for a system undergoing mild tidal distortion as it orbits a more massive host. They present a possible orbital solution for M33 in which it passed in front of M31 at a pericentric distance of $\sim 50$\,kpc a few Gyrs ago. It continued in its orbit to apocenter and is now observed infalling into M31. This solution was consistent with the known proper motion of M33 (\citealt{brunthaler2005}), as well as the positional and radial velocity data for M33 and M31. The reasonably close pericentric approach heats the outskirts of the M33 disk sufficiently to produce a tidal feature that approximates the observed substructure; a much closer pericenter would destroy the M33 disk. Independently, \cite{putman2009} argue that an interaction of M33 with M31 is required in order to explain the extreme warp in the HI disk (e.g., \citealt{rogstad1976, corbelli1997}). 

The hypothesised pericentric passage that perturbs the M33 disk also perturbs the outskirts of the M31 disk. In this respect, HST stellar population studies of the disks of these galaxies have proved particularly interesting. \cite{bernard2012} obtained two HST fields located at $\sim 26$\,kpc from the center of M31; one of these was badly affected by differential extinction in M31, but for the other they found it underwent roughly constant star formation until around 4.5 Gyrs ago, at which point there was a decline. However, a period of intense star formation began around 3 Gyrs ago, peaking about 1.5 Gyrs ago. \cite{bernard2012} hypothesize that the onset of this star formation episode coincided with an interaction with M33. To test this, they reanalyse a M33 outer disk field from \cite{barker2011} and derive its star formation history in the same way as their M31 fields. They find that there is a similar burst of star formation in M33 that is of similar duration and exactly coeval with the M31 burst, lending credence to an interaction hypothesis. 

It is worth noting that further evidence for a widespread burst of star formation around 2\,Gyrs ago in M31 is presented in \cite{bernard2015b} and \cite{williams2015}, the latter from data using the PHAT survey (\citealt{dalcanton2012}). However, a further two HST fields in the outer disk of M31 along the south-west semi-major axis observed by \cite{bernard2015a} do not show this signature. This is potentially not inconsistent with the interaction hypothesis; \cite{davidge2012a} show that the AGB population of M31 is lop-sided, even though some of these stars have ages of a few billion years, and it may be the other disk stellar populations are not well mixed.  Independent from HST surveys, analysis of the M31 Planetary Nebulae populution by \cite{balick2013} and \cite{corradi2015} suggest that there was a global period of star formation approximately 2 Gyrs ago, that they attribute to an M31 -- M33 interaction.

The measurement of the proper motion of M31 using HST ACS observations by \cite{sohn2012} led to dynamical modeling of the Local Group by \cite{vandermarel2012a,vandermarel2012b}. These authors conclude that the proper motion measurement of M31 is consistent with the type of orbit for M33 presented in \cite{mcconnachie2009b}. More recently, however, \cite{patel2017} finds analogs of the M31 -- M33 subsystem in the Illustris cosmological simulations (\citealt{genel2014}) and examine the types of orbits that these galaxies are on. They conclude that an orbit such as that proposed by \cite{mcconnachie2009b} is highly improbable based on the statistics of M31 -- M33 analogs in the cosmological simulations. They further argue that, through an analytic argument that considers the proper motions of the two systems, it is unlikely that M31 and M33 could have passed sufficiently close to have had a mild tidal interaction. During the preparation of this paper, \cite{vandermarel2018} present first estimates of the proper motion of M31 and M33 based on Gaia data. They conclude that these estimates are consistent with the earlier estimates, but tend towards values that favor orbital solutions where M33 is on its first infall. If these studies are correct, this would require a significant re-interpretation of the available observational data on the M31 -- M33 system, its star formation history and its tidal features. Clearly, more analysis of the orbital history of M31 and M33, as well as the stellar and gasous ``warps'' around M33, are required.

\end{itemize}

The above is a complete list of all the stellar substructures that are found at large radius in the halo of M31 which have previously been discussed in the literature and are labelled in Figures~\ref{schematic} and \ref{profile}. However, there are also a few stellar substructures at reasonably large radius that have previously been discussed in the literature but which are not labelled in Figure~\ref{schematic}. These are:

\begin{itemize}

\item {\bf Major axis diffuse structure:}  \cite{ibata2007} identified what they called ``Major axis diffuse structure'', that appeared in their data as an overdensity of stars along the edge of the quadrant for which they had data, aligned with the major axis. In Figure~\ref{adapt} the full extent of this feature (or features) can be seen. In projection at least, it appears to overlap with the SW Cloud. It may be part of a broader complex that includes the very extended Andromeda~XIX dwarf galaxy; the discovery paper for this object showed how Andromeda~XIX may be surrounded by tidal debris, although a physical association between the dwarf and the surrounding stars  has not been robustly established (\citealt{mcconnachie2008b}). 

\item {\bf The NGC205 Loop:}  \cite{mcconnachie2004b} identified the NGC205 loop based on the INT WFC of the inner regions of the M31 halo. This feature emanates from the north of NGC205, and loops round to the east. It is only $\sim 1$\,degree in extent, and the immense number of stars resolved by PAndAS in the inner regions of the survey obscure it from view in Figure~\ref{adapt}.  The ground-based CMD of the NGC205 loop is consistent with the stellar populations of NGC205, and initial estimates of its velocity dispersion from spectroscopic studies are of order $10$\,km/s, suggesting it is dynamically cold. \cite{mcconnachie2004b} therefore suggest it is a tidal stream from NGC205. 

An additional Keck/DEIMOS field centered on the loop is presented in \cite{ibata2005}, and these authors suggest there is no evidence based on the kinematics to link the loop to NGC205. However, \cite{mcconnachie2005c} analyse the same field and identify the expected signature of the loop, causing them to reinforce their conclusion that the loop is linked to NGC205. HST ACS stellar population studies of the NGC205 loop by \cite{ferguson2005} and \cite{richardson2008} do not clearly associate it preferentially with NGC205, the GSS, or the M31 disk (i.e., these authors label this field a ``composite''). 

Dynamical modeling of the orbit of NGC205 by \cite{howley2008}, based on internal kinematics of the central regions of the dwarf, lead them to dispute the conclusions of \cite{mcconnachie2004b} and \cite{mcconnachie2005c}. They suggest that NGC205 is on its first passage around M31 and not associated with the loop. If correct, an alternative origin of the loop needs to be considered. Further, the dE morphology of NGC205 needs to be able to be explained without recourse to the environmental influence of M31, since the prevailing concept for how dE galaxies obtain their morphology is through interactions with massive companions (e.g., \citealt{faber1983, moore1998}). But if NGC205 is on its first infall into M31, then clearly it cannot have undergone significant interactions with this galaxy.

\item {\bf Stream B:} Stream B is a feature that \cite{ibata2007} identified in the south-west quadrant of M31, between Streams A and C centered around $(\xi, \eta) \simeq (4.5, -3.0)$\,degrees, approximately 75\,kpc from M31. In Figure~\ref{adapt}, an amorphous overdensity of stars is visible in this region, but it is hard to delineate this feature from the surrounding halo, especially Stream C and the Giant Stream, and we have not tried to do so in Figure~\ref{schematic}. Whatever the nature of this feature, it is substantial;  \cite{ibata2007} estimate its luminosity to be in excess of $10^7\,L_\odot$. 

\item {\bf Stream E:} \cite{tanaka2010} conduct a survey of M31 out to large radius in a narrow region along the minor axis, concentrating on the north-west side of the galaxy. There, they discover a couple of new overdensities, that they dub Stream E and Stream F. Stream F has subsequently been shown to be part of the NW stream. Stream E, located at $(\xi, \eta) \simeq (-3.5, 3.0)$\,degrees, is not visible in the PAndAS data. The CMD of this feature in \cite{tanaka2010} shows that the overdensity is due to relatively bright sources that PAndAS should have detected, although they do not clearly follow a stellar isochrone (in contrast to Stream F). From this, we conclude that Stream E is non-existent, and the previous claim was based on mis-classified sources (possibly background galaxies).
\end{itemize}

\subsubsection{Substructure in the dwarf galaxy population}

\begin{table*}
\begin{center}
\begin{tabular*}{0.8\textwidth}{lllccccc}
\hline
Galaxy    &  R.A. & Dec. & $D_{MW}$ & $D_{M31}$ & $M_V$ &  $M_\star$ & Notes \\
&  $^{h m s}$& $^\circ$' " & kpc & kpc & mags  & $10^6 M_\odot$ & \\
\hline
M32                & 0 42 41.8 & +40 51 55 & 809 &  22 &$ -16.4 \pm  0.2$& 382.99\\
Andromeda IX       & 0 52 53.0 & +43 11 45 & 770 &  40 &$  -8.8 \pm  0.3$&   0.35\\
NGC205            & 0 40 22.1 & +41 41  7 & 828 &  41 &$ -16.5 \pm  0.1$& 401.03\\
Andromeda I        & 0 45 39.8 & +38  2 28 & 748 &  58 &$ -11.9 \pm  0.1$&   5.69\\
Andromeda XVII     & 0 37  7.0 & +44 19 20 & 731 &  69 &$  -7.7 \pm  0.3$&   0.12\\
Andromeda XXVII    & 0 37 27.1 & +45 23 13 & 832 &  74 &$  -7.9 \pm  0.5$&   0.15\\
Andromeda III      & 0 35 33.8 & +36 29 52 & 751 &  75 &$ -10.2 \pm  0.3$&   1.20\\
Andromeda XXV      & 0 30  8.9 & +46 51  7 & 816 &  88 &$  -9.2 \pm  0.3$&   0.51\\
Andromeda XXVI     & 0 23 45.6 & +47 54 58 & 765 & 102 &$  -5.9 \pm  0.7$&   0.02\\
{\it Andromeda V}        & 1 10 17.1 & +47 37 41 & 777 & 109 &$  -9.5 \pm  0.2$&   0.67\\
Andromeda XI       & 0 46 20.0 & +33 48  5 & 738 & 110 &$  -6.3 \pm  0.4$&   0.03\\
Andromeda XIX      & 0 19 32.1 & +35  2 37 & 823 & 113 &$ -10.1 \pm  0.3$&   1.09\\
Andromeda XXIII    & 1 29 21.8 & +38 43  8 & 774 & 126 &$  -9.8 \pm  0.2$&   0.88\\
Andromeda XX       & 0  7 30.7 & +35  7 56 & 744 & 129 &$  -6.4 \pm  0.4$&   0.04\\
Andromeda XIII     & 0 51 51.0 & +33  0 16 & 843 & 132 &$  -6.8 \pm  0.4$&   0.05\\
Andromeda X        & 1  6 33.7 & +44 48 16 & 674 & 133 &$  -7.4 \pm  0.3$&   0.10\\
Andromeda XXI      &23 54 47.7 & +42 28 15 & 830 & 133 &$  -9.1 \pm  0.3$&   0.44\\
NGC147            & 0 33 12.1 & +48 30 32 & 715 & 120 &$ -15.8 \pm  0.1$& 206.62\\
{\it Andromeda XXXII}          & 0 35 59.4 & +51 33 35 & 780 & 140 &$ -12.3 \pm  0.7$&   8.15\\
Andromeda XXX            & 0 36 34.9 & +49 38 48 & 686 & 147 &$  -8.2 \pm  0.3$&   0.19\\
Andromeda XIV      & 0 51 35.0 & +29 41 49 & 798 & 161 &$  -8.7 \pm  0.3$&   0.31\\
Andromeda XII      & 0 47 27.0 & +34 22 29 & 932 & 178 &$  -7.1 \pm  0.5$&   0.07\\
Andromeda XV       & 1 14 18.7 & +38  7  3 & 629 & 178 &$  -8.0 \pm  0.4$&   0.16\\
Andromeda II       & 1 16 29.8 & +33 25  9 & 656 & 184 &$ -12.6 \pm  0.2$&  10.94\\
NGC185            & 0 38 58.0 & +48 20 15 & 623 & 185 &$ -15.5 \pm  0.1$& 156.74\\
{\it Andromeda XXIX}     &23 58 55.6 & +30 45 20 & 733 & 188 &$  -8.3 \pm  0.5$&   0.22\\
Triangulum         & 1 33 50.9 & +30 39 37 & 809 & 206 &$ -18.8 \pm  0.1$&3525.18\\
Andromeda XXIV     & 1 18 30.0 & +46 21 58 & 604 & 208 &$  -7.6 \pm  0.3$&   0.11\\
{\it Andromeda VII}      &23 26 31.7 & +50 40 33 & 764 & 218 &$ -13.2 \pm  0.3$&  19.73\\
{\it IC 10}              & 0 20 17.3 & +59 18 14 & 798 & 252 &$ -15.0 \pm  0.2$& 102.61\\
{\it Andromeda XXXI}           &22 58 16.3 & +41 17 28 & 760 & 262 &$ -11.7 \pm  0.7$&   4.91\\
{\it LGS 3}              & 1  3 55.0 & +21 53  6 & 773 & 268 &$ -10.1 \pm  0.1$&   1.16\\
{\it Andromeda VI}       &23 51 46.3 & +24 34 57 & 785 & 268 &$ -11.5 \pm  0.2$&   3.97\\
Andromeda XXII     & 1 27 40.0 & +28  5 25 & 925 & 273 &$  -6.8 \pm  0.4$&   0.05\\
Andromeda XVI      & 0 59 29.8 & +32 22 36 & 480 & 323 &$  -7.3 \pm  0.4$&   0.08\\
{\it Andromeda XXXIII}         & 3  1 23.6 & +40 59 18 & 779 & 349 &$ -10.3 \pm  0.7$&   1.40\\
{\it Andromeda XXVIII}   &22 32 41.2 & +31 12 58 & 660 & 367 &$  -8.5 \pm  0.6$&   0.26\\
\hline 
Andromeda XVIII    & 0  2 14.5 & +45  5 20 &1216 & 452 &$  -9.2 \pm  0.4$&   0.50\\
{\it Pegasus dIrr}       &23 28 36.3 & +14 44 35 & 921 & 474 &$ -12.2 \pm  0.2$&   7.93\\

\end{tabular*}
\caption{Probable (above horizontal line) and possible ($D_{M31} < 500$\,kpc; below horizontal line) satellite galaxies of M31, sorted by distance from M31. Objects in italics are not in the PAndAS footprint. Stellar masses are calculated assuming a stellar mass-to-light ratio of $\Upsilon_\star = 1.2\,M_\odot/L_\odot$ (\citealt{mcgaugh2014}). Distances and magnitudes are from the updated compilation of \cite{mcconnachie2012}.}
\label{dwarfs}
\end{center}
\end{table*}

A total of 30 dwarf galaxies are known to lie within the PAndAS footprint (including M33), of which 19 were discovered using PAndAS data. These are shown in Figure~\ref{schematic} as blue ellipses, where the shape and orientation of the ellipses reflects the measured position angles and ellipticities of the dwarfs. These are taken from the updated compilation by \cite{mcconnachie2012}, that includes many values from the systematic analysis by \cite{martin2016} based on PAndAS observations. For clarity, the area of the ellipses are all equal. Table~\ref{dwarfs} lists all candidate M31 dwarf satellites; those listed in italics are not within the PAndAS footprint; those that are listed below the horizontal line are candidate members only, based on their very large separation from M31.

The available positional and dynamical data for the dwarf galaxies have caused several authors to postulate the existence of several sub-groups of dwarf galaxies in the M31 system (possible substructures in the dwarf galaxy distribution):

\begin{itemize}

\item {\bf The M33 satellite system:} M33 is the third most massive galaxy in the Local Group. It has recently become clear that there are a large number of faint satellites of the Milky Way in the general vicinity of the fourth most massive galaxy, the Large Magellanic Cloud (LMC; \citealt{bechtol2015,drlicawagner2015,drlicawagner2016,kim2015a,kim2015b,kim2015c,koposov2015,koposov2018,martin2015,laevens2015a,torrealba2016a,torrealba2016b,torrealba2018}), in addition to the bright Small Magellanic Cloud. This has caused speculation that many of these systems are in fact dwarf galaxy satellites of the LMC, i.e., satellites of a satellite (\citealt{deason2015,yozin2015,jethwa2016,sales2017}). In this respect, it should not be unexpected that M33 has its own dwarf galaxy satellites. Only one clear candidate exists, however, and that is Andromeda XXII; all other candidates, such as Andromeda II, do not lie close enough to M33 compared to M31 to account for the much larger mass of M31. We also note that satellites of M33 as faint as those recently discovered around the LMC are sufficiently faint that they would be extremely challenging to detect in PAndAS imaging.

\cite{chapman2013} examine Andromeda XXII in the context of a M31 -- M33 interaction. They conclude that the dynamical model presented in \cite{mcconnachie2009b} would cause many of the outer dynamical tracers of M33, such as distant dwarfs, globular clusters and/or an M33 stellar halo, to be stripped. Andromeda XXII, therefore, would be a fortunate survivor of this process. Of course, this interpretation is not possible if the orbital history of M33 as suggested by \cite{patel2017} is correct. A close encounter potentially also explains the absence of a spatially extended population of globular clusters around M33 (\citealt{cockcroft2011}), as well as an absence of any obvious M33 stellar halo component (\citealt{cockcroft2013, mcmonigal2016}). It also means that it is possible that some of the globular clusters and dwarf galaxies currently associated with M31 were in fact originally associated with M33. 

\item {\bf The NGC147/185 subgroup}: \cite{vandenbergh1998} argue that these two bright dE companions of M31 form a binary pair. As well as being separated by only $\sim 1$\,degree in projection, they have similar distances ($D_{147} = 676 \pm 28$\,kpc, $D_{185} = 617 \pm 26$\,kpc; \citealt{mcconnachie2012}) and radial velocities ($v_{r, 147} = -193 \pm 1$\,km/s, $v_{r, 185} = -204 \pm 1$\,km/s). \cite{fattahi2013} independently identify this pair as a tight grouping in distance and velocity space, over and above that which which would be expected for a ``random'' population of cosmological sub-halos. Subsequently, Andromeda XXX (Cassiopeia II) was discovered in PAndAS data and has a very similar distance and radial velocity with respect to the two dEs ($D_{XXX} = 681^{+32}_{-78}$\,kpc, $v_{r, XXX} = -139.8^{+6.0}_{-6.6}$\,km/s; \citealt{conn2012, collins2013}). This implies that these three M31 satellites are in fact a sub-sub-group of the Local Group.

\cite{arias2016} examine this curious dynamical group in more detail, to determine if it is possible for this group of galaxies to be physically associated, given that they have to survive the tidal effects of orbiting M31, and that one of them (NGC147) is clearly in the process of tidal disruption. They find that it is probable that this group is no longer bound, but that they could originally have been physically associated. Indeed, it is even possible to find orbital solutions in which NGC147 is disrupted even though the other companions are not. \cite{watkins2013} analyse the orbital properties of all the M31 dwarfs taking into account the highly incomplete nature of the data, and find that Andromeda XXX is probably on a similar orbit to NGC185, but that neither is on a very similar orbit to NGC147. Given that \cite{arias2016} do not conclude that the whole sub-group is bound (and hence all on similar orbits), it seems plausible that these findings are consistent with one another.

\item {\bf Andromeda I and III:} As part of the same analysis that identified NGC147 and NGC185 as a statistically significant pairing of galaxies in distance/velocity space, \cite{fattahi2013} also identify Andromeda I and III as the only other significant pairing of (similar luminosity) dwarf galaxies in the M31 system. Their physical separation is of order 30\,kpc and their radial velocities are similar to within $\sim 30$\,km/s.

\item {\bf The Plane of Satellites:} \cite{ibata2013} noticed that a large number of M31 dwarf satellites appeared to align in a roughly north-south direction through the center of M31. In three dimensions, 15 out of the then known 27 satellites form a plane which has an rms scatter of only $\sim14$kpc (\citealt{conn2013}). In addition, 13 of the 15 satellites in this plane appear to exhibit coherent motion, insofar as those to the north of M31 are moving away from us with respect to M31, and those in the south are moving towards us.  Viewed in this way, the NGC147/185 subgroup and the Andromeda I/III pair are finer substructures within this dominant ``plane of satellites''. \cite{collins2015} show that there are no apparent differences between the dwarf galaxies that are associated with the plane and those that are not. In some respects, this plane bears some similarity to claims of a similar feature around the Milky Way (\citealt{lyndenbell1976, kroupa2005}), and recently the satellites of Centaurus A have also been claimed to lie in a narrow plane with coherent velocities (\citealt{muller2018}). A recent review of much of the relevant literature on this topic can be found in \cite{pawlowski2018}.

The M31 satellite plane has been argued to be extremely rare within cosmological simulations (see \cite{ibata2014b}, but see also \citealt{bahl2014} and \citealt{buck2015}). \cite{cautun2015} reanalyse the statistical significance of this feature, and consider the ``look elsewhere effect''; namely, that when searching a large region of parameter space for unusual features, it is necessary to consider that ``unexpected'' features will be present by chance. Examined in this way, these authors suggest that the plane of satellites is unusual, but not particularly rare (approximately 10\% of their cosmological simulations have features as ``unusual'' as the M31 plane of satellites).

A physical explanation for the plane is still lacking; several ideas exist, including triaxial halos (\citealt{bowden2013}), preferential accretion along filaments (\citealt{tempel2015}), accretion of a galaxy with its satellites (\citealt{smith2016}) tidal dwarf formation (\citealt{hammer2013}), potentially in a modified gravity framework (\citealt{zhao2013}). A basic issue facing all these ideas is the extreme thinness of the plane; given its rms thickness of 14kpc as derived in \cite{ibata2013}, and given that a velocity of 1km/s is equivalent to 1kpc/Gyr, then any significant out-of-plane velocity dispersion would quickly cause the plane to puff up and not be observable. This argues that the feature may be a relatively short, potentially transient, phenomenon (e.g., see \citealt{buck2016} and \citealt{fernando2017}). 

\end{itemize}

\subsubsection{Substructure in the outer globular cluster population}

The outer globular cluster system of M31 has grown dramatically as a result of discoveries in PAndAS and the precusor INT WFC survey, as well as results from SDSS, and a comprehensive study of these objects can be found in \cite{mackey2010,huxor2011,huxor2014,ditulliozinn2013,ditulliozinn2014,ditulliozinn2015,veljanoski2014}. Mackey et al. (2018, submitted) list a total of 92 clusters in M31 beyond a projected radius of 25kpc, of which 87 are more than 2 degrees from M31. This latter population are shown as red points in Figure~\ref{schematic} and their positions and luminosities are listed in Table~\ref{gcs}. Also included in the figure and table are the globular cluster populations of NGC147 and NGC185, to which PAndAS has contributed several new discoveries (\citealt{veljanoski2013a}). The other satellites of M31 that are known to have at least one globular cluster are Andromeda~I (\citealt{caldwell2017} and references therein) and the Pegasus dIrr (\citealt{cole2017} and references therein). Andromeda XXV has also recently been claimed to host a globular cluster (\citealt{cusano2016}). The Andromeda~I globular cluster is also indicated in Figure~\ref{schematic}, as are the 6 outer globular clusters of M33 that are listed in \cite{cockcroft2011}.

\cite{mackey2010}, Mackey et al. (2018) and \cite{veljanoski2013b, veljanoski2014} discuss several possible associations of M31 outer halo globular clusters found using both position and velocity data. Many of these coincide with stellar substructures that have been discussed in Section 4.1.1: Mackey et al. (2018) estimate between 35 -- 60\% of outer globular cluster lie on top of stellar substructures. A complete list of the possible groupings identified by these authors is summarised here for completeness:

\begin{itemize}
\item {\bf East cloud:} Two globular clusters are robustly associated with the substructure (PA-57 and PA-58; \citealt{huxor2014, veljanoski2014}), and both have similar velocities to within $\sim 20$\,km/s. \cite{mackey2018} suggest that PA-56 is third highly-likely member. \cite{mcmonigal2016} identify a further 3 globular clusters that are potentially associated with this substructure that may be extensions to the East Cloud (and which highlight the difficulty of identifying boundaries to these features).

\item {\bf Stream C:} Three GCs lie on top of the southern end of this feature, but they do not form a tight kinematic group. One of these - the well-studied HEC12 (also known as EC4; \citealt{huxor2005, collins2009, veljanoski2014}) - lies on top of, and shares a similar velocity to, the metal-poor component of Stream C. Another appears to match the velocity of the metal-rich component of Stream C. A further nine GCs lie on the northern part of stream C where it overlaps with Stream D. This group of GCs appears to split into two kinematic groupings of GCs, one containing 5 GCs and the other 3 GCs, with the last GC not being clearly associated with either group. 

\item {\bf Stream D:}  Three GCs lie on top of the southern end of this feature, but they do not form a tight kinematic group. The radial velocity signature of Stream D is uncertain, and so it is not possible to robustly associate any of these GCs to Stream D using kinematics.  As noted above, a further nine GCs lie on the northern part of stream D where it overlaps with Stream C. This group of GCs appears to split into two kinematic groupings of GCs, one containing 5 GCs and the other 3 GCs, with the last GC not being clearly associated with either group. 

\item {\bf South West Cloud:} Two globular clusters that are coincident in projection with the SW Cloud have similar metallicities to the structure, as measured by \cite{mackey2013}. A third globular cluster is identified by \cite{veljanoski2014} to overlap with the southern part of the feature and have a velocity that is similar to the other two (all three clusters taken together appear to trace a velocity gradient across the feature). Their velocities closely match that of the stars within the substructure (\citealt{mackey2014}, Mackey et al. 2018). 

\item {\bf North West Stream:} Seven GCs lie on top of this feature (the K2 component labelled in Figure~\ref{schematic}, and the 6 most distant from M31 show a clear trend in velocity suggesting they are all related (the seventh does not match this velocity trend).

\item {\bf Association 2:} \cite{mackey2010} and \cite{veljanoski2014} propose that a group of 11 GCs near the western major axis are a distinct subgroup of GCs (the highest density grouping of GCs in the outer parts of M31, at $(\xi \simeq 2, \eta \simeq -1)$ in Figure~\ref{schematic}). \cite{veljanoski2014} split Association 2 into two distinct subgroups (each with 4 clusters, with the membership of the three remaining clusters being ambiguous). One of these groups has a velocity close to the expected value derived from extrapolation of the velocity trend seen in the NW Stream GCs, suggesting that they could be associated with this feature. The second subgroup has a similar velocity compared to the outer disk in this region (\citealt{ibata2005}), suggesting a disk-origin for this subgroup.

\item {\bf Binary globular clusters:} While not discussed explicitly in the literature before, PAndAS-53 and PAndAS-54 lie within 130 arcsecs of each other at a distance of 7 degrees (96\,kpc) from the center of M31. They are visible as the ``single'' dot in Figure~\ref{schematic} close to Andromeda XV and XXIII, next to the dashed circle representing  100kpc from M31. Their extreme proximity in projection to each other suggests they may in fact form a binary pair. However, velocities from \cite{veljanoski2014} imply that they are not currently bound to each other ($v_{r, 53} = -253 \pm 10$\,km/s, $v_{r, 54} = -336 \pm 8$\,km/s), although perhaps they may have been associated in the past. We note \cite{holland1995} highlighted another potential binary globular cluster in M31, albeit at much smaller radii from the main galaxy.
\end{itemize}

\subsection{Connecting accretion events}

Nearly 30 different potential substructures -- arising as features in the stellar populations, in the dwarf galaxy population and in the globular cluster population -- are discussed in the previous subsection. In the stellar distribution alone, there are at least 13 different, distinct, structures visible in Figures~\ref{adapt}, \ref{schematic} and \ref{profile}. Under the reasonable assumption that substructures such as these are remnants of earlier accretion events, exactly how many accretion events are required to explain what PAndAS has detected in the vicinity of M31? Here, we summarise some relevant considerations.

\subsubsection{Clues from HST stellar population studies}

With respect to the stellar substructures in the inner ($R_p\lesssim 3$\,degrees) regions of M31, definitive studies of their stellar populations have been completed by \cite{ferguson2005, richardson2008} and \cite{bernard2015b} using deep HST ACS fields. These analyses provide compelling evidence of a causal connection between many of the structures labelled in Figure~\ref{schematic}. They analyze the CMDs of fields placed on the NGC205 loop, the Eastern Shelf, the Western Shelf, the Giant Stellar Stream, the G1 clump, Andromeda North-East, and other fields around the disk and in the halo of M31. As previously discussed, they identify several fields (including the G1 clump and Andromeda North-East) that have disk-like stellar populations, and several fields (the Giant Steam, but also the Western and Eastern Shelves) with stellar populations that look like the Giant Stellar Stream. This leads them to associate several substructures with debris from the progenitor of the stellar stream. They also associate several substructures with heating/disruption processes in the disk. For the latter, interactions between M31 and another system could potentially cause the disruption to the disk necessary to create these features (potentially a merger event or a fly-by interaction, e.g., with M33). Such a merger event could also produce the asymetric AGB distribution discussed in \cite{davidge2012a} and the substructure discussed in \cite{davidge2012b}.

In the scenario laid out above, the event that heats the M31 disk could plausibly be the progenitor of the GSS. If this was the case then, with reference to Figure~\ref{schematic}, it would imply the Giant Stream, the Western Shelf, the Eastern Shelf, Andromeda North-East and the G1 clump (5 of the 13 highlighted features) are all causally connected via a single accretion event (the first three as remnants of the merged satellite, and the latter 2 as debris from the disk of M31 as it responded to the induced gravitational perturbation). It is worth noting that recent work by \cite{hammer2018} consider a relatively major (4:1) merger in Andromeda about 2 - 3 Gyrs ago, and they conclude that it is possible that the majority of inner halo substructures may have been formed in a single event.

\subsubsection{Clues from the globular cluster population}

\cite{huxor2011} have already noted the excellent agreement between the shape of the radial profile of the globular clusters and that of the stellar halo that is visible in Figure~\ref{profile}. Combined with the obvious correlation between the spatial locations of the globular clusters and many of the stellar substructures, summarised earlier and discussed at length in \cite{mackey2010}, Mackey et al. (2018), and \cite{veljanoski2014}, the accretion events that put the stellar halo in place must also put in place the outer globular cluster population. 

We suggest a third consideration regarding the globular clusters that is relevant to the accretion history of M31. Namely, the sheer number of globular clusters that can be associated with stellar substructures. 

There are 5 stellar substructures that are apparently associated with globular clusters, and in each case there are at least 2 globular cluster companions identified. The integrated luminosities of each of these 5 stellar substructures are in the approximate range $-13 \lesssim M_V \lesssim -10$, and a total of $\sim 30$ globular clusters are potentially associated with them. The dwarf galaxies in the same luminosity interval in the Local Group are Sculptor, Leo I, Andromeda I, II, III,VI, VII,  XXIII, LGS3, Cetus, Pegasus dIrr, Leo A, Aquarius, Sag DIG and UGC4879. Of these, Andromeda I and the Pegasus dIrr have a single globular cluster each (\citealt{caldwell2017, cole2017} and references therein). At even fainter magnitudes, Eridanus II (\citealt{koposov2015, crnojevic2016}) and Andromeda XXV (\citealt{cusano2016}) may also each have a single globular cluster. The lowest luminosity Local Group galaxies with multiple confirmed globular clusters are the Sagittarius dSph ($M_v \simeq -13.5$; up to 9 globular clusters have been associated with this galaxy, see \citealt{bellazzini2003b, law2010}) and Fornax ($M_v \simeq -13.4$; 5 known globular clusters). However, the total number of globular clusters associated with relatively faint dwarf galaxies falls well short of the numbers that appear necessary to explain the frequency of globular clusters in the outer halo of M31.

It is considerably easier to reconcile the apparent number of globular clusters associated with the stellar substructures in the M31 halo with our knowledge of the globular cluster populations of Local Group dwarf galaxies if the progenitors of the substructures were considerably more massive than their present luminosities suggest. \cite{bate2014} and  \cite{mcmonigal2016} make the argument that the photometric metallicities measured by the East Cloud and the SW Cloud, respectively, imply that their original stellar masses were considerably larger (potentially by a factor of 2 or so). 

A more extreme version of this hypothesis is also possible, and that is that several of the stellar substructures with known globular clusters are actually from a {\it much} larger, single, progenitor (see also \citealt{veljanoski2014} and \citealt{ferguson2016}). 

The East Cloud and SW Clouds are both located  at large distances from M31 (of order 100\,kpc in projection). Further,  their long dimension is roughly tangential to the radius vector connecting them to M31. If the long dimension indicates the path of the stream, then this imples these systems are on roughly tangential orbits (although we caution that a stream does not follow the progenitor's orbit exactly). It is difficult to disrupt dwarf galaxies at such large radius unless they happen to be at apocenter on an otherwise quite radial orbit (which we stress is entirely possible given the data available).

\begin{figure*}
  \begin{center}
    \includegraphics[angle=270, width=15cm]{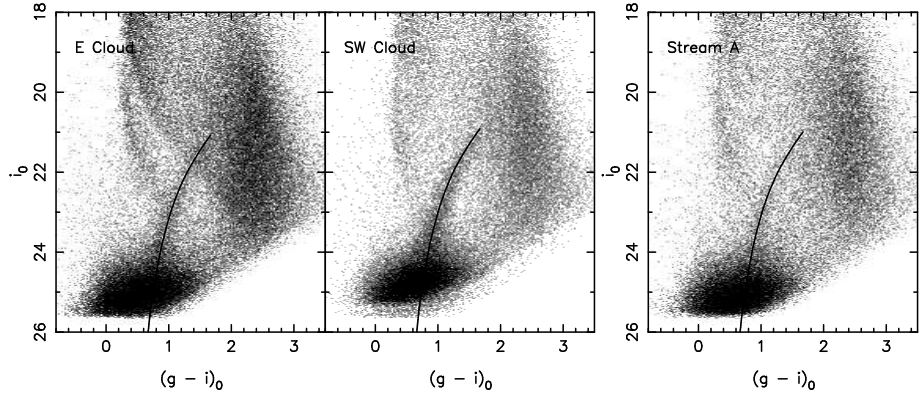}
    \caption{CMDs for the East Cloud, South-West Cloud, and Stream A. Only stars within the relevant polygons shown in  Figure~\ref{schematic} are plotted. The isochrone overlaid in each panel is shifted to the distance of M31 and corresponds to a metallicity of [Fe/H] $= -1.4$ dex. Based on these CMDs, it is reasonable to postulate that all three structures - and at least the East and South-West Clouds - have a similar origin as shell-debris from a single accretion event.}\label{shells}
  \end{center}
\end{figure*}

An alternative scenario to consider is that the East Cloud and SW Cloud (and possibly also Stream A given its broadly similar morphology and distance from M31), are in fact ``shells'' that have been created by a more massive system on a much more radial orbit that ejects material (stars and globular clusters) out to large radius. In this scenario, the Clouds are the large-radius analogs to the Eastern and Western Shelves (although we are not suggesting that they were created by the same progenitor). 

Such an interpretation is appealing since it explains the existence of ``tangential streams'' at large radii. Further, it explains why we are not able to continue tracing them over a larger area (as is the case for the K2 segment of the NW Stream), since we instead only see that part of the feature where the stars have ``piled up'' at apocenter. Critically, this interpretation provides a natural explanation why such faint structures should be associated with a (relatively) large number of globular clusters, and why the photometric metallicities of these faint structures are surprisingly high.  

If the above interpretation is to be credible, the stellar populations of these features should be similar. Figure~\ref{shells} shows their CMDs, here selected from inside the relevant polygons shown in Figure~\ref{schematic}. \cite{mcmonigal2016} and \cite{bate2014} derive distances ($814^{+20}_{-9}$ kpc and $757^{+9}_{-30}$ kpc) and metallicities ([Fe/H] $\simeq -1.2$ and $-1.4$ dex) to the East Cloud and SW Cloud, respectively. No distance or metallicity estimate has been published for Stream A. Figure~\ref{shells} shows an isochrone corresponding to [Fe/H] $= -1.4$\,dex overlaid on each CMD, moved to the correct distance for each feature (in the absence of other information, we adopt a distance equal to that of M31 for Stream A). This isochrone is consistent with the earlier estimates, and demonstrates that all three features have plausibly similar stellar populations. With current data, it therefore seems possible that some or all of these structures could have a common origin as debris from a single accretion event.

\subsubsection{The best-guess accretion history of M31}

What is the best guess accretion history of M31 that results from all the recent discoveries and investigations of the outer halo that PAndAS has made or motivated? As the previous discussions will have made clear, there are still a lot of uncertainties and unknowns when it comes to interpreting what we observe in M31's halo. Nevertheless, the multitude of work and results discussed up to this point appear consistent with the following broad-brush merger history for M31:

\begin{itemize}
\item M31 accreted up to about one-quarter of its stellar halo at early times. This is the fraction of stars in the stellar halo that we now identify as contributing to the ``smooth'' stellar halo of M31 (see Section~5.2 and \citealt{ibata2014a}). This was long enough ago that the individual objects are no longer recognizable as distinct substructures based on their spatial morphologies. We cannot determine whether these stars have a common origin, possibly in a major merger, or were the result of a larger number of disparate processes.
\item About 3 Gyrs ago, it is possible that M33 had a pericentric passage with M31, triggering the formation of tidal tails around M33 and disrupting the HI disk of M33. Possibly, M33 lost much of its own extended halo, and especially globular clusters and satellite galaxies, in this encounter, and would imply some of the M31 satellites are in fact accreted M33 satellites. However,  see \cite{patel2017} and \cite{vandermarel2018} for an alternative M33 orbital history;
\item About 2 -- 3 Gyrs ago, the NGC147/185 subgroup had a pericentric passage with M31; the orbital characteristics of the subgroup meant that NGC147 passed closer to M31 and formed a tidal stream, whereas NGC185 appears less perturbed. The subgroup became unbound by the encounter, but continued on similar orbits;
\item If the Plane of Satellites has a causal physical origin, and since it contains the NGC147/185 subgroup, it follows that all these galaxies were accreted into the M31 subgroup at around the time of the accretion of the NGC147/185 subgroup, 2 -- 3 Gyrs ago.;
\item About 1 Gyr ago, a galaxy slightly smaller than the mass of the LMC fell in on a nearly radial orbit and had a destructive pericentric passage with M31, forming the GSS, the Eastern and Western Shelves, and potentially disrupting a lot of the outer extended disk of M31. In principle, this disk disturbance could explain the star forming ring in M31, as well as the origin of the G1 clump, the NE Structure, and the substructure discussed in \cite{davidge2012b}
\item At some point in the last few Gyrs, the object that we identify as Andromeda XXVII was accreted into the M31 satellite system. During the disruption of the progenitor, the stream we identify as K1 was likely formed. Of the remaining substructures, it is possible that at least the East Cloud and SW Cloud originated in a single event. In principle, this could be the same event that produced K1, and at most could maybe even explain Streams A, C and D. Detailed modeling needs to be conducted to see just how plausible this is. Kinematics, however,  suggest that the K1 structure is not associated with K2 (Preston et al., {\it in preparation}), and so this would imply the need for at least a second independent event. 
\item Several of the progenitor systems must have possessed significant globular cluster systems of their own, and were likely dwarf galaxies with the approximate stellar mass of Fornax, or larger. Specifically, the East Cloud and the SW Cloud may have been formed as shell structures in an event or events involving a progenitor with several globular clusters. Likewise, the progenitor of Streams C, D and K2 appear to have had significant globular cluster populations.
\item The minimum number of events that could have produced the features we see in the PAndAS map is 5 (the M33 interaction, the NGC147/185 interaction, the GSS event, the K1 event and the K2 event, assuming the latter two also produce the remaining features in some combination). At the other extreme, each of the remaining features could be the result of separate events. In this scenario, we would require a maximum of 12 events. 
\item The connection of M32 to any of the substructures surrounding M31 is unclear. The majority of compact elliptical galaxies so far identified are associated  with interactions with a more massive companion, and so this implies that M32 is responsible for at least some of the accreted stars or substructures in M31's stellar halo. It is important to note that M32 overlaps in projection with the GSS, but \cite{ibata2004} show that the kinematics of the stream and M32 are inconsistent with the two being directly associated in a simple way. It has not therefore been possible to firmly link, or disprove, a connection of any of the substructures in M31's halo to M32. \citealt{block2006}, however, have demonstrated the possible connection of M32 to features in the M31 disk. Immediately prior to submission of this manuscript, a new paper by \cite{dsouza2018} propose that M32 is in fact the remains of the  progenitor of the GSS (presumably on a different wrap around M31 in order to allow consistency with the kinematics) and is responsible for many of the stars in the inner halo. They further propose that it is this interaction that leads to the enhancement of star formation approximately 2\,Gyrs ago in M31.
\end{itemize}

The possible M33 interaction, NGC147-subgroup interaction, the GSS-progenitor interaction, the possible M32 interaction, as well as potentially several smaller interactions, make clear that the last few Gyrs of M31's history has been anything but dull. Clearly, the individual substructures that are most clearly defined in PAndAS tend towards younger ages. It makes sense that the most obvious features we can observe are relatively dynamically young, and it points to the need to obtain deeper photometry to identify fainter, potentially dynamically older, accretion events. Further, it highlights the need to obtain extensive kinematic information for the M31 halo, to try to discern accretion events via their dynamics (since the overall phase-space density is conserved).

\section{The stellar mass budget of the M31 halo}

We now present the break-down of the stellar mass budget of the M31 halo. Specifically, we estimate the stellar mass found in dwarf galaxies in comparison to globular clusters in comparison to obvious stellar substructures in comparison to the stellar mass in the rest of the halo (the ``smooth'' component). 

\subsection{Integrated luminosity estimates of stellar substructures}
\label{n147}

\begin{figure}
  \begin{center}
    \includegraphics[angle=270, width=12cm]{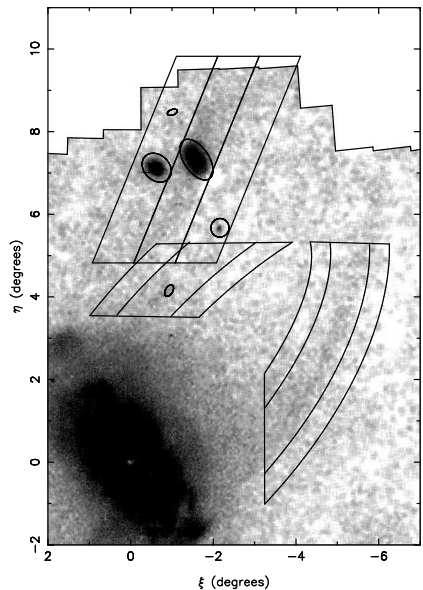}
    \caption{Zoomed version of Figure~\ref{adapt} focusing on the north-western regions of the survey. The polygons mark the regions used to estimate the luminosities of the NGC147 stellar stream and both segments of the northwest stream, along with neighbouring regions used to estimate the background contamination (foreground stars, background galaxies, and M31 field populations). Ellipses demarcate five half-light radii from the centers of (from left to right) NGC185, Cassiopeia II, NGC147, Andromeda XXV and Andromeda XXVII. Stars within these regions are omitted from the analysis.}\label{polygons}
  \end{center}
\end{figure}

\begin{figure*}
  \begin{center}
    \includegraphics[angle=0, width=12cm]{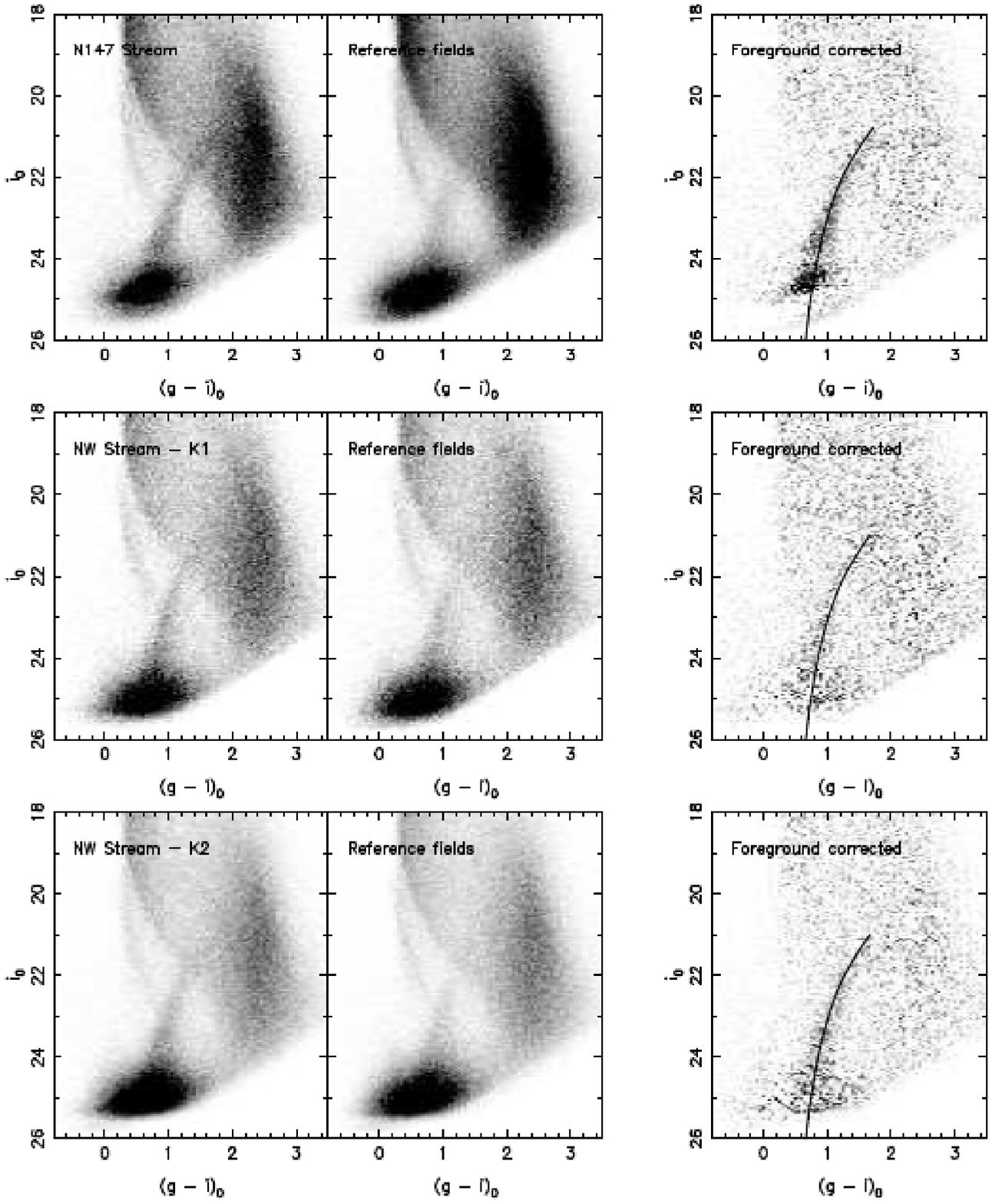}
    \caption{Hess CMDs of the NGC147 stream (top panels), the North West K1 stream (middle panels), and the North West K2 stream (bottom panels). These CMDs are constructed from stars within the areas shown in Figure~\ref{polygons}. The left panels show the field centered on the main substructure, the middle panel shows the reference fields, and the right panels show the subtracted CMDs. For the latter panels, an isochrone is overlaid: for the NGC147 stream, this is shifted to the distance modulus of NGC147 and has [Fe/H] $= -1.3$ dex; for the other two substructures, these are shifted to the M31 distance modulus and have [Fe/H] $= -1.4$ dex.}\label{newcmds}
  \end{center}
\end{figure*}

A few of the major substructures indicated in Figure~\ref{schematic} have not previously had their luminosities estimated. In order to construct a stellar mass budget, we first provide estimates of the integrated luminosities of the NGC147 stellar stream (L1 and L2 in Figure~\ref{schematic})  and the NW Stream (K1 and K2 in Figure~\ref{schematic}).

Figure~\ref{polygons} shows a zoomed version of the stellar density map shown in Figure~\ref{adapt} that focuses on the north-western region of the survey. For the ensuing analysis, we select stars within polygons from the original stellar catalog (not the foreground subtracted catalog) that are positioned so as to contain the majority of the stars that define the NGC147 stream and both segments of the NW stream as indicated in the figure. 

For each of the three main science regions, we also select regions on either side of these polygons as reference fields. These reference fields straddle the main science regions in order to try to provide as accurate an empirical measurement of the contamination in the main science region (from foreground stars, background galaxies and the M31 field population) as is possible, while still avoiding any other prominent substructures. Each individual reference field is scaled to the same area as the science region, and the average value of the two applicable reference regions is used as an estimate of the contamination in the science region. Ellipses are drawn for each of the five dwarf galaxies in this region (NGC185, Cassiopeia II, NGC147, Andromeda XXV and Andromeda XXVII) at five half-light radii from the centers of these systems, and all stars within these ellipses are ignored in the creation of the science and reference regions (we adopt 8 half-light radii from NGC185, since we found that a smaller radius did not fully remove the stellar signature of this galaxy from our reference field).

Figure~\ref{newcmds} shows, for each of the substructures, Hess CMDs of stars in the main science regions, the (scaled) reference CMDs, and the foreground-subtracted CMDs. These are shown with a linear scaling and only positive residuals are shown for the foreground-subtracted CMDs for clarity. In each case, a residual RGB is clearly present.

\subsubsection{The NGC147 stream}

For the NGC147 stream, an isochrone is overlaid on the residual Hess diagram in Figure~\ref{newcmds} corresponding to an age of 12\,Gyr and [Fe/H] $= -1.3$\,dex at the distance of NGC147. If the age and distance assumptions are reasonable, then this is consistent with the implied photometric metallicity of the main body of NGC147 as shown in \cite{crnojevic2014}. These authors do find a slight metallicity gradient in NGC147, and the outermost metallicity distribution function that they analyse (at a distance of $\sim 0.8$\,degrees) shows a shift to more metal-poor values, with a modal metallicity of [Fe/H] $\simeq -1.3$\,dex consistent with the apparent photometric metallicity of the stream. 

To estimate the luminosity of the NGC147 stream, we follow \cite{mcconnachie2010b} and others, and first estimate the $g$-band luminosity of all stars within the top 2.5\,mags of the stream CMD (corrected for the luminosity of stars in the same luminosity range in the reference fields). We then empirically correct this ``RGB luminosity'' ($m_{RGB}$) to a total luminosity by comparison to features for which we have direct estimates of their integrated magnitude.

Specifically, we use the PAndAS data to calculate $m_{RGB}$ for Andromeda II, III, NGC147 and 185, since these are within our footprint and have luminosity estimates based on integrated light measurements (\citealt{mcconnachie2006b} for the dwarf spheroidals and \citealt{crnojevic2014} for the dwarf ellipticals). We do not use Andromeda I due to the high background (in the resolved stellar populations) caused by the presence of the Giant Stellar Stream. We compare the $m_{RGB}$ values for these galaxies to their integrated apparent magnitudes in the V band, and find that $\Delta\,m = (2.8, 3.1, 3.2, 2.9)$ mags for Andromeda II, III, NGC147 and 185, respectively. These values include an offset from the $g$ to $V$ bands. There is some inherent uncertainty in this transformation caused by the unknown luminosity function of the streams relative to dwarf galaxies, as well as uncertainties in the original measurements of the integrated magnitudes of these galaxies. However, the relative consistency between these 4 measurements suggests that these systematic uncertainties are at the level of 0.2\,mags or thereabouts. For the ensuing analysis, we adopt $\Delta\,m =  3.0$ mags. We also note that in conducting this analysis, we discovered an error in the reported magnitudes of NGC147 and NGC185 in \cite{crnojevic2014} upon integration of the S{\'e}rsic profiles for these galaxies. In particular, we calculate that $M_g^{N147} = -15.4$ and $M_g^{N185} = -15.1$. Following \cite{crnojevic2014}, this corresponds to $(M_V, M_I)_{N147} = (-15.8, -17.1)$ and $(M_V, M_I)_{N185} = (-15.5, -16.8)$ for an adopted value of $(g-i)_0 = 1.2$ for both systems.

We estimate that the NGC147 stream has $m_{RGB} \simeq 15.0$\,mags. Correcting for $\Delta m$ and assuming that the stream is at the same distance of as the main body of NGC147, we find that $M_V = -12.2 \pm 0.5$ mags. Here, the uncertainty is estimated based on the dominant systematic uncertainties of our technique (the unknown distance of the stream, the unknown luminosity function of the stream, and the uncertainty in the integrated magnitudes that are used to estimate $\Delta m$). Overall, it appears that the stream contains $\sim 4$\% of the total stellar mass of NGC147.

\subsubsection{The North West stream(s)}

A direct connection between the two ``segments'' of the northwest stream has previously been proposed given that they appeared to align along a large ellipse. However, the reprocessed data casts this original claim into doubt, and kinematics being analysed by our team (Preston et al., {\it in preparation})  demonstrate that a connection between these segments is unlikely. Thus, we analyse these segments as distinct features. 

The rightmost panels of the middle and lower rows of Figure~\ref{newcmds} show the CMDs for the K1 and K2 features, respectively (using the naming from Figure~\ref{schematic}). A 12\,Gyr, [Fe/H] $= -1.4$\,dex isochrone is overlaid in both panels, assuming a distance to each feature equal to the distance of M31. We stress that, given the vast area over which these features extend, it is not clear that either of them will be at the same distance as M31. 

We estimate the ``RGB-luminosity'' of K1 and K2 in a similar way as for the NGC147 stream, where we assume they are both at the same distance as the main body of M31. We measure  $m_{RGB,K1} = 16.9$ and $m_{RGB,K2} = 15.2$, corresponding to $M_{V,K1} = -10.5 \pm 0.5$ and $M_{V,K2} = -12.3 \pm 0.5$. 

Andromeda XXVII (\citealt{richardson2011}) sits right on top of the K1 stream, and is considerably fainter than the stream that surrounds it ($M_V = -7.9$). Kinematic analysis by \cite{collins2013} suggests that the galaxy is not in dynamical equilibrium. This is backed up by \cite{martin2016} and \cite{cusano2017}, the latter of whom suggests that the galaxy is completely destroyed. It seems reasonable, therefore, to link Andromeda XXVII as the progenitor of the K1 stream.

No progenitor for the K2 stream has been identified, although it is associated with a large number of globular clusters. We note that K2 is as bright as the NGC147 stream, and is approximately the same luminosity as Andromeda II.

\subsection{The break-down of stellar mass in the M31 outer halo}

\begin{figure*}
  \begin{center}
    \includegraphics[angle=270, width=18cm]{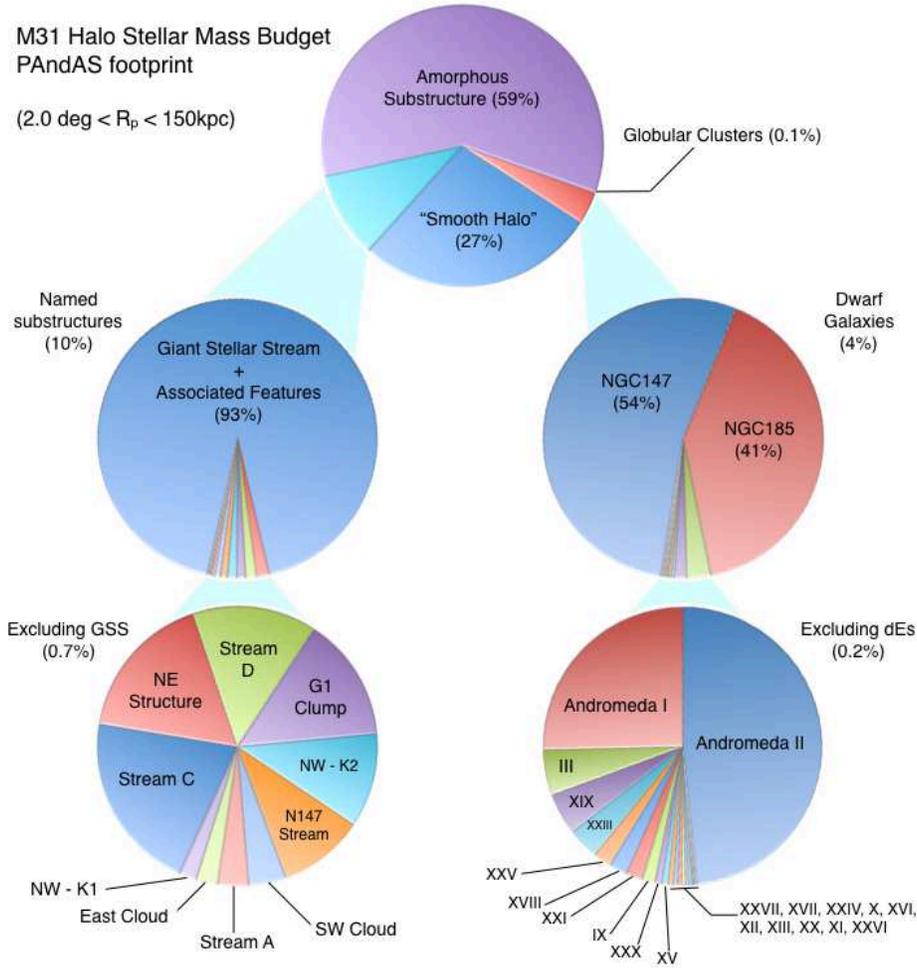}
    \caption{The stellar mass budget in the outskirts of M31, as estimated from PAndAS data. We only consider objects in the projected radial range between 27.2 kpc (2 degrees) and 150\,kpc from M31, and which lie in the PAndAS footprint. The ``halo'' and ``Amorphous substructure'' elements are estimated in Paper I; the remaining elements are from Tables~\ref{dwarfs}, \ref{gcs} and \ref{substrlum}.}\label{pie}
  \end{center}
\end{figure*}

\begin{figure}
  \begin{center}
    \includegraphics[angle=270, width=8cm]{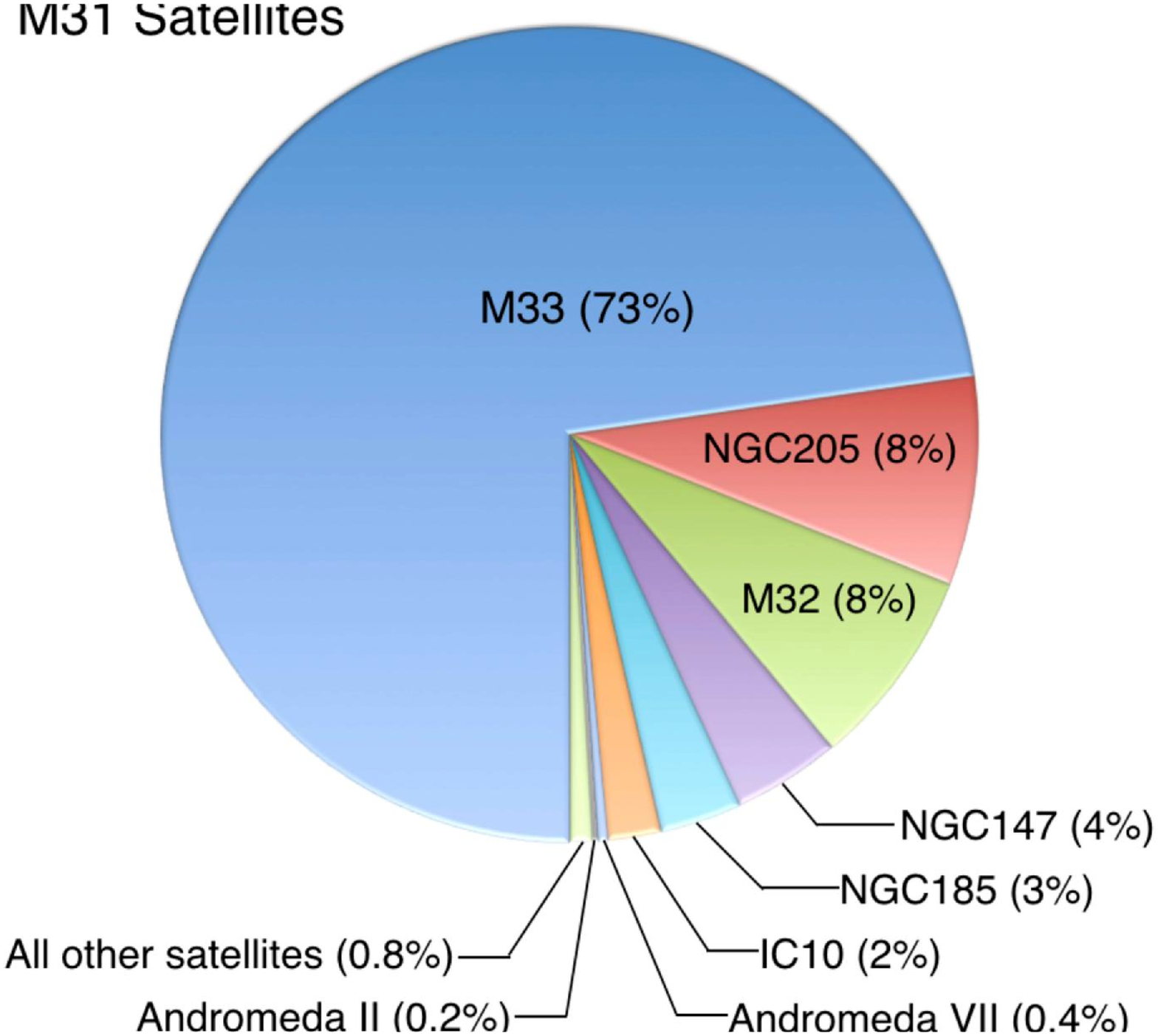}
    \caption{The distribution of stellar mass between all M31 satellites (all objects in Table~\ref{dwarfs}, regardless of their position relative to M31).}\label{pie_sats}
  \end{center}
\end{figure}

Tables~\ref{dwarfs}, \ref{gcs} and \ref{substrlum} list all outer halo globular clusters, dwarf galaxies and stellar halo substructures that have been identified in the surroundings of M31 and which are visible in the PAndAS footprint.
For each of the dwarf galaxies, globular clusters and substructures, we have converted luminosities to stellar masses adopting a stellar mass-to-light ratio of $\Upsilon_\star \approx 1.2\,M_\odot/L_\odot$, which \cite{mcgaugh2014} deem a suitable crude approximation for the optical bands if multiple colors are not available ($K-$band is particularly useful here). We note that this is in general most appropriate for older stellar populations. For a few of the dwarf galaxies -- notably M33 and IC10 which have significant young populations -- this is not a great approximation. However, for the purpose of the ensuing analysis, adopting a constant $\Upsilon_\star$ is a convenient approximation that does not change the results of this qualitative comparison. For the dwarf galaxies in Table~\ref{dwarfs}, distances and magnitudes are taken from the updated compilation of \cite{mcconnachie2012}\footnote{http://www.astro.uvic.ca/$\sim$alan/Nearby\_Dwarf\_Database.html}. Many of the adopted values come originally from studies based on PAndAS data, especially \cite{conn2012} and \cite{martin2016}.

The outer halo globular clusters in Table~\ref{gcs} are taken from the compilation of Mackey et al. (2018). In addition, we have included the NGC147/NGC185 clusters from \cite{veljanoski2013a}, as well as the Andromeda I globular cluster described in \cite{caldwell2017}. For completeness, we have also included the six M33 outer halo globular clusters as listed in Table 3 of \cite{cockcroft2011}. We have not included the candidate cluster in Andromeda XXV (\citealt{cusano2016}) since its nature is still ambiguous.

For the stellar substructures in Table~\ref{substrlum}, unless an independent distance to the structure has been derived, we adopt the M31 distance modulus (\citealt{mcconnachie2005a}). For the NE Structure, we have converted the original estimate made in $g_{SDSS}$ by \cite{zucker2004b} to the $V$ band\footnote{http://www.sdss3.org/dr8/algorithms/sdssUBVRITransform.php}. For the G1 Clump, we have converted the Isaac Newton Telescope $V'$ band estimate by \cite{ferguson2002} to the standard $V$ using transformations provided by \cite{mcconnachie2005a}. 

We estimate the stellar mass budget of the M31 halo by comparing the estimated stellar masses of the objects in Tables~~\ref{dwarfs}, \ref{gcs} and \ref{substrlum} to the halo stellar mass budgets presented in Tables 4 and 5 of Paper I. These authors calculate the stellar mass of M31's halo in the projected radial range between 27.2 kpc (2 degrees) and 150\,kpc. Thus, we also only include those objects from the tables that fall within this range. They find a total stellar mass in this range of $10.5 \times 10^9 M_\odot$, that is reduced to $3 \times 10^9 M_\odot$ when they automatically mask out regions with significant positive residuals, that they identify as substructure. This implies a total mass in substructure in this region that is greater than the total mass in ``named substructures'' in Table~\ref{substrlum}. For the purposes of this budget, we shall call this excess substructure ``amorphous substructure'', that is not attributed to any specific feature. This highlights the difficulty in coherently defining and quantifying stellar substructure given a dataset such as PAndAS, and we endeavor to develop a more satisfying quantification of M31's stellar halo substructure in the next section.

In the meantime, the stellar mass budget of the M31 halo between 27.2 kpc (2 degrees) and 150\,kpc is shown in Figure~\ref{pie}. The top chart shows the split between the ``smooth'' halo, the stellar substructure, the dwarf galaxies, and the globular clusters. The substructure and dwarf galaxies sectors of pie are then exploded to show the break-down between the individual features in each category. For the substructures, we show the pie chart corresponding to all the relevant substructures, in addition to a second exploded pie of everything except the GSS. For the dwarf galaxies, a second exploded pie shows the break-down between all the dwarfs after exclusion of the two dwarf ellipticals.

For the dwarf galaxies in particular, the imposed radial range removes some very significant galaxies from consideration (the most massive of which are M33, M32 and NGC205). Figure~\ref{pie_sats} therefore shows the stellar mass budget of all the known satellites of M31, irrespective of their projected position relative to M31. 

Figures~\ref{pie} and \ref{pie_sats}  are an attempt to synthesize a very large and sometimes irregular body of literature. The stellar mass estimates on which they are based are necessarily uncertain, and the quantification and definitions of substructure are, at best, subjective. Even with these extensive caveats, however, these figures are a revealing visual summary of the stellar mass distribution in the surroundings of M31. Numerous aspects are worth highlighting:

\begin{enumerate}
\item A majority of the stellar halo of M31 is in the form of stellar substructures;
\item To within a factor of a few, the overall stellar mass of the smooth halo (potentially completely destroyed dwarf galaxies), halo substructures (more recently destroyed dwarf galaxies), and ``surviving'' dwarf galaxies, are comparable;
\item The contribution to the stellar mass budget of halo substructures and dwarf galaxies are, in both cases, dominated by the one or two most massive systems (the GSS and associated structures, and the NGC147/185 pairing, respectively); 
\item Globular clusters contribute a negligible fraction of the stellar mass of the surrounding of M31. This is true also for all but the most massive of the dwarf galaxies and stellar substructures;
\item Excluding the most massive substructures and dwarf galaxies, the contribution to the stellar mass budget of dwarf galaxies, ``named'' substructures, and globular clusters, are comparable;
\end{enumerate}

\section{The hierarchy of structure in the M31 stellar halo}

Figure~\ref{pie} and the associated discussion highlights a fundamental issue with the current analysis of M31's stellar halo, that extends more broadly into discussions of halo substructure in general. Namely, when is a clump not a clump? When instead is it a cloud, or a stream, or a dwarf galaxy, or a star cluster, or a tidal remnant, or a dissolving dwarf, or just a common or garden overdensity? And how far does a stream extend? How can we best define its boundaries? And how can we do this consistently for all ``substructures'' in a single halo and between different halos? Indeed, is it even meaningful to do this given the obvious interconnections between substructures as has been discussed throughout this paper? The fact that most of the mass in M31's halo is, according to Figure~\ref{pie}, categorized as ``amorphous substructures'', implies that our (subjective) methods for the identification of substructures are not comprehensive or complete.

Taking a step back and re-examining the stellar distribution of M31's halo, for example in Figure~\ref{adapt}, it is clear that the stellar halo contains structures on all spatial scales i.e., up to scales of many degrees across, and potentially down to scales set by the seeing of the site (an arcsecond or so). In addition, whatever structures are present come with a range of shapes and (projected) density profiles. Finally, the structures are potentially hierarchical (with the term being  used with reference to morphology, not formation). For example, NGC147 is a distinct entity, but it is also an entity within the larger structure that includes its stellar stream, and that is an entity within a larger structure that includes the entire NGC147/185 subgroup. 

How can we robustly and  consistently quantify the clustering of a set of points, where the clusters may be any given shape, and given that the points may cluster on different scales, with potentially the same points clustering on a range of different scales, i.e., hierarchically? Further, how can we do so in a relatively simple way that is flexible enough to be used on different datasets, and where the derived description is conceptually simpler and less complex than the original set of points that are being described? After all, one of the major purposes of coming up with such a description of the data is to be able to use quantifiable metrics to objectively compare and contrast different systems without having to tune the metrics to very specific (and potentially subjective) situations.  The authors on this paper have contemplated this complex issue for several years with relation to M31's halo and halo substructure in general, but we note that this issue is fundamental to many different areas of astronomy (e.g., galaxy clustering, N-body simulations and halo finding, any multi-parameter clustering problem).

As part of our investigation into this problem, we have searched the non-astronomy literature for insights into hierarchical clustering problems. In particular, we have investigated the use of the OPTICS algorithm (\citealt{ankerst1999}). OPTICS appears to be used most in the general fields of computer science and analytics. It is an extension of DBSCAN (\citealt{ester1996}). Descriptions of the DBSCAN and OPTICS algorithms are given in the Appendix. In brief, OPTICS provides an objective quantification of the clustering of an N-dimensional dataset by determining the spatial scale necessary to associate any given point to a cluster (termed the ``reachability distance'' of the point).

\cite{zhang2013} use OPTICS to help develop a way to identify and characterize geographical neighbourhoods based on social media “check-ins”. This included developing a technique that accurately predicts the home neighbourhoods of Twitter users. It turns out that this same algorithm appears to shows considerable promise in helping to understand the hierarchical structure of stellar halos.

DBSCAN has more than 12000 citations, and  OPTICS has more than 3000 citations\footnote{From Google Scholar}. It was our understanding during most of the preparation of this work that none of these citations came from astronomy. However, during the finalization of this manuscript, we were made aware of a previous application of OPTICS in astronomy - specifically to this same problem of stellar halos - by \cite{sansfuentes2017} . At around the same time, a very recent paper appeared on the arXiV that applies DBSCAN to Gaia data (\citealt{castroginard2018}).

We note that another variant of DBSCAN, ``Hierarchical DBSCAN'' (HDBSCAN; \citealt{campello2013}) could also be quite successful for this problem, although here we only discuss OPTICS. 

\subsection{The application of OPTICS to the M31 halo}

\begin{figure}
  \begin{center}
    \includegraphics[angle=270, width=8cm]{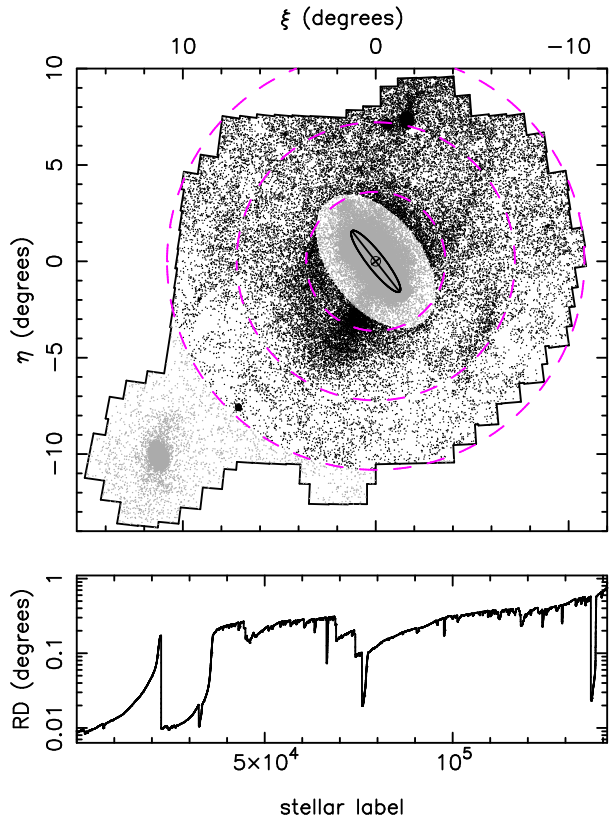}
    \caption{Top panel: distribution of stars on which we run the OPTICS algorithm (black points), compared to the full footprint (gray points). Only stars in the foreground-subtracted dataset are analysed, and we apply a magnitude cut at $i = 22.4$ in order to create as large a dataset as can be processed (141 117 stars) using our current implementation of the algorithm. For clarity, only one-third of the stars are plotted (only 10\% within the inner large ellipse). Bottom panel: the reachability diagram for the M31 stellar halo, with the minimum number of points required to define a cluster set at 0.1\% of the total number of points. Stellar substructures (clusters) are valleys in these diagrams. See the Appendix for a description of the OPTICS algorithm.}\label{optics}
  \end{center}
\end{figure}

We apply the OPTICS algorithm to the M31 dataset; more specifically, we apply it to the foreground-subtracted version of the dataset used to make Figure~\ref{adapt}. However, the calculation of the reachability diagram for OPTICS scales as $N^2$ in our current implementation of the algorithm. Thus, we are limited in the number of stars for which it is practical to run the algorithm. We therefore choose to apply a cut to our data beyond 150\,kpc in projection (i.e., removing the M33 region), and we also excise the central regions of M31 that is dominated by disk-related structures (defined by a large ellipse with position angle of $38.1^\circ$, an ellipticity of 0.4, and a semi-major axis of 4 degrees). These spatial cuts are shown in the top panel of Figure~\ref{optics}; gray dots show the full survey area, and black dots show those regions satisfying our spatial cuts. Even then, however, there are too many points for us to successfully run OPTICS, and so we apply an additional magnitude cut of $i = 22.4$. This selects 141,117 stars; we found this was essentially the maximum number of stars we could run our implementation of OPTICS for, using a  250GB RAM machine (the actual calculation is relatively rapid, requiring only $\sim 40$\,minutes on the 16-core machine). For clarity, only one-third of these points are plotted in the upper panel of Figure~\ref{optics}, except in the inner gray ellipse, where only 10\% of points are plotted. 

OPTICS formally requires two parameters as input, but in practice only one parameter matters. The primary parameter is the minimum number of points required for an object to be considered a cluster, and the secondary parameter is the maximum spatial scale to explore ($\epsilon$, and the maximum value of $l$, respectively, using the nomenclature introduced in the Appendix). However, by setting this second parameter to something larger than the maximum spatial scale probed in the data, OPTICS essentially requires only a single parameter. Large values of $\epsilon$ cause the reachability diagrams to appear relatively smooth compared to smaller values, with the small structures (valleys) that appear for small $\epsilon$ being absent. If $\epsilon$ is too small, however, then it becomes difficult to distinguish real structures from effects due to shot noise. Our initial presentation of OPTICS in this paper is to demonstrate its characteristics and usefulness in the taxonomy of stellar halos and their substructure (and indeed for other branches of astronomy). Given this purpose, we present results for only a single value of $\epsilon$, corresponding to $0.1\%$ of the total number of points. This value appears sensible, given that Figure~\ref{pie} suggests that many of the ``obvious'' substructures are $\sim 0.1 - 1$\% by mass of the total stellar mass. 

The lower panel of Figure~\ref{optics} is the reachability diagram derived using OPTICS for the stellar distribution shown in the top panel. Individual stars are ordered along the x-axis; the y-axis is the reachability distance of each individual star, that is the value of $l$ that would need to be adopted to ensure that the star was considered a member of a cluster with at least 141 members, using the criteria of DBSCAN.  Clusters of points (stellar substructures) appear as valleys in these diagrams. 

\begin{figure*}
  \begin{center}
    \includegraphics[angle=270, width=15cm]{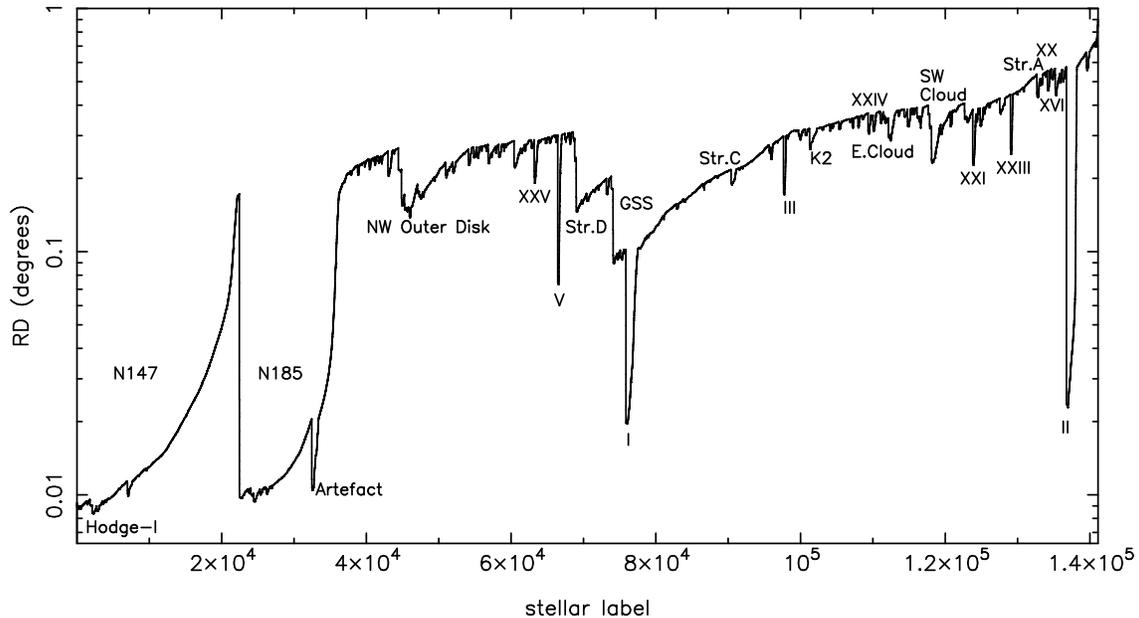}
    \caption{The reachability diagram for the M31 stellar halo, same as the lower panel of Figure~\ref{optics}, with the minimum number of points required to define a cluster set at 0.1\% of the total number of points. The labels are a result of cross-matching any notable valleys in this diagram with the previously known stellar substructures in the halo of M31. Note that the size and extent of these features in this diagram may not correspond directly to their previously-accepted (subjective) boundaries.}\label{labels}
  \end{center}
\end{figure*}

Figure~\ref{labels} shows the same reachability diagram as the bottom panel of Figure~\ref{optics}. Here, we have examined the points corresponding to all of the notable valleys in the diagram, and cross-matched these with known features of M31's halo. It is impressive to see that OPTICS identifies a large number of the prominent -- and not so prominent! -- substructures that were previously identified. This includes structures of vastly different morphology and luminosity. The bright dwarf galaxies -- NGC147, 185, Andromeda I, II, II and V -- all stand out extremely clearly in this diagram, and it is curious to see that a dip at the bottom of the valley corresponding to NGC147 coincides in position with the location of the NGC147 globular cluster Hodge I (although possibly this is a coincidence). A prominent dip in the valley corresponding to NGC185 actually corresponds to a spurious stellar detection caused by an artefact in the data. Note that objects have been labeled in this diagram if any of the points in the corresponding valley are in common with the previously known substructures. Therefore, the overall shape of the features labelled in these diagrams need not correspond exactly to the (subjective) boundaries of the substructures as previously identified.

\subsection{Automatic identification of clusters in reachability diagrams}\label{auto}

Figures~\ref{optics} and \ref{labels} show that OPTICS is potentially useful at identifying known substructures. However, it is also important to objectively and robustly define clusters in the reachability diagram without using prior knowledge. Unlike many clustering algorithms, OPTICS does not do this automatically, since it is intended primarily as a means of visualizing the dataset.

To automatically identify clusters within reachability plots, we define a custom algorithm inspired by, but different from, the algorithms of \cite{sander2003} and \cite{zhang2013}.

First, we attempt to reduce the impact of shot noise by smoothing the reachability diagram using a Gaussian with a FWHM of $\epsilon/2.5$. The factor of 2.5 is introduced so that clusters of interest (those with at least $\epsilon$ members) are still sampled at the Nyquist level. In our experiments, we found (as expected) that the smoothing factor is most useful when $\epsilon$ is very small, and that it is not really relevant if $\epsilon$ is large. 

Since clusters are valleys in reachability plots, our algorithm is based on the premise that at least one edge of the cluster is bounded by a peak. Conceptually, we consider an imaginary horizontal line that descends the y-axis of the reachability plot. Each time a peak is encountered, the line splits, and each split of the line represents a new cluster that is a child of the cluster represented by the line before it split. Defined in this way, we note that some of our clusters will possibly contain a few ``stray'' members, that are those points immediately after the first peak or immediately before the second peak that have reasonably large reachability distances compared to the rest of the cluster.

We proceed as follows: we identify all peaks within the smoothed reachability plot. Each peak represents one boundary of a potential cluster to its left, and a second potential cluster to its right (unless the points are at the extreme end of the diagram, in which case there is only one potential cluster to consider). To find the other boundary, we draw a horizontal line to the left and right of this peak; the point where this next intercepts the reachability diagram is the other boundary to the potential cluster.  All points in between these boundaries are members of these potential clusters.

We then consider each potential cluster in turn. We discard all potential clusters that do not have at least $\epsilon$ members, since these are too small to be considered significant.  We also compare the median reachability distance of all the members of the potential cluster to the reachability distance of the boundary points. If the boundary value is at least a factor of $\Delta$ larger than the median of the points contained within the boundaries, then the cluster is considered significant.  Results are not particularly sensitive to the precise value of $\Delta$, but in general if $\Delta$ is too large then obvious clusters are missed, and if $\Delta$ is too small then features that do not appear to represent any obvious substructures are selected. We set $\Delta = 1.1$ and obtain good results on both simulated and real data.

Finally, we sort the clusters by size, and find all clusters that are subsets (i.e., children) of larger clusters. Any child clusters that are not at least $\epsilon$ points smaller than their parent are considered to be just smaller versions of their parent and are deleted. Further, parents must have at least 2 children. Otherwise, the lone child cluster is considered to be just a smaller version of the parent and we delete the child. We experimented with several different criteria during the development of this algorithm, including variants on the idea of requiring a child cluster to be a considerably different size that its parent, or for the two to have considerably different median reachability distances. However, the most satisfactory solution that we found was to require there to be multiple children. This is a reasonable requirement, and in contrast to the other variants we tried, it prevents a large number of clusters being identified that are not very different from their parents. However, we note that this requirement also prevents Andromeda~I from being identified as a distinct entity separate from the Giant Stellar Stream based on projected positions alone. 

\subsection{The anatomy of a stellar halo}

\begin{figure*}
  \begin{center}
    \includegraphics[angle=270, width=15cm]{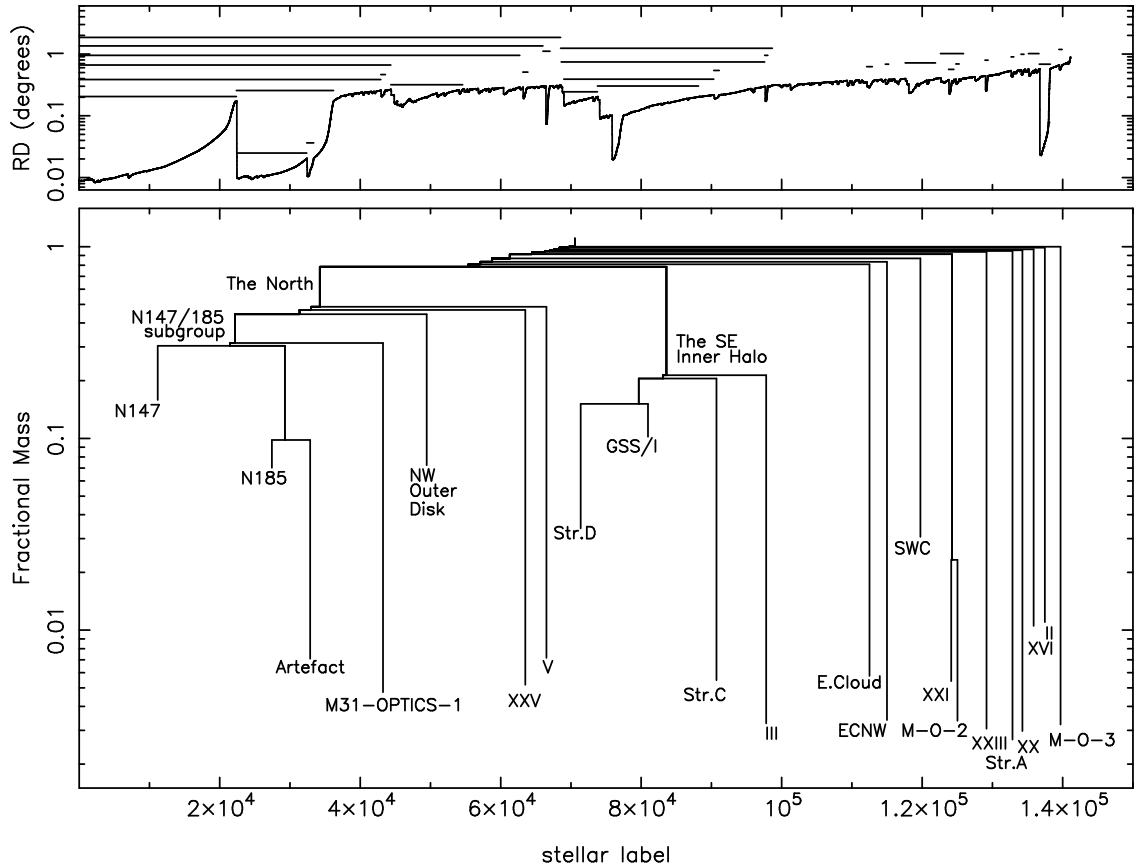}
    \caption{Top panel: the M31 stellar halo reachability diagram (as in Figures~\ref{optics} and \ref{labels}), with horizontal lines highlighting clusters that are found automatically as described in the text. Bottom panel: tree diagram, where each cluster is represented by a vertical line centered on the middle of the cluster on the x-axis. The starting (lower) position on the y-axis is the fractional mass the cluster contains relative to the total mass. Horizontal lines connecting clusters indicate the merging of multiple clusters into their parent cluster.}\label{tree}
  \end{center}
\end{figure*}

\begin{figure*}
  \begin{center}
    \includegraphics[angle=270, width=15cm]{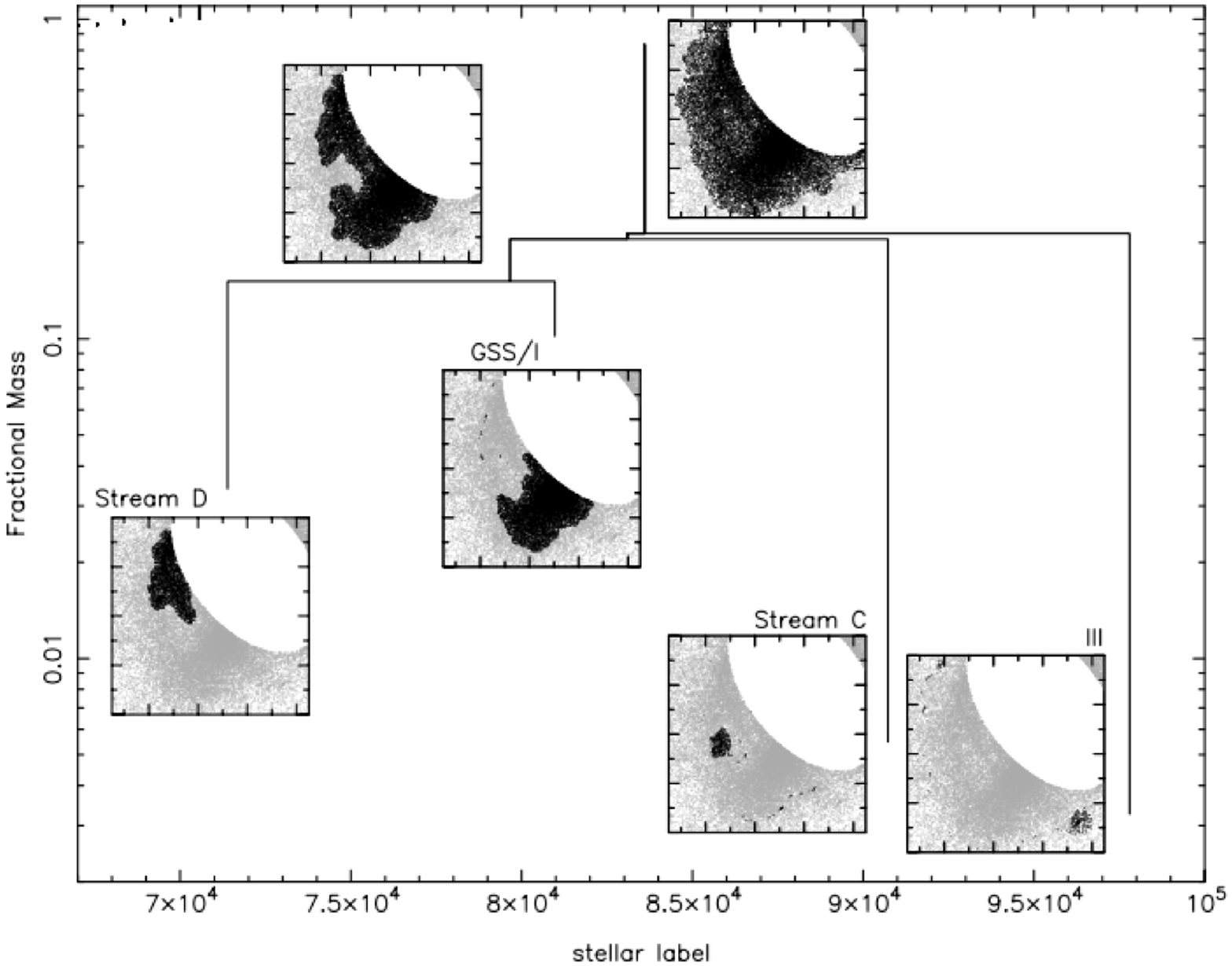}
    \caption{A closer look at the branch of the tree labelled the SE inner halo in Figure~\ref{tree}. For each sub-branch, we highlight in black the points that have been identified, compared to all the points present in the region shown with gray.}\label{anatomy}
  \end{center}
\end{figure*}

The top panel of Figure~\ref{tree} shows the same reachability diagram as in Figures~\ref{optics} and \ref{labels}. Also highlighted by horizontal lines are the locations of all the clusters that are less than 50\% of the total mass of the halo and  that were found by application of the algorithm described in Section~\ref{auto}. 

The lower panel of Figure~\ref{tree} contains the same information as the upper panel, but now displayed as a tree diagram. Each cluster identified in the upper panel is represented by a vertical line centered on the middle of the cluster on the x-axis. The starting (lower) position on the y-axis is the fractional mass the cluster contains relative to the total mass (i.e., the number of stars in the cluster relative to the total number of stars analysed). The end (upper) position on the y-axis is the relative mass of its parent. Horizontal lines connecting clusters indicate that the merging of multiple clusters into their parent cluster. Clusters are labeled when they contain points that coincide with previously identified structures. 

Figure~\ref{tree} provides a quantitative representation of the complex structure of a stellar halo. There are essentially two main branches in the M31 halo, namely the set of structures in the north of the survey, and the set of structures in the south-east of the survey. All other substructures in M31's halo are not (spatially) associated with any major ``mega-structures''; their immediate parent is the M31 halo as a whole. 

Figure~\ref{anatomy} takes a closer look at the branch of the tree labelled the SE inner halo, to illustrate the different substructures present and how they combine to form the parent ``mega-structure''. For each sub-branch, we highlight in black the points that have been identified, compared to all the points present in the region shown with gray.  The SE inner halo ``mega-structure'' and the Northern ``mega-structure''  together account for a majority of the mass of M31's halo.

\begin{figure*}
  \begin{center}
    \includegraphics[angle=270, width=8cm]{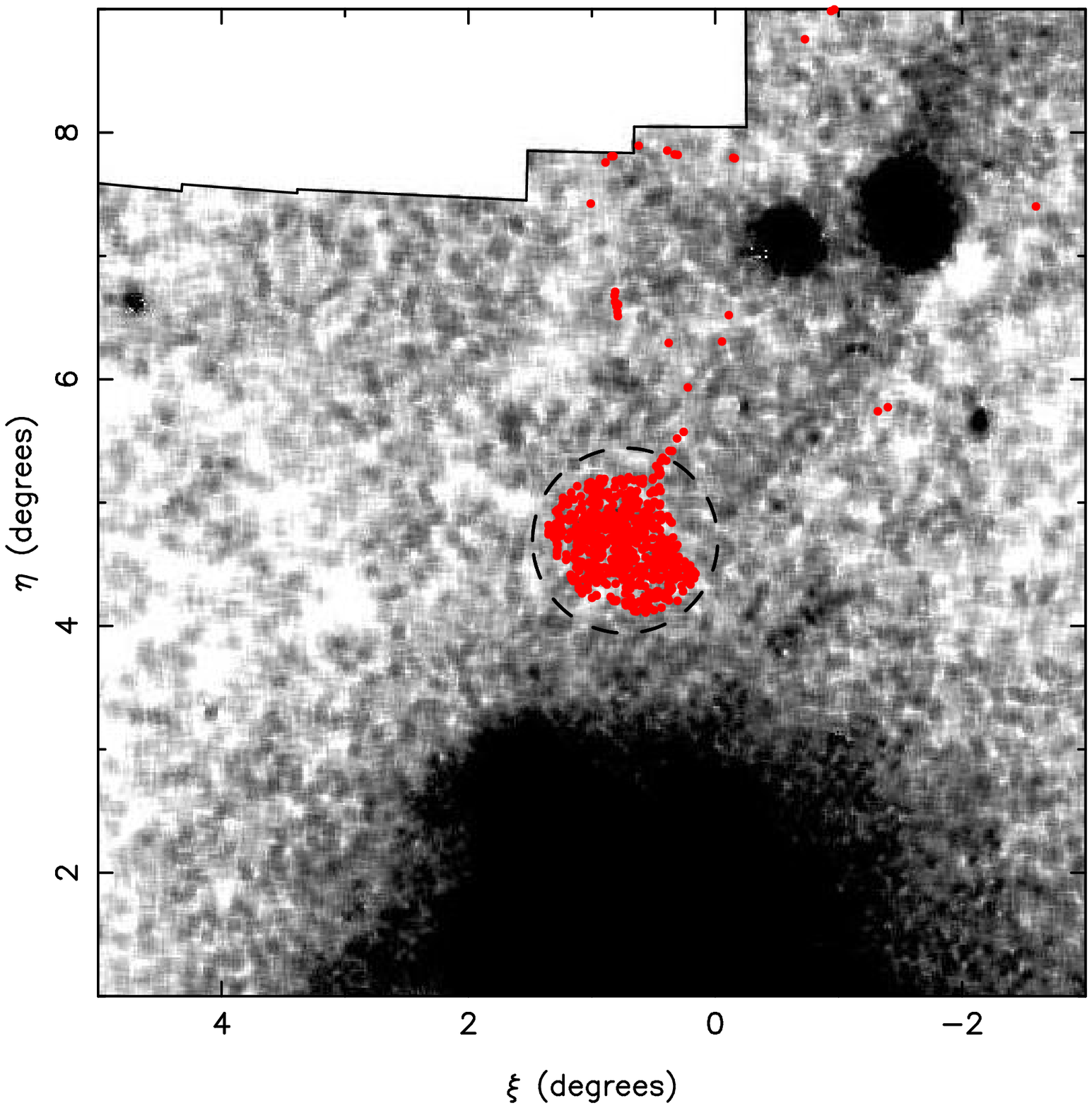}
    \includegraphics[angle=270, width=8cm]{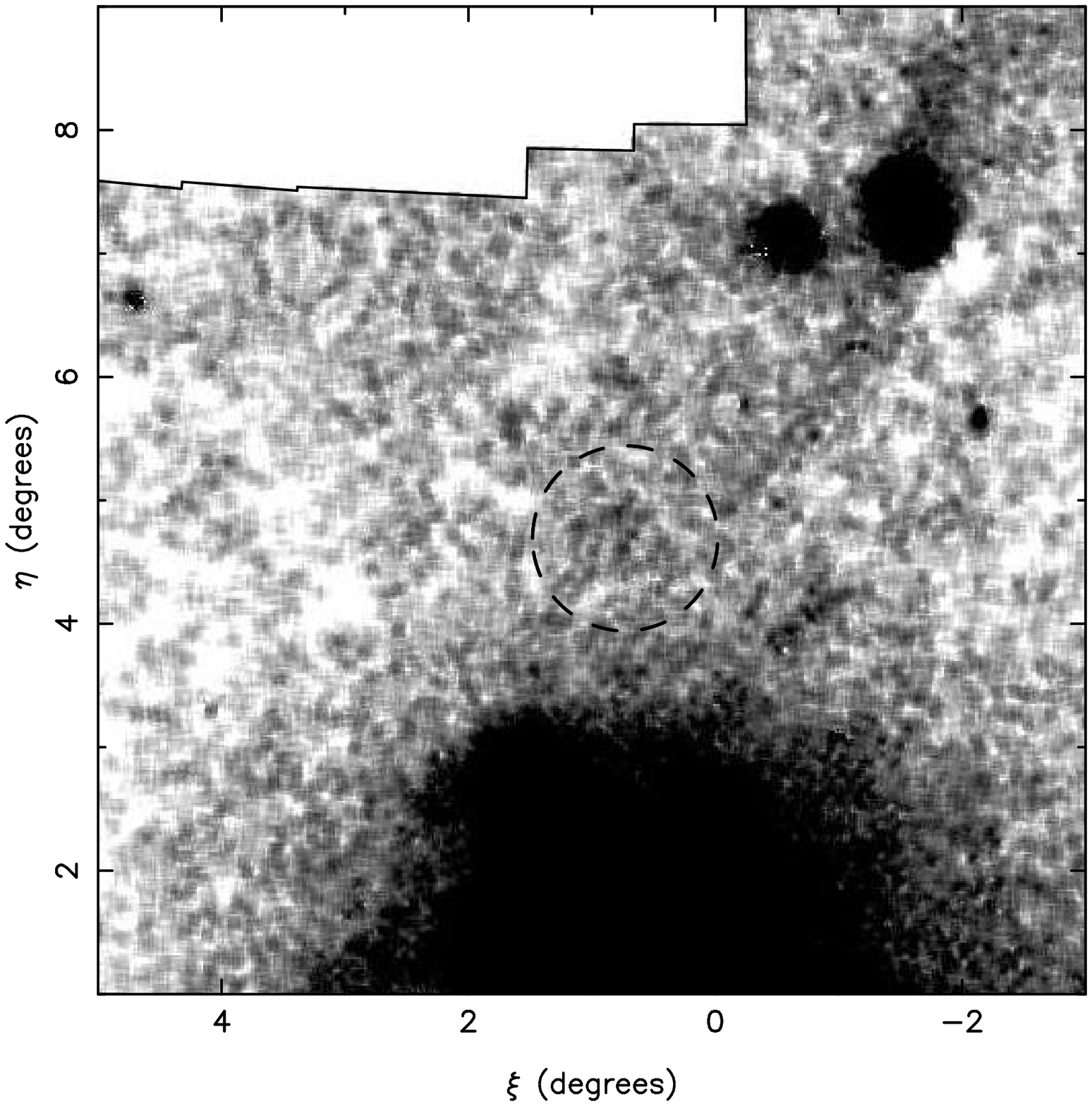}
    \includegraphics[angle=270, width=8cm]{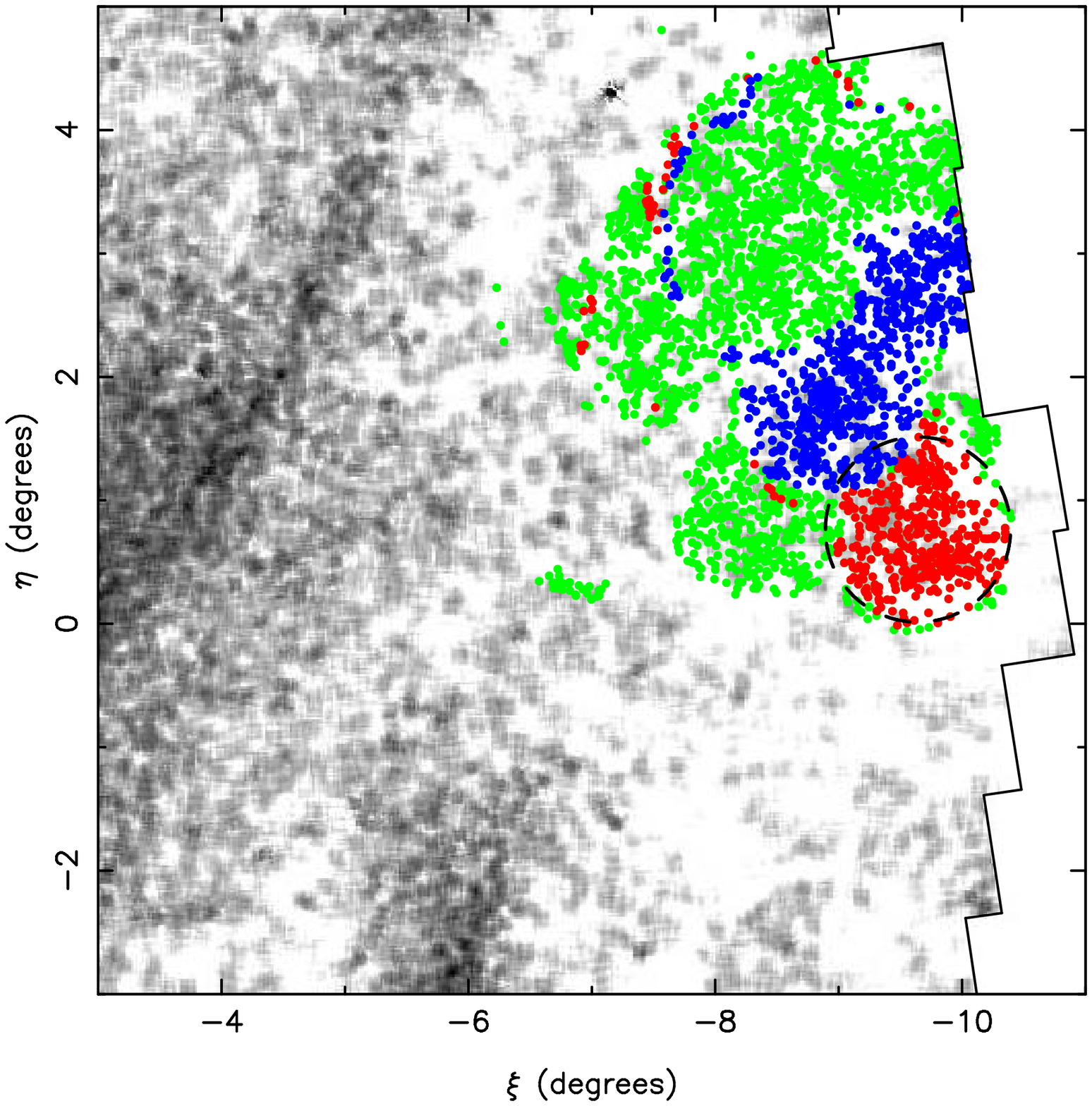}
    \includegraphics[angle=270, width=8cm]{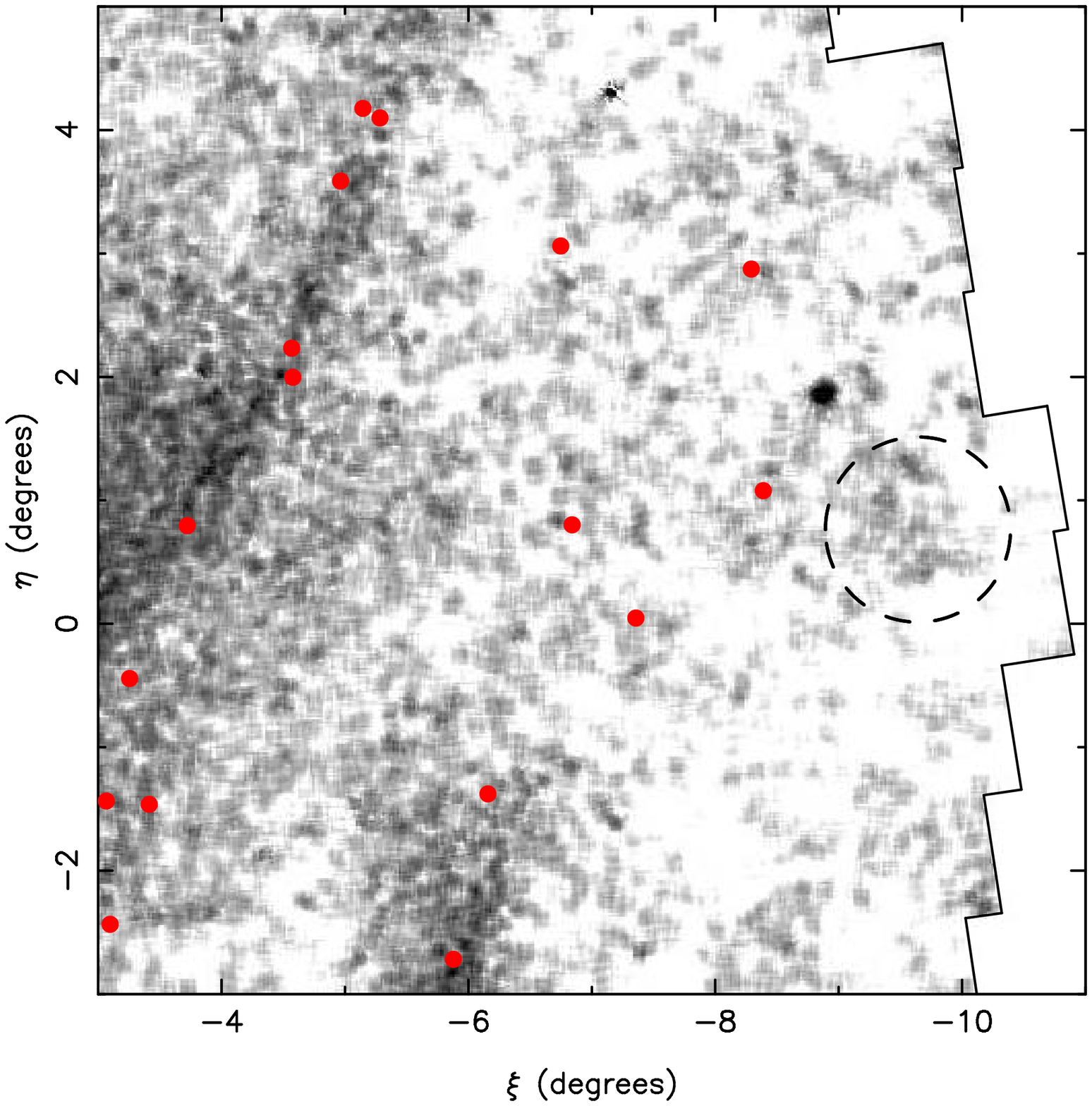}
    \includegraphics[angle=270, width=8cm]{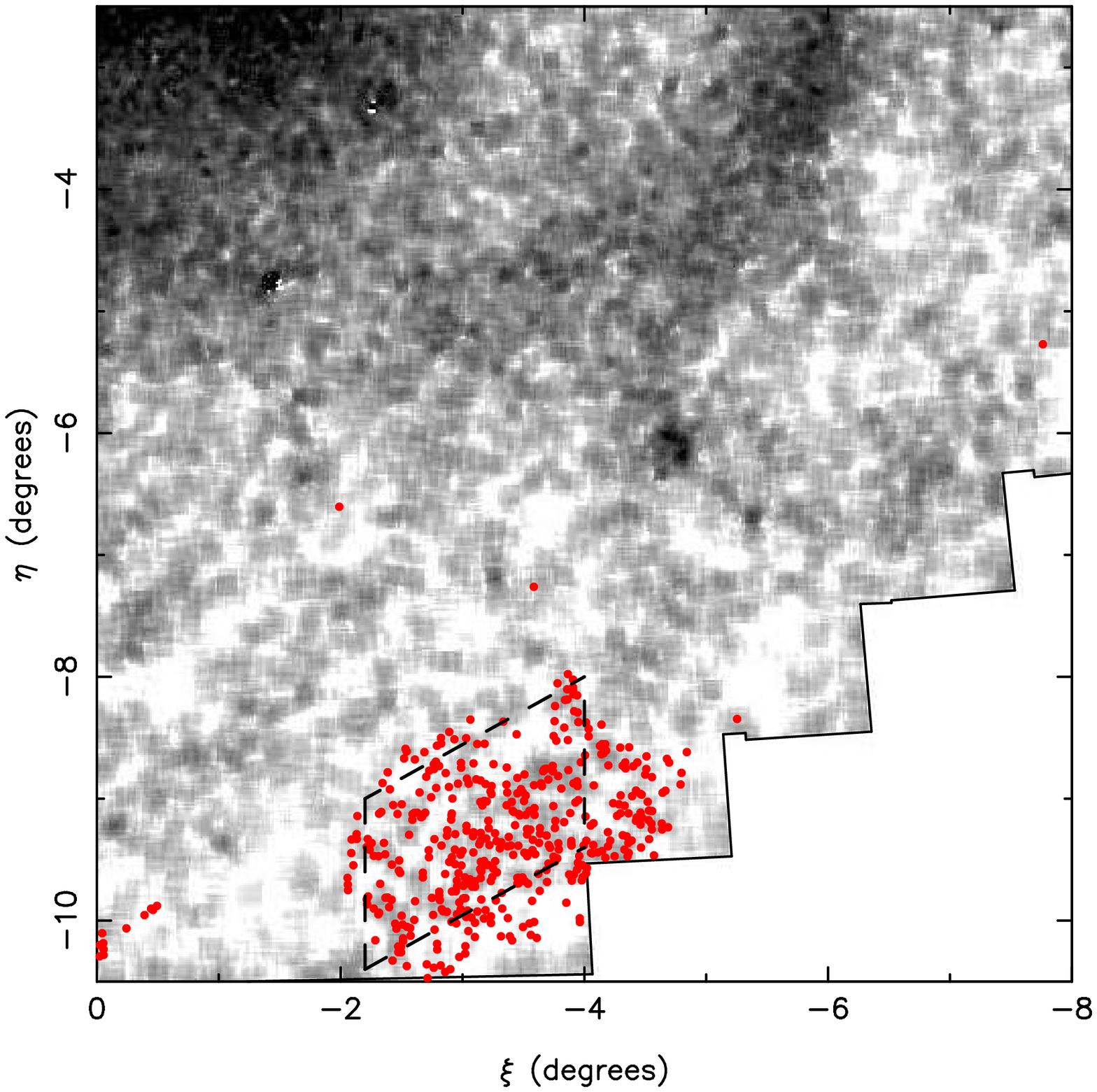}
    \includegraphics[angle=270, width=8cm]{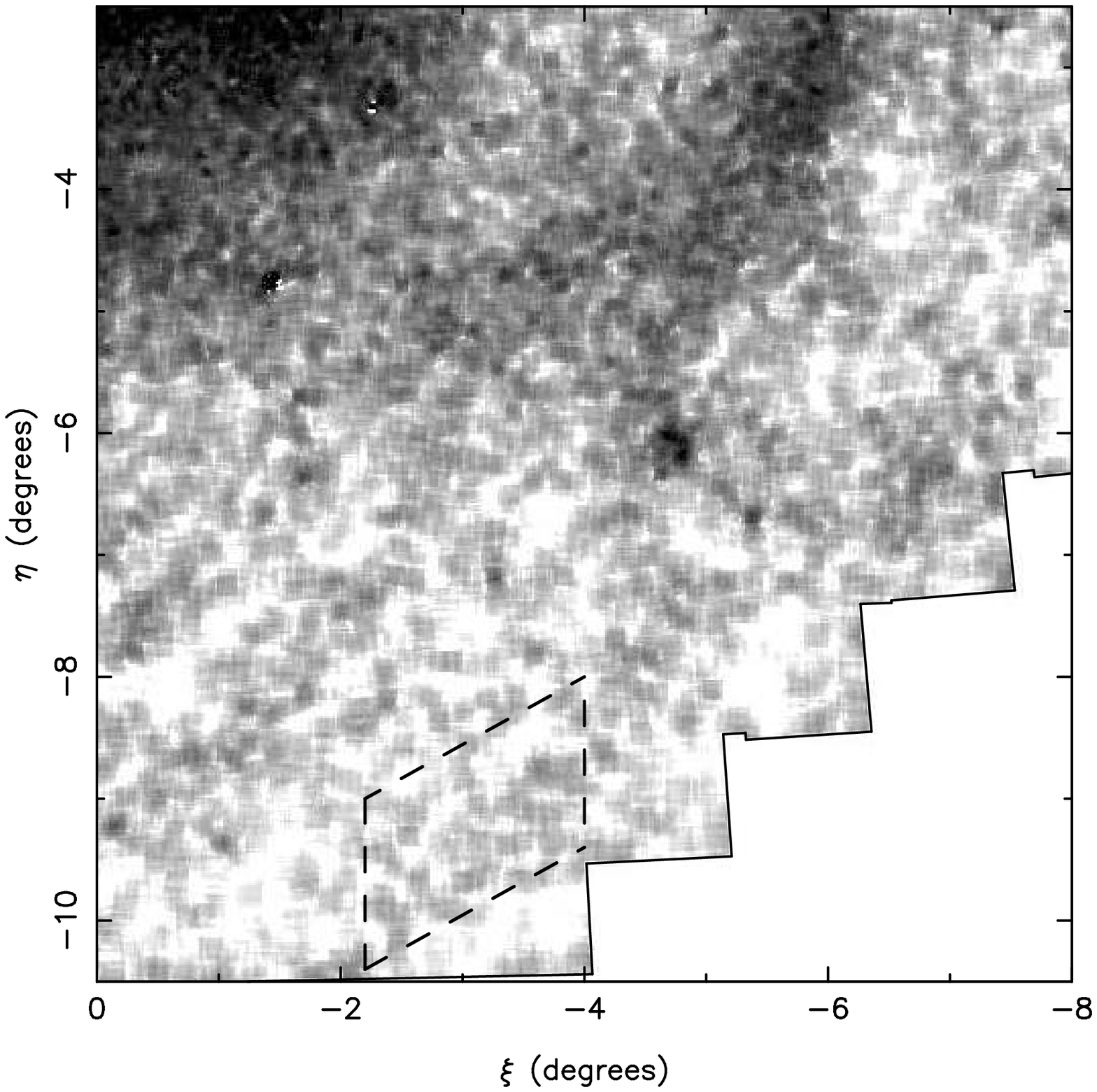}
    \caption{Zoom-in of Figure~\ref{adapt} in the regions surrounding the three new structures identified by the OPTICS algorithm, rescaled as necessary to show the overdensities to which these correspond. The top row shows the region surrounding M31-OPTICS-1, the middle row shows the region surrounding M31-OPTICS-2, and the bottom row shows the region surrounding M31-OPTICS-3. The left panels show the positions of the points identified as belonging to these substructures in red, and the right panels show the same regions without the points. Dashed lines indicate approximate boundaries to the structures. For M31-OPTICS-2, we also show in the left panel those points corresponding to a substructure containing Andromeda~XXI (blue points), and the parent substructure that includes both M31-OPTICS-2 and Andromeda~XXI (green points). In the right panel for M31-OPTICS-2, we additionally show the locations of the outer halo globular clusters as red dots; some of these appear to overlap with the parent structure indicated in the left panel.}\label{newstructures}
  \end{center}
\end{figure*}

\begin{figure*}
  \begin{center}
    \includegraphics[angle=0, width=15cm]{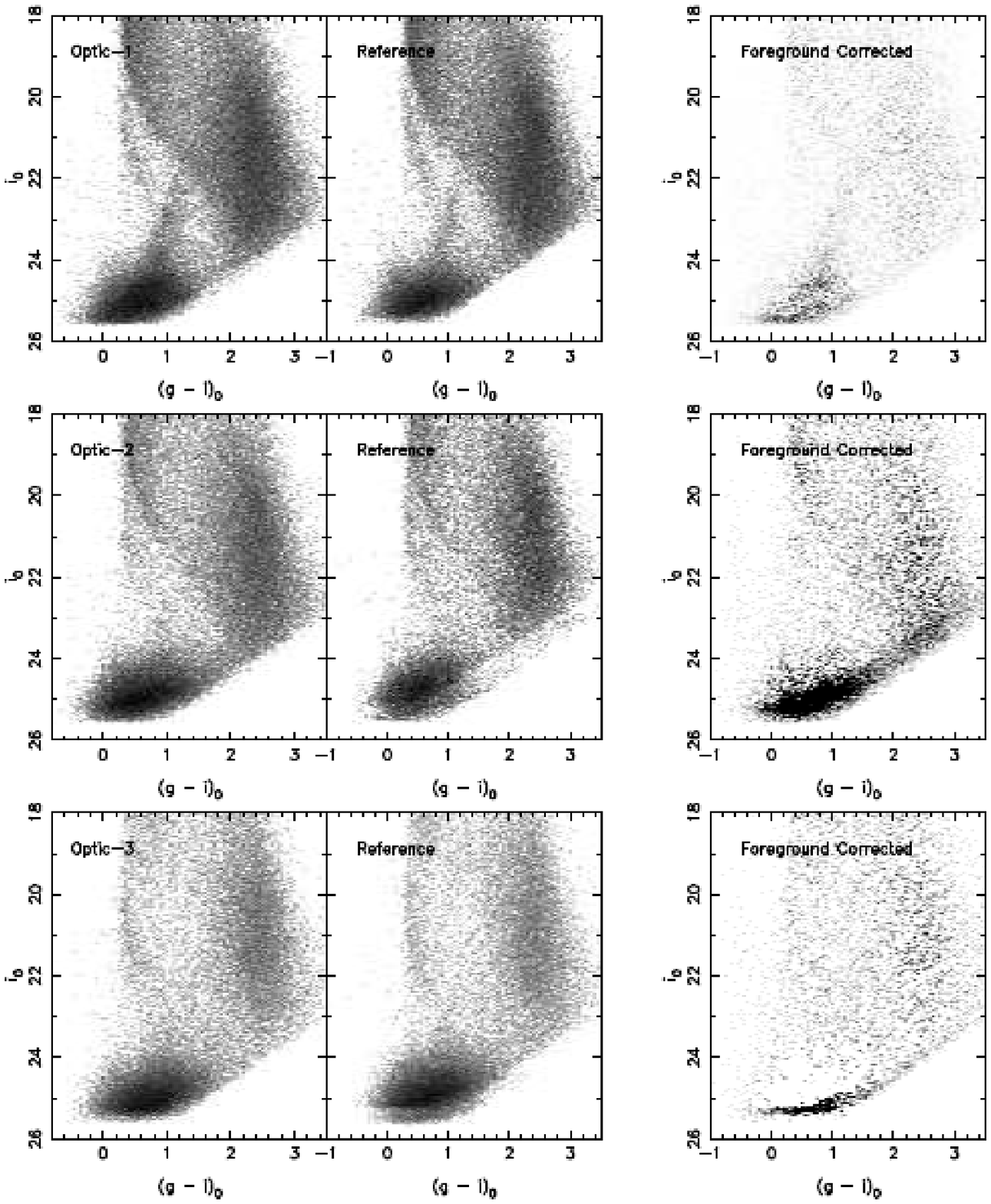}
    \caption{CMDs in the regions of the new substructures M31-OPTICS-1 (top row), M31-OPTICS-2 (middle row) and M31-OPTICS-3 (bottom row). The first panel in each row shows the CMD of all stars bounded by the dashed lines in Figure~\ref{newstructures}. The middle panel shows CMDs of appropriate references areas, the locations of which are described in the text. Both the left and middle panels are shown on a logarithmic scale. The right panel shows the subtracted CMD, where only positive residuals are shown on a linear scale. The stretch has been modified in the middle and bottom rows to try to show up faint features in the subtracted CMD. }\label{optcmds}
  \end{center}
\end{figure*}

A few other interesting features in Figure~\ref{tree} are worthy of comment:

\begin{itemize}
\item \cite{mcmonigal2016} study the East Cloud, and note that it contains some possible extensions, away from the main core. They name one of these extensions as the East Cloud, North-West (ECNW). Impressively,  ECNW is identified by OPTICS as a separate structure immediately adjacent to the East Cloud;
\item Three new substructures are identified with this algorithm, that we call M31-OPTICS-1, -2, -3. The points corresponding to each of these features are shown in Figure~\ref{newstructures} (ordered top to bottom), overlaid on a zoomed and rescaled version of Figure~\ref{adapt}. For M31-OPTICS-2, we additionally overlay the outer halo globular clusters as red dots in the right panel. The CMDs corresponding to the spatial regions enclosed by the dashed lines in Figure~\ref{newstructures} (that mark the approximate limits of each of these structures) are shown in Figure~\ref{optcmds}, along with reference areas and the subtracted CMD. In each row, the first two figures are shown with logarithmic scaling, and the third panel is shown with linear scaling (only positive values are shown). 

Each of these new features identified by the algorithm is large (more than a degree across), diffuse and faint, and each expands across at least a couple of the MegaCam pointings that make up the survey area. Two of these structures (M31-OPTICS-2 and -3) were likely identified because they lie in otherwise extremely empty regions of M31's halo.

\item {\bf M31-OPTICS-1:} This is the most obvious of the new structures. The dashed circle in the top panels of Figure~\ref{newstructures} has a diameter of 1.5 degrees and is centered at ($\xi=0.730^\circ,\eta=4.692^\circ$); the reference CMD in Figure~\ref{optcmds} is from an area of the same shape offset 1.5 degrees to the east. The subtracted CMD shows a clear RGB, indicating that it is certainly a real (projected) overdensity of stars. It is not too far in projection from the far outer parts of the disk of M31 (e.g., \citealt{ibata2005}), and so whether it is physically associated with the halo - and potentially the NGC147/185 sub-group - or whether it is associated with the outer disk, is ambiguous based on projected data alone. After all, these structures all connect into the same ``mega-structure'' as shown in Figure~\ref{tree}. We note that M31-OPTICS-1 overlaps with some of the extended HI distribution in M31 (see Figure~2 of \citealt{lewis2013}) although there is no HI concentration at these coordinates implying its own HI reservoir.

\item {\bf M31-OPTICS-2:} Intriguingly, M31-OPTICS-2 is a very low mass, but very diffuse, structure that the algorithm identifies as a child of a spatially extended overdensity that includes the dwarf galaxy Andromeda~XXI. The left panel of the middle row of Figure~\ref{newstructures} shows M31-OPTICS-2 as the red points, the blue points show the structure that includes Andromeda~XXI, and the green points show the parent structure that includes both M31-OPTICS-2 and Andromeda~XXI. The parent structure is very extended and includes most of the region at the extreme west of the survey area. It is fascinating to note that this extended parent region overlaps with two globular clusters in the far outer halo of M31 (PAndAS-01, PAndAS-02), shown as red dots in the right panel of the middle row of Figure~\ref{newstructures}, and an additional three outer halo globular clusters lie very close to this extended region (PAndAS-03, PAndAS-05, PAndAS-06). There is no HI detected in this region (\citealt{lewis2013}).

The dashed circle that picks out the region occupied by M31-OPTICS-2 has a diameter of 1.5 degrees and is centered at ($\xi=-9.643^\circ,\eta=0.764^\circ$). The reference CMD is an equivalent area 1.5 degrees south. There is tentative evidence for an extremely weak excess RGB population by comparison of these CMDs  and in the subtracted CMD (which has a stretched scaling to try to enhance faint features).

Andromeda~XXI was discovered by \cite{martin2009} and is one of the larger dwarf galaxies in the Local Group, with a half-light radius of order 1\,kpc. Within the Local Group, such extended dwarfs are typically only found in the vicinity of M31 (e.g., \citealt{mcconnachie2006b}). It may be that tides have played an important role in the evolution of this galaxy, and it could be that the diffuse structures identified by OPTICS in its surroundings are tidal debris from the galaxy. The spatial coincidence of at least a few globular clusters with the parent structure, even though the globular clusters were not considered by the OPTICS algorithm, is intriguing when viewed in this context.  It is equally possible, however, that the structures are physically unassociated with each other, given that we are applying this algorithm in projected space only. The distance to Andromeda~XXI is $\sim 830$\,kpc, but we clearly cannot get a distance to M31-OPTICS-2 with the current data.

\item {\bf M31-OPTICS-3:} The most diffuse of the new detections, the position of the points corresponding to M31-OPTICS-3 are shown in the bottom row of Figure~\ref{newstructures}. They span an area of order 6 square degrees, making this structure -- if it is real -- of comparable size to the east or south-west clouds. There is no HI detected in this region of the M31 halo (\citealt{lewis2013}).
 The center of this structure (the median position of the points) is ($\xi=-3.252^\circ,\eta=-9.332^\circ$). The CMD of all stars within the parallelogram in Figure~\ref{newstructures} is shown in the bottom row of Figure~\ref{optcmds}. The reference area is the mirror image of this area, reflected about the left-hand edge of the parallelogram. The subtracted CMD has the same stretch as used for M31-OPTICS-2. However, there is nothing clearly visible in the subtracted CMD; this is perhaps not too surprising given that it is of broadly the same stellar mass as M31-OPTICS-2, but spread out over an area that is 3 or 4 times as large. It also indicates that we are currently pushing OPTICS to the limit given the input dataset.

\end{itemize}

\subsection{OPTICS and the hierarchical clustering of points in astronomy}

Our initial experiments with OPTICS suggest that it is a powerful algorithm that could be of considerable use for any astronomical problem in which we are dealing with the hierarchical structure of discrete datapoints, such as observational galaxy clustering, dark matter N-body modeling, or near-field cosmology (including the Milky Way as well as nearby galaxies).  

We are using OPTICS only on the $(x, y)$ spatial data in PAndAS, and so $l = \sqrt{x^2 + y^2}$. However, OPTICS can be used to examine the clustering of any number of parameters, some or none of which can be spatial coordinates. For PAndAS, the obvious extension to our current analysis is to consider photometric metallicity estimates in  addition to spatial information. We have already noted that our process for identifying clusters does not separate the Giant Stellar Stream from Andromeda I using only the spatial data (this is because Andromeda~I would be the sole ``child'' of the Giant Stellar Stream, and single child clusters are not allowed in our algorithm). Of course, we expect Andromeda~I would stand out from the GSS if we used photometric metallicity as well as spatial coordinates. In addition, kinematics and other spectroscopic measurements could be used if and when the relevant datasets become extensive enough.

We summarise the key aspects of OPTICS with respect to the quantification of stellar halos, below:

\begin{enumerate}
\item OPTICS works on the discrete points (stars), and so does not require the pixelisation or smoothing of data that removes information on small scales;
\item OPTICS identifies clusters of any shape and size;
\item OPTICS is easily extendable to datasets of any dimensionality;
\item Essentially, only a single parameter with a well defined physical interpretation ($\epsilon$, the minimum number of members in a cluster) is required to run OPTICS (the second parameter, the maximum value of $l$, can be set to a value much larger than the largest scale probed by the dataset); 
\item The (ordered) reachability distance is a simple, one-dimensional description of the clustering structure of the higher-dimesionality dataset;
\item The reachability diagram is a conceptually simple visualization of clustering in the data on all scales; tree-diagrams can be easily derived from reachability diagrams and these too explicitly demonstrate the hierarchy of structures in the data;
\item The reachability diagram allows for robust, repeatable, objective definitions of substructure. Further, any custom metrics or analysis can be applied to the reachability distances to quantify and compare different datasets.
\end{enumerate}

There are also a few areas in which OPTICS presents challenges:

\begin{enumerate}
\item OPTICS does not define clusters automatically, and custom routines need to be written. It appears as if this issue has prevented OPTICS from being used as frequently as other routines in other areas of science and computing;
\item The current OPTICS implementation used by these authors is an $O(N^2)$ routine that is memory intensive, and prevents millions of stars from being analysed simultaneously. However, this criticism primarily reflects the software-development capabilities of the lead author, and is not expected to be a fundamental issue in its future use.
\end{enumerate}

\section{Summary}

In this paper, we have attempted to collate and synthesize the extensive information available on the outer halo stellar substructures in M31 observed with PAndAS. We estimate that the 13 distinct stellar substructures discussed in various papers in the literature represent at least 5 distinct accretion events, and likely more. We hypothesize that at least a few of the widely separated, far outer halo structures that have globular clusters associated with them may in fact be shells produced from the same accretion event.

We estimate luminosities of a few of the substructures for which previous estimates were not available, and determine an approximate mass budget for the stellar halo. Approximately one quarter of the stellar halo beyond 2 degrees and within a projected radius from M31 of 150\,kpc  is apparently ``smooth'' down to the limits of PAndAS; about 70\% stellar halo is in the form of substructure, and the remaining 5\% of the mass of the stellar halo is in the dwarf galaxies. Globular clusters contribute a negligible fraction of the stellar mass; however, this negligible fraction is in fact comparable to the mass in dwarf galaxies and stellar substructures excluding the one or two most massive examples.

We conclude this paper by quantifying the (projected) stellar density distribution of M31 in terms of a hierarchy of structures. Specifically, we use the OPTICS hierarchical clustering algorithm, and demonstrate its utility for both visualising and quantifying the structure of complex datasets of points that cluster in a diverse range of shapes and on a wide range of scales. OPTICS identifies a few new structures in the halo of M31 whose physical natures remain to be determined. We show that M31's halo, in projection, is dominated by two ``mega-structures'', that can be considered as the two most significant branches of a merger tree that is produced by breaking M31's stellar halo into smaller and smaller structures based on the spatial clustering of the stars. More generally, OPTICS provides a means to simplify the visualisation and interpretation of highly structured data, and to create metrics to objectively describe and robustly compare different datasets. We suggest that OPTICS is a powerful algorithm that may prove useful not just for near field cosmology, but for other areas of astronomy as well.

Finally, the publication of this paper coincides with the public release of all high-level data products from the PAndAS collaboration. Access to these data is available through the CADC. 

\section*{Acknowledgements}

Based on observations obtained with MegaPrime/MegaCam, a joint project of CFHT and CEA Saclay, at the Canada-France-Hawaii Telescope (CFHT) which is operated by the National Research Council (NRC) of Canada, the Institut National des Science de l'Univers (INSU) of the Centre National de la Recherche Scientifique (CNRS) of France, and the University of Hawaii.

This research used the facilities of the Canadian Astronomy Data Centre operated by the National Research Council of Canada with the support of the Canadian Space Agency.

This work was supported in part by the Canadian Advanced Network for Astronomical Research (CANFAR), which has been made possible by funding from CANARIE under the Network-Enabled Platforms program. 

AWM thanks the organizers and attendees at the Lorentz Center Workshop on ``Large Surveys of the Great Andromeda Galaxy'' in July 2017, at which many aspects of this paper were discussed and refined.

BM acknowledges the support of an Australian Postgraduate Award. THP acknowledges support by FONDECYT Regular Project No.$\sim$1161817 and the BASAL Center for Astrophysics and Associated Technologies (PFB-06).

We thank the referee for a useful and supportive set of comments.

\appendix

\section{The characterization of hierarchical clustering using OPTICS}

Here, we provide a summary of the concept surrounding our application and usage of the OPTICS algorithm. This necessitates first describing the DBSCAN algorithm.

\subsection{DBSCAN}

``Density Based Spatial Clustering of Applications with Noise'' (DBSCAN) is a clustering algorithm introduced by \cite{ester1996}. It is designed to find clusters of arbitrary shape by looking for overdensities of points that are associated by their mutual proximity, and to be able to do so even in the presence of a significant background (noise or other contamination). Conceptually, it bears a strong resemblance to the ``Friends of friends'' (FoF) algorithm that is more commonly used in astronomy (\citealt{huchra1982}).

Searches for clusters using DBSCAN require the user to specify two inputs, a ``MinPts'' ($\epsilon$) parameter and an ``Eps'' ($\l$) parameter. The DBSCAN algorithm takes a point in the dataset and asks if it is a potential ``core'' point. That is, is the point in an overdense region such that there are at least $\epsilon$  points within a radius of $l$ from it? If the point is not a core point, it is ignored and the algorithm moves to the next point. If it is a core point, then a cluster is formed that consists of this point and all its neighbours within a radius of $l$. Then, the algorithm seeks to ``grow'' the cluster by finding all points that are within a distance of $l$ of any point that has been identified as being part of the cluster. It does this recursively until no more points are added to the cluster. At this point, the algorithm then moves on to find the next core point to act as a seed for the next cluster. 

The case of $\epsilon = 1$ describes the standard astronomy FoF algorithm. However, the difference with DBSCAN when $\epsilon>1$ is that not every point can act as a ``seed'' from which one can grow a cluster. Instead, points can only act as seeds if they are identified to be in an overdense region. Further, once the algorithm has completed, every point in FoF is considered to be a potential cluster member (albeit perhaps with only one or a few members). In DBSCAN, however, some points will not have been assigned membership of any cluster. The key parameter is $l$; in FoF, $l$ describes only the search radius within which to look for friends, but in DBSCAN it additionally characterises the spatial density scale of the ``core'' that seeds the cluster. The concept of a core defined in this way is particularly well suited to the physics of our specific problem, since  a given number of stars (points) within a certain area corresponds to a surface brightness. Thus, by specifying $\epsilon$ and $l$, we are essentially specifying the core surface brightness of the substructures we are interested in finding. This appealing characteristic led us to investigate DBSCAN further. 

The critical issue with DBSCAN for our problem is that, as with FoF and many other clustering algorithms, it is only able to find clusters once a given length scale has been provided. Identifying the value of $l$ is user dependent and affects the types of clusters that will subsequently be identified (large and diffuse or small and compact). However, when clustering occurs on different spatial scales in the same data, the use of a single value of $l$ is unsatisfactory since results will not represent the hierarchy of structure that is actually present in the data.\footnote{We stress again that the term ``hierarchy'' refers generally only to the distribution of structure in a dataset, and should not be confused with a specific formation scenario for the structure.} OPTICS and HDBSCAN are extensions of DBSCAN designed to overcome this limitation.

\subsection{OPTICS}

``Ordering points to identify the clustering structure'' (OPTICS; \citealt{ankerst1999}) is based on the DBSCAN concept. Unlike DBSCAN, it does not actually identify clusters. Rather, it provides a way of visualising the clustering structure of the dataset so that one can identify the value of $l$ required so that a given point is identified as being part of a cluster.

OPTICS requires the user to specify values for $\epsilon$ and also $l$, but here $l$ corresponds to the maximum scale on which to look for clusters. Thus, by setting $l$ to be larger than the maximum scale probed by the dataset, OPTICS essentially requires only a single parameter. OPTICS returns a ``reachability plot'' that is a type of dendogram displaying the clustering structure of the dataset. The y-value shows the ``reachability distance'' for each point in the dataset, that has been ordered in a certain way on the x-axis. A reachability plot is easily understood for the case of $\epsilon = 2$. In this case, for a given point in the dataset, the reachability distance is the distance to its closest neighbour that has not already been examined i.e., it is the minimum distance that would be necessary to specify for $l$, to satisfy the clustering condition for this point using the DBSCAN algorithm when $\epsilon = 2$. For this new point, the procedure is repeated, continually until all points in the dataset have been examined. The x axis of the reachability diagram shows the order in which the points in the dataset have been examined.

The critical feature in reachability plots are the valleys. The valleys represent a group of points that are all in the general vicinity of each other (hence have broadly comparable reachability distances). An upward-hill at the end of a valley represents the situation where the next nearest neighbours to these points are actually quite distant, hence the need for a large reachability distance to connect to them. Once the algorithm has moved to the new point situated far from these original points, the next nearest neighbour might be reasonably close to it, and so a peak is produced in the reachability plot. Valleys are therefore the clusters within the dataset, and it is possible to have valleys within valleys, indicating hierarchical structuring. This basic description holds even for the case for which $\epsilon > 2$.

\begin{longtable}{lllcccc}
\hline
Glob. Clus.   &  R.A. & Dec. & \multicolumn{2}{c}{$R_{M31}$} & $M_V$ & $M_\star$\\
&  $^{h m s}$& $^\circ$' " & deg & kpc &  & $10^4M_\odot$\\
\hline
M31:             &           &           &        &        &        &       \\ 
B514             & 0 31  9.8 & +37 54  0 &  4.030 &   55.1 &   -8.9 &  37.60\\ 
B517             & 0 59 59.9 & +41 54  6 &  3.297 &   45.1 &   99.9 &   0.00\\ 
dTZZ-05          & 0 36  8.6 & +39 17 30 &  2.328 &   31.8 &   -7.0 &   6.66\\ 
dTZZ-21          & 1 28 49.2 & +47  4 21 & 10.200 &  139.4 &   -7.2 &   8.15\\ 
G001             & 0 32 46.5 & +39 34 40 &  2.529 &   34.6 &  -10.8 & 212.41\\ 
G002             & 0 33 33.8 & +39 31 18 &  2.458 &   33.6 &   -8.9 &  37.95\\ 
G339             & 0 47 50.2 & +43  9 16 &  2.127 &   29.1 &   -7.6 &  11.05\\ 
H1               & 0 26 47.8 & +39 44 46 &  3.386 &   46.3 &   -8.7 &  30.99\\ 
H2               & 0 28  3.2 & +40  2 55 &  3.035 &   41.5 &   -7.5 &  10.26\\ 
H3               & 0 29 30.1 & +41 50 31 &  2.547 &   34.8 &   -6.5 &   4.16\\ 
H4               & 0 29 45.0 & +41 13  9 &  2.443 &   33.4 &   -7.8 &  13.78\\ 
H5               & 0 30 27.3 & +41 36 19 &  2.330 &   31.8 &   -8.4 &  24.39\\ 
H7               & 0 31 54.6 & +40  6 47 &  2.347 &   32.1 &   -7.2 &   7.57\\ 
H8               & 0 34 15.4 & +39 52 53 &  2.114 &   28.9 &   -5.7 &   1.97\\ 
H9               & 0 34 17.3 & +37 30 43 &  4.085 &   55.8 &   99.9 &   0.00\\ 
H10              & 0 35 59.7 & +35 41  3 &  5.739 &   78.4 &   -8.9 &  35.91\\ 
H11              & 0 37 28.0 & +44 11 26 &  3.099 &   42.3 &   -7.9 &  14.56\\ 
H12              & 0 38  3.9 & +37 44  0 &  3.635 &   49.7 &   -8.2 &  19.37\\ 
H15              & 0 40 13.2 & +35 52 36 &  5.412 &   74.0 &   -6.6 &   4.48\\ 
H17              & 0 42 23.7 & +37 14 34 &  4.014 &   54.9 &   -7.2 &   8.00\\ 
H18              & 0 43 36.1 & +44 58 59 &  3.742 &   51.1 &   -8.1 &  17.67\\ 
H19              & 0 44 14.9 & +38 25 42 &  2.839 &   38.8 &   -7.3 &   8.46\\ 
H22              & 0 49 44.7 & +38 18 37 &  3.237 &   44.2 &   -7.7 &  11.78\\ 
H23              & 0 54 25.0 & +39 42 55 &  2.702 &   36.9 &   -8.1 &  17.67\\ 
H24              & 0 55 43.9 & +42 46 15 &  2.856 &   39.0 &   -7.1 &   7.10\\ 
H25              & 0 59 34.6 & +44  5 38 &  4.210 &   57.5 &   -7.9 &  15.25\\ 
H26              & 0 59 27.5 & +37 41 30 &  4.814 &   65.8 &   -7.4 &   9.36\\ 
H27              & 1  7 26.3 & +35 46 48 &  7.334 &  100.2 &   -8.4 &  23.29\\ 
HEC1             & 0 25 33.9 & +40 43 38 &  3.285 &   44.9 &   -5.8 &   2.18\\ 
HEC2             & 0 28 31.5 & +37 31 23 &  4.638 &   63.4 &   -5.6 &   1.78\\ 
HEC3             & 0 36 31.7 & +44 44 16 &  3.673 &   50.2 &   -5.4 &   1.43\\ 
HEC6             & 0 38 35.4 & +44 16 51 &  3.128 &   42.7 &   -5.9 &   2.39\\ 
HEC7             & 0 42 55.1 & +43 57 27 &  2.710 &   37.0 &   -6.6 &   4.36\\ 
HEC10            & 0 54 36.5 & +44 58 44 &  4.320 &   59.0 &   -6.1 &   2.93\\ 
HEC11            & 0 55 17.4 & +38 51  1 &  3.399 &   46.5 &   -6.7 &   4.69\\ 
HEC12            & 0 58 15.4 & +38  3  1 &  4.385 &   59.9 &   -6.2 &   2.99\\ 
HEC13            & 0 58 17.1 & +37 13 49 &  5.034 &   68.8 &   -5.5 &   1.69\\ 
MGC1             & 0 50 42.5 & +32 54 58 &  8.546 &  116.8 &   -9.6 &  70.34\\ 
PAndAS-01        &23 57 12.0 & +43 33  8 &  8.775 &  119.9 &   -7.5 &  10.07\\ 
PAndAS-02        &23 57 55.7 & +41 46 49 &  8.459 &  115.6 &   -6.8 &   5.49\\ 
PAndAS-03        & 0  3 56.4 & +40 53 19 &  7.357 &  100.5 &   -4.2 &   0.48\\ 
PAndAS-04        & 0  4 42.9 & +47 21 42 &  9.210 &  125.9 &   -7.1 &   7.03\\ 
PAndAS-05        & 0  5 24.1 & +43 55 35 &  7.410 &  101.3 &   -5.1 &   1.07\\ 
PAndAS-06        & 0  6 11.9 & +41 41 20 &  6.888 &   94.1 &   -8.0 &  16.56\\ 
PAndAS-07        & 0 10 51.3 & +39 35 58 &  6.310 &   86.2 &   -5.0 &   1.03\\ 
PAndAS-08        & 0 12 52.4 & +38 17 47 &  6.477 &   88.5 &   -5.4 &   1.48\\ 
PAndAS-09        & 0 12 54.7 & +45  5 55 &  6.687 &   91.4 &   -6.8 &   5.14\\ 
PAndAS-10        & 0 13 38.7 & +45 11 11 &  6.627 &   90.6 &   -5.4 &   1.52\\ 
PAndAS-11        & 0 14 55.6 & +44 37 16 &  6.124 &   83.7 &   -6.7 &   5.10\\ 
PAndAS-12        & 0 17 40.1 & +43 18 39 &  5.085 &   69.5 &   -5.3 &   1.39\\ 
PAndAS-13        & 0 17 42.7 & +43  4 31 &  4.994 &   68.2 &   -6.5 &   4.05\\ 
PAndAS-14        & 0 20 33.9 & +36 39 34 &  6.319 &   86.4 &   -7.0 &   6.53\\ 
PAndAS-15        & 0 22 44.1 & +41 56 14 &  3.807 &   52.0 &   -5.0 &   1.06\\ 
PAndAS-16        & 0 24 59.9 & +39 42 13 &  3.715 &   50.8 &   -8.4 &  24.39\\ 
PAndAS-17        & 0 26 52.2 & +38 44 58 &  3.940 &   53.8 &   -8.2 &  19.02\\ 
PAndAS-18        & 0 28 23.3 & +39 55  4 &  3.034 &   41.5 &   -5.3 &   1.42\\ 
PAndAS-19        & 0 30 12.2 & +39 50 59 &  2.763 &   37.8 &   -4.7 &   0.80\\ 
PAndAS-20        & 0 31 23.7 & +41 59 20 &  2.245 &   30.7 &   -5.4 &   1.52\\
PAndAS-21        & 0 31 27.5 & +39 32 21 &  2.747 &   37.5 &   -7.1 &   6.84\\
PAndAS-22        & 0 32  8.4 & +40 37 31 &  2.097 &   28.7 &   -6.2 &   3.04\\
PAndAS-23        & 0 33 14.1 & +39 35 15 &  2.457 &   33.6 &   -5.0 &   1.05\\
PAndAS-24        & 0 33 50.6 & +38 38 28 &  3.119 &   42.6 &   -4.7 &   0.76\\
PAndAS-25        & 0 34  6.2 & +43 15  6 &  2.562 &   35.0 &   -5.2 &   1.25\\
PAndAS-26        & 0 34 45.1 & +38 26 38 &  3.200 &   43.7 &   -5.1 &   1.13\\
PAndAS-27        & 0 35 13.5 & +45 10 37 &  4.166 &   56.9 &   -7.7 &  12.22\\
PAndAS-30        & 0 38 29.0 & +37 58 39 &  3.377 &   46.2 &   -5.4 &   1.51\\
PAndAS-33        & 0 40 57.3 & +38 38 10 &  2.638 &   36.0 &   -5.4 &   1.47\\
PAndAS-36        & 0 44 45.6 & +43 26 34 &  2.226 &   30.4 &   -7.3 &   8.53\\
PAndAS-37        & 0 48 26.5 & +37 55 42 &  3.503 &   47.9 &   -7.3 &   8.94\\
PandAS-38        & 0 49 45.7 & +47 54 33 &  6.807 &   93.0 &   -4.5 &   0.65\\
PAndAS-41        & 0 53 39.6 & +42 35 14 &  2.434 &   33.3 &   99.9 &   0.00\\
PAndAS-42        & 0 56 38.0 & +39 40 25 &  3.081 &   42.1 &   -6.6 &   4.44\\
PAndAS-43        & 0 56 38.8 & +42 27 17 &  2.859 &   39.1 &   -5.3 &   1.32\\
PAndAS-44        & 0 57 55.9 & +41 42 57 &  2.886 &   39.4 &   -7.7 &  12.57\\
PAndAS-45        & 0 58 38.0 & +41 57 11 &  3.057 &   41.8 &   -4.1 &   0.43\\
PAndAS-46        & 0 58 56.4 & +42 27 38 &  3.255 &   44.5 &   -8.7 &  30.14\\
PAndAS-47        & 0 59  4.8 & +42 22 35 &  3.251 &   44.4 &   -5.7 &   1.88\\
PAndAS-48        & 0 59 28.3 & +31 29 10 & 10.438 &  142.6 &   -4.7 &   0.80\\
PAndAS-49        & 1  0 50.1 & +42 18 13 &  3.539 &   48.4 &   -4.8 &   0.86\\
PAndAS-50        & 1  1 50.7 & +48 18 19 &  7.873 &  107.6 &   -6.4 &   3.66\\
PAndAS-51        & 1  2  6.6 & +42 48  6 &  3.924 &   53.6 &   99.9 &   0.00\\
PAndAS-52        & 1 12 47.0 & +42 25 24 &  5.736 &   78.4 &   -7.6 &  11.05\\
PAndAS-53        & 1 17 58.4 & +39 14 53 &  7.047 &   96.3 &   -9.1 &  44.38\\
PAndAS-54        & 1 18  0.1 & +39 16 59 &  7.041 &   96.2 &   -8.6 &  27.74\\
PAndAS-56        & 1 23  3.5 & +41 55 11 &  7.610 &  104.0 &   -7.6 &  11.57\\
PAndAS-57        & 1 27 47.5 & +40 40 47 &  8.582 &  117.3 &   -5.7 &   1.96\\
PAndAS-58        & 1 29  2.2 & +40 47  8 &  8.808 &  120.4 &   -6.2 &   3.01\\
\hline
NGC147:          &           &           &        &        &        &       \\
HodgeI           & 0 33 12.2 & +48 30 32 &  7.494 &  102.4 &   -7.4 &   9.36\\
HodgeII          & 0 33 13.6 & +48 28 49 &  7.465 &  102.0 &   -6.7 &   5.05\\
HodgeIII         & 0 33 15.2 & +48 27 23 &  7.440 &  101.7 &   -8.2 &  19.73\\
HodgeIV          & 0 33 15.0 & +48 32 10 &  7.519 &  102.7 &   -5.8 &   2.10\\
PA-N147-1        & 0 32 35.3 & +48 19 48 &  7.344 &  100.4 &   -7.8 &  13.16\\
PA-N147-2        & 0 33 43.3 & +48 38 45 &  7.609 &  104.0 &   -7.4 &   9.27\\
PA-N147-3        & 0 34 10.0 & +49  2 39 &  7.989 &  109.2 &   -6.9 &   5.80\\
SD-GC5           & 0 32 22.9 & +48 25 49 &  7.451 &  101.8 &   -6.7 &   4.73\\
SD-GC7           & 0 32 22.2 & +48 31 27 &  7.544 &  103.1 &   -7.8 &  13.40\\
SD-GC10          & 0 32 47.2 & +48 32 11 &  7.538 &  103.0 &   -4.9 &   0.97\\
\hline
NGC185:          &           &           &        &        &        &       \\
FJJI             & 0 38 42.7 & +48 18 40 &  7.134 &   97.5 &   -6.3 &   3.27\\
FJJII            & 0 38 48.1 & +48 18 16 &  7.125 &   97.4 &   -6.0 &   2.48\\
FJJIII           & 0 39  3.8 & +48 19 58 &  7.149 &   97.7 &   -8.0 &  15.82\\
FJJIV            & 0 39 12.2 & +48 22 48 &  7.195 &   98.3 &   -6.6 &   4.44\\
FJJV             & 0 39 13.4 & +48 23  5 &  7.199 &   98.4 &   -7.8 &  14.03\\
FJJVII           & 0 39 18.4 & +48 23  4 &  7.198 &   98.4 &   -5.8 &   2.24\\
FJJVIII          & 0 39 23.7 & +48 18 45 &  7.124 &   97.3 &   -6.9 &   6.01\\
PA-N185          & 0 38 18.8 & +48 22  4 &  7.198 &   98.4 &   -5.6 &   1.70\\
\hline
Andromeda I:     &           &           &        &        &        &       \\
AndI-GC1         & 0 45 42.9 & +38  1 54 &  3.273 &   44.7 &   -3.4 &   0.24\\
\hline
M33:             &           &           &        &        &        &       \\
M33-A            & 1 35 41.7 & +28 49 15 & 16.921 &  231.2 &   -5.7 &   2.03\\
M33-B            & 1 36  2.1 & +29 57 49 & 16.006 &  218.7 &   -7.0 &   6.72\\
M33-C            & 1 37 14.5 & +31  4 27 & 15.301 &  209.1 &   -6.4 &   3.87\\
M33-D            & 1 35  2.2 & +31 14 21 & 14.827 &  202.6 &   -3.6 &   0.29\\
M33-E            & 1 35 22.7 & +32  4 32 & 14.241 &  194.6 &   -4.9 &   0.97\\
M33-F            & 1 32 58.5 & +29 52  3 & 15.635 &  213.7 &   -6.0 &   2.67\\
\hline

\caption{M31 globular clusters with $D_{proj} > 2^\circ$, including globular clusters belonging to NGC147, NGC185, Andromeda~I and M33 systems.  Stellar masses are calculated assuming a stellar mass-to-light ratio of $\Upsilon_\star = 1.2\,M_\odot/L_\odot$.}
\label{gcs}
\end{longtable}

\begin{table*}
\begin{center}
\begin{tabular*}{0.9\textwidth}{lccp{0.25\textwidth}p{0.25\textwidth}}
\hline
Substructure   &   $M_V$&   $M_\star$ & Notes & References\\
\hline
Stream A & -11.1 & $2.8 \times 10^6$ & &\cite{ibata2007}\\
East Cloud & -10.7& $1.96 \times 10^6$ & & \cite{mcmonigal2016}\\
Stream C & -13.0 &$1.68 \times 10^7$ & &\cite{ibata2007}\\
Stream D & -12.6 & $1.14 \times 10^7$ & &\cite{ibata2007}\\
Giant Stream/\\
East, West Shelves & -- & $10^9M_\odot$ & Progenitor mass, based on dynamical modelling& \cite{fardal2006}\\
NE Structure & -12.8 &$1.35 \times 10^7$ & Original measurement in $g_{SDSS}$ &\cite{zucker2004b}\\
G1 Clump& -12.6 &$1.03 \times 10^7$ & Original measurement in INT $V'$ &\cite{ferguson2002}\\
SW Cloud & -11.3 & $7.1 \times 10^6$ & &\cite{mcmonigal2016, bate2014}\\
NW Stream - K1 & -10.5  & $9.4 \times 10^5$ & &This paper\\
NW Stream - K2 & -12.3 & $8.5 \times 10^6$& &This paper\\
NGC147 Stream&-12.2 & $6.5 \times 10^6$ &&This paper\\
M33 Stream&-12.7 & $1.23 \times 10^7$ &&\cite{mcconnachie2010b}\\
\hline
\end{tabular*}
\caption{Prominent halo substructures around M31. If no independent distance estimate exists, we assume they are at the same distance as the main body of M31. Stellar masses are calculated from luminosities assuming a stellar mass-to-light ratio of $\Upsilon_\star = 1.2\,M_\odot/L_\odot$. }
\label{substrlum}
\end{center}
\end{table*}

\end{document}